\documentclass[%
 reprint,
superscriptaddress,
nofootinbib,
 amsmath,amssymb,
 aps,
prc,
floatfix,
]{revtex4-2}
\usepackage[justification=raggedright,singlelinecheck=false]{caption}
\usepackage{graphicx}% Include figure files
\usepackage{dcolumn}% Align table columns on decimal point
\usepackage{bm}% bold math
\usepackage{tikz}
\usetikzlibrary{positioning, calc}
\usepackage{afterpage}
\usepackage{amssymb}
\usepackage{amsfonts}
\usepackage{mathtools}
\usepackage{enumitem} 
\newtheorem{definition}{Definition}[section]
\newtheorem{theorem}{Theorem}[section]
\newtheorem{proposition}[theorem]{Proposition}
\newtheorem{corollary}[theorem]{Corollary}

\usepackage[pdftex,colorlinks,linkcolor = blue,urlcolor  = blue,citecolor = blue]{hyperref}
\usepackage[capitalize]{cleveref}
\crefdefaultlabelformat{\textbf{#2}}
\crefname{equation}{Eq.}{Eqs.}
\crefname{section}{Sec.}{Secs.}
\crefname{figure}{Fig.}{Figs.}
\Crefname{figure}{Figure}{Figures} % capitalized form
\crefdefaultlabelformat{#2#1#3}
\usepackage{changes}
\usepackage{siunitx}
\usepackage[normalem]{ulem}
\usepackage{lineno}
\usepackage{csquotes}
\usepackage{placeins}
\usepackage[version=4]{mhchem}
\usepackage{xstring}
\usepackage{algorithm}
\usepackage{algpseudocode}
\usepackage[framemethod=TikZ]{mdframed}
\usepackage{xparse}
\usepackage{csquotes}
\usepackage{bbm}
\usepackage{overpic}
\definecolor{mseagreen}{RGB}{46, 139, 87}

\newcommand{\rom}[1]{\mathcal{#1}}
\newcommand{\rmu}{\rom{M}}
\newcommand{\rnu}{\rom{N}}
\newcommand{\reta}{\rom{H}}
\newcommand{\rtau}{\rom{T}}

\newcommand{\vecc}[1]{\boldsymbol{#1}}
\newcommand{\vmu}{\vecc{\mu}}
\newcommand{\vnu}{\vecc{\nu}}
\newcommand{\veta}{\vecc{\eta}}
\newcommand{\hnu}{\hat{\vnu}}
\newcommand{\heta}{\hat{\veta}}
\newcommand{\hmu}{\hat{\vmu}}
\newcommand{\vn}{\mathbf{y}}
\newcommand{\vtau}{\vecc{\tau}}
\newcommand{\htau}{\hat{\vtau}}
\newcommand{\sol}[1]{\hat{{#1}}_{\text{sol}}}

\newcommand{\snu}{\sol{\vecc{\nu}}}
\newcommand{\seta}{\sol{\vecc{\eta}}}
\newcommand{\vxi}{\vecc{\xi}}
\newcommand{\hxi}{\hat{\vxi}}
\newcommand{\con}{\xi}
\newcommand{\vcon}{\vxi}
\newcommand{\hcon}{\hat{\vcon}}
\newcommand{\vcons}[1][i]{\left\{\vcon_#1\right\}}
\newcommand{\hcons}[1][i]{\left\{\hcon_#1\right\}}

\newcommand{\hbeta}{\hat{\vecc{\beta}}}
\newcommand{\B}{\matt{B}}
\newcommand{\Ps}{\matt{P}}
\newcommand{\vbeta}{\vecc{\beta}}
\newcommand{\vpi}{\vecc{\pi}}
\newcommand{\sbeta}{\hat{\beta}_\text{sol}}
\newcommand{\matt}[1]{\mathbf{#1}}
\newcommand{\mc}[2][k]{{#2}^{*(#1)}}
\newcommand{\mcbox}[1]{\left\{\mc{#1}\right\}}
\newcommand{\poi}[1]{\sim \text{Poisson}\left(#1\right)}
\newcommand{\Eg}{E_{\gamma}}
\newcommand{\Ex}{E_\text{in}}
\newcommand{\vEx}{\vecc{\Ex}}

\newcommand{\Gg}{\matt{G}_{\gamma}}
\newcommand{\Gin}{\matt{G_{\text{in}}}}
\newcommand{\vfe}{\vecc{p}_{\text{f}}}
\newcommand{\vse}{\vecc{p}_{\text{s}}}
\newcommand{\vde}{\vecc{p}_{\text{d}}}
\newcommand{\vap}{\vecc{p}_{\text{a}}}
\newcommand{\mcb}{\matt{P}_{\text{c}}}
\newcommand{\mfe}{\matt{P}_{\text{f}}}
\newcommand{\mse}{\matt{P}_{\text{s}}}
\newcommand{\mde}{\matt{P}_{\text{d}}}
\newcommand{\map}{\matt{P}_{\text{a}}}
\newcommand{\mpeak}{\matt{P}_{\text{p}}}
\newcommand{\mcom}{\matt{P}_{\text{c}}}
\newcommand{\my}{{\matt{Y}}}

\newcommand{\Gk}{\matt{G}_\kappa}
\newcommand{\Gl}{\matt{G}_\lambda}
\newcommand{\nobold}[1]{%
  \ensuremath{\mathnormal{#1}}
}

\newcommand{\Rlk}{\matt{R}_{\lambda,\kappa}}

\newcommand{\retak}{\mathcal{H}_{\lambda,\kappa}}

\newcommand{\elemof}[2]{{\nobold{#1}}_{#2}}
\newcommand{\usflag}{
\begin{tikzpicture}[scale=0.02]
    \foreach \i in {0,2,4,6,8,10,12} {
        \fill[red] (0,\i) rectangle (19,\i+1);
    }
    
    \fill[blue] (0,7) rectangle (7.6,13);
    
    \foreach \i in {0,...,5} {
        \foreach \j in {0,...,8} {
            \ifthenelse{\isodd{\i}}{
                \fill[white] (0.85+\j*0.75,12.25-\i*0.85) circle (0.15);
            }{
                \fill[white] (0.475+\j*0.75,12.25-\i*0.85) circle (0.15);
            }
        }
    }
\end{tikzpicture}
}

\newenvironment{notationsummary}{%
    \begingroup
    \small
    \begin{mdframed}[
        linewidth=0.2pt,
        topline=true,
        bottomline=true,
        leftline=true,
        rightline=true,
        backgroundcolor=gray!0,
        innertopmargin=5pt,
        innerbottommargin=5pt,
        innerleftmargin=10pt,
        innerrightmargin=10pt,
        roundcorner=5pt,          % This creates the rounded corners
        skipabove=0.5em,         % Space before the box
        skipbelow=0.5em          % Space after the box
    ]
    \textbf{Notation Summary:}\\
}{%
    \end{mdframed}
    \endgroup
}

\NewDocumentCommand{\notation}{m m}{%
    \(\begin{aligned}[t] #1 \end{aligned}\) & #2 \\
}

\NewDocumentCommand{\NotationBlock}{+m}{%
    \begin{notationsummary}
    \begin{tabular}{@{}>{\raggedright\arraybackslash}p{0.2\linewidth}@{~}>{\raggedright\arraybackslash}p{0.8\linewidth}@{}}
    #1
    \end{tabular}
    \end{notationsummary}
    \vspace{1em}
}

\begin{document}

\preprint{APS/123-QED}

\title{Regularized Unfolding of gamma-ray Spectra for Nuclear Physics Applications}
%\author{Authors}
%\affiliation{Department of Physics, University of Oslo, N-0316 Oslo, Norway}

\author{E.~Lima}
\affiliation{Department of Physics, University of Oslo, N-0316 Oslo, Norway}
\affiliation{Norwegian Nuclear Research Center, Norway}
\email{erlend.lima@fys.uio.no; a.c.larsen@fys.uio.no}

\author{L.~L.~Braseth}
\affiliation{Department of Physics, University of Oslo, N-0316 Oslo, Norway}
\email{ lasselb@fys.uio.no }

\author{A.~H.~Mj{\o}s}
\affiliation{Department of Physics, University of Oslo, N-0316 Oslo, Norway}
\affiliation{Norwegian Nuclear Research Center, Norway}
%\email{a.h.mjos@fys.uio.no}

\author{M.~Hjorth-Jensen}
\affiliation{Department of Physics, University of Oslo, N-0316 Oslo, Norway}
\affiliation{Center for Computing in Science Education, University of Oslo, N-0316 Oslo, Norway}
%\email{morten.hjorth-jensen@fys.uio.no}

\author{A.~Kvellestad}
\affiliation{Department of Physics, University of Oslo, N-0316 Oslo, Norway}
\email{ anders.kvellestad@fys.uio.no }

\author{A.~C.~Larsen}
\affiliation{Department of Physics, University of Oslo, N-0316 Oslo, Norway}
\affiliation{Norwegian Nuclear Research Center, Norway}
\email{a.c.larsen@fys.uio.no}

\date{\today}% It is always \today, today,

\begin{abstract}
Reconstructing gamma-ray spectra from detector measurements is an ill-posed
inverse problem. Standard methods, such as Folding Iteration with Compton
Subtraction (FICS), provide point estimates but lack calibrated uncertainties and
may bias the spectrum. We introduce an unfolding framework based on regularized
maximum-likelihood estimation (RMLE) that enforces non-negativity and
detector-response constraints while explicitly modeling background and
contaminant contributions. Simulations and analytical results show that RMLE
yields smoother reconstructions with well-calibrated confidence intervals and
outperforms existing techniques for low-complexity spectra. Although high-complexity
data remain challenging, the intervals produced by RMLE maintain correct
coverage.
\end{abstract}

\maketitle

%\tableofcontents

%%%%%%%%%%%%%%%%%%%%%%%%%%%%%%%%%%%%%%%%%%%%%%%%%%%%%%%%%%%%%%%%%%

%START SECTION
\section{Introduction}
\label{sec:intro}
% Cecilie: I started writing something in the intro. Please feel free to modify and/or rewrite!

%In particle physics experiments,
%the distortion is usually comparatively simple and is due to smearing because of the
%finite detector resolution, limited acceptance and reduced efficiency, and
%non-linear detector response (see, e.g.,
%Refs.~\cite{blobel2002unfolding,Cowan:2002in}). 

%\anders{Småplukk: Hva er policy på plassering av fotnoter? Jeg er vant til at fotnotesymbolet (når det er på slutten av en setning) plasseres rett etter punktum, heller enn rett før.}
%\cecilie{Dette vil vanligvis Phys. Rev. C fikse selv i proof-runden.}

%\anders{Småplukk 2: Jeg synes vi litt for ofte starter nye avsnitt. Jeg har slått sammen avsnitt noen steder, men dere må gjerne slå sammen noen flere dersom dere er enige.}

%\anders{Småplukk 3: Kan vi bruke forkortelsen ``Figs.'' når vi henviser til flere figurer, eller er det tidsskriftets stil som gjør at vi alltid bruker ``Fig.''?}

%\cecilie{Phys. Rev. C bruker forkortelsen ``figs.'' for flere figurer, men om det er i starten av setningen så tror jeg at ordet skal skrives ut (altså ``Figures'').}

% Point: Introduce the problem
The accurate reconstruction of gamma-ray spectra from detector measurements is a
fundamental challenge in nuclear physics experiments. 
The intrinsic physical processes of gamma-ray
detection --- notably Compton scattering, pair production, and backscattering --- transform the true spectrum.
Recovering the original spectrum from these transformed measurements constitutes
an ill-posed inverse Poisson problem that demands statistical treatment.

% Point: Contextualize the problem(s)
Inverse Poisson problems of similar complexity are encountered in various
scientific fields, such as medical imaging, astronomy, and particle physics,
where advanced methods have been developed. Techniques such as
expectation-maximization in PET imaging, Richardson-Lucy deconvolution in
astronomy, and singular value decomposition in particle physics have proven effective in their
respective domains. However, these approaches fall short when applied
to nuclear physics because of three key challenges. First, unlike the predominantly
geometric or smooth response functions in other fields, nuclear detector
responses are highly complex, combining multiple physical interaction mechanisms
that create correlations across the spectrum. Second, the measured spectra in
nuclear experiments are highly diverse, ranging from sharp discrete peaks to
broad continuous distributions, often in the same spectrum. Finally, the
optimization processes often involve millions of parameters, rendering many
standard algorithms computationally infeasible.

%\anders{Do we need the sentence about particle physics? It may seem a bit arbitrary or out of place to most readers, since this paragraph now has the structure "nuclear physics, particle physics, back to nuclear physics", instead of just focusing on our topic of nuclear physics.}
%\cecilie{I agree, and commented out the particle-physics stuff.}
%$\mathcal{M} \mathcal{N}$

% Point: Specify the problem in nuclear physics
The gamma-ray spectra of primary interest in this work are those used as input data for
the \textit{Oslo Method}~\cite{GUTTORMSEN1996371,GUTTORMSEN1987518,SCHILLER2000498}. 
The goal of the Oslo Method is to experimentally extract nuclear
level densities (NLD) $\rho$ and gamma-ray transmission coefficients $\mathcal{T}$~\cite{Rekstad1983}. %\erlend{cite rekstad}. %Cecilie: added the ref!
This is achieved
by simultaneously measuring the gamma-ray energies $\Eg$ and the initial excitation
energy $\Ex$ of gamma cascades of the nucleus under investigation. The first step in the data
analysis involves correcting for the detector response, i.e.,\ \textit{unfolding} the spectrum.

% Point: Introduce previous methods
To date, this unfolding has been performed using an iterative technique called
\textit{Folding Iteration with Compton Subtraction}
(FICS)~\cite{GUTTORMSEN1996371}. This approach is a variant of Richardson's
method\cite{richardson1911finite} and incorporates two regularization strategies: early stopping
and the Compton subtraction method. However, FICS has several limitations. It
produces only a point estimate for each bin in the unfolded spectrum without
quantifying uncertainties. Furthermore, the method is prone to overfitting
statistical noise and can introduce artificial structures into the reconstructed
spectrum.

In the original software for decomposing $\rho$ and $\mathcal{T}$, Schiller
\textit{et al.}~\cite{SCHILLER2000498} estimated the unfolding uncertainty as
the ratio between the detector array’s solid-angle coverage and its full-energy
peak efficiency at a gamma-ray energy of 1.33~MeV~\cite{CACTUS1990}. Schiller
\textit{et al.} nevertheless cautioned that this estimate is \enquote{quite uncertain}.

Midtb{\o} \textit{et al.}~\cite{MIDTBO2021107795} developed a software
implementation of the Oslo Method to estimate total statistical uncertainties
using Monte Carlo simulations. In this approach, the data is resampled under the
assumption of a Poisson distribution and processed through the standard Oslo
Method. Confidence intervals are then derived from the mean and standard
deviation of the resulting ensemble. However, this procedure relies on two
critical assumptions: that the FICS unfolding procedure is unbiased and that the resulting ensemble is approximately Gaussian.
As we will show, FICS is biased, and the Monte
Carlo ensemble exhibits non-negligible higher-order moments, rendering
confidence intervals based solely on the mean and standard deviation unreliable.

A fundamental property of the unfolding problem is that it is not identifiable.
The detector response folds the underlying spectrum in a way that renders the
inverse problem non-unique. Consequently, infinitely many mathematically
distinct spectra are consistent with the same measured data. Any unfolding
method must therefore select one solution among many, implicitly or explicitly
through its regularization. Failure to recognize this can lead to
over-interpretation of noise or regularization artifacts as genuine spectral
structure.

% Point: Explain what we develop
In this article, we present a novel approach to the unfolding problem. Our
method provides well-motivated and transparent uncertainty estimates for
individual bins in gamma-ray spectra, offering unbiased results with
confidence intervals calibrated to their stated confidence levels. 

%While we developed and validated this method specifically for
%scintillator-type gamma-ray detectors, its versatility allows for adaptation to
%similar detectors.

% Point: Explain how the method works and its benefits

Our approach leverages regularized maximum likelihood estimation (RMLE) to
recover the expectation value of the unfolded spectrum under physical
constraints, rather than seeking arbitrary mathematical solutions that merely
satisfy the folding relations. By prioritizing physically meaningful solutions,
the method yields spectra that are robust against statistical noise. Its
flexibility allows for principled statistical treatment of both
background spectra and contaminants. Importantly, this robustness comes without
computational cost: the method efficiently unfolds spectra with millions of bins
in seconds on a standard desktop GPU.

% Point: Explain the drawbacks
Our analysis demonstrates that the spectral complexity of the input spectrum
significantly influences the unfolding process. We establish a mathematical
framework to classify spectra as either low- or high-complexity. For low-complexity
spectra, our regularization approach yields reliable results with expected
coverage levels. Overall, the method produces smoother spectra and achieves more
accurate confidence interval coverage than the ensemble method of
Midtb\o~\cite{MIDTBO2021107795}. While high-complexity spectra pose fundamental
challenges to regularization that our current methodology cannot fully address,
they can still be unfolded in practice, though users should expect greater
uncertainty and interpret potential spurious peaks with caution. The treatment
of high-complexity spectra remains an open area for future research.

% Point: Paper roadmap
The article is organized as follows. In Sec.~\ref{sec:unfprob}, we present a
theoretical framework for the unfolding problem. The detector response functions
are discussed in Sec.~\ref{sec:response}, followed by theoretical and numerical
implications of the ill-posedness and strategies for addressing it in
Sec.~\ref{sec:unfoldingspaces}. Our unfolding methodology is described in
Sec.~\ref{sec:modelest}, and the uncertainty quantification procedure is
detailed in Sec.~\ref{sec:uncquant}. Results on simulated data are presented in
Sec.~\ref{sec:sims}. A comparison to FICS is given in Sec.~\ref{sec:compfics}.
Finally, Sec.~\ref{sec:sumout} provides a summary and outlook.

The extensive theoretical treatment in these sections addresses a critical gap
in the field, as many of the relevant mathematical and physical principles have
previously been scattered across disparate papers in nuclear physics and other
domains. By consolidating them into a unified framework, we aim to provide a
solid foundation for future research and make the concepts more accessible to
practitioners.

\Cref{app:math} provides relevant mathematical background, while \cref{app:smoothness} gives a mathematical description of fluctuations in gamma-ray spectra. Convergence conditions are presented in \cref{sec:convergence}. \Cref{app:margsimci} outlines the theory behind marginal and simultaneous confidence intervals, and \cref{sec:appcoverage} discusses the mathematical basis of coverage probability. In~\cref{app:exp}, we explain the Oslo Method and introduce typical experimental measurements using various reactions and detector types. \Cref{sec:regularization_systematics} presents systematic results on regularization selection. Finally, the FICS method is derived and discussed in~\cref{app:fics}.

This paper presents a complete theoretical framework from fundamentals through
implementation. Recognizing that readers may have different needs, we suggest
the following reading paths:

\begin{enumerate}
    \item For practitioners mainly interested in implementing the method: Secs. \ref{sec:modelest} to \ref{sec:sims} provide a nearly self-contained description
    of the algorithm, implementation, and numerical results. 
    \item For those interested in method derivation without full theoretical foundations,
    Secs. \ref{sec:response} to \ref{sec:unfoldingspaces} develop the method for Oslo-type data.
    \item Readers interested in the complete theoretical background should start with~\cref{sec:unfprob}.
\end{enumerate}

%%%%%%%%%%%%%%%%%%%%%%%%%%%%%%%%%%%%%%%%%%%%%%%%%%%%%%%%%%%%%%%%%%
%START SECTION
\section{The unfolding problem}
\label{sec:unfprob}
%[\lasse{This is just a first draft.}]

\subsection{The Poisson distribution and counting experiments}\label{sec:def of unfolding}

% \anders{In this subsection we discuss \textit{intensity functions}, which is never used elsewhere in the paper. Should we add a short paragraph (or large footnote?) that explains how the intensity function of a Poisson process is connected to the Poisson probability distribution? Or, could we potentially rephrase this subsection in terms of probability distributions directly, rather than introduce the concept of intensity functions?}
%To justify this model, it is useful to first consider inferences made within small regions where
%individual events occur, and then generalize to account for the effects of an
%imperfect detector over larger regions. To this end, a review of some basic
%properties of the Poisson distribution is necessary. Broadly speaking, whenever
%one counts the occurrence of events within a given region, and the average rate
%of occurrence is constant over that region, the Poisson distribution provides a
%natural model.

Experiments on atomic nuclei that produce gamma-ray emissions are commonly modeled as counting processes. These are mathematically described by Poisson point processes, which provide a natural framework for capturing the stochastic nature of event detection.

To develop a rigorous foundation for this model, we adopt a measure-theoretical
framework where regions of observation are treated as
\textit{measurable}\footnote{The term \textit{measure} used here refers to the
mathematical concept in measure theory, specifically the Lebesgue measure, which provides a systematic way to
assign \textquote{sizes} to sets, generalizing concepts like length, area, and volume. It
should not be confused with physical measurements of energy or other detector
observables.} subsets of the event space. We begin by examining individual
events at a point-level, then extend the analysis to account for detector \textit{smearing}\footnote{In this work, \textit{smearing} refers to any process that redistributes or spreads out true values into observed measurements due to detector effects or measurement uncertainties. When specifically referring to smearing by a Gaussian distribution, we will explicitly use the term \textit{Gaussian smearing}.} of the signal. When the average rate of occurrence remains constant, the Poisson distribution emerges as the natural model.

The Poisson distribution is defined as follows: Let $Y$ be a random variable representing the number of events occurring in a fixed region. 
\begin{definition}[Poisson Distribution]
A random variable $Y$ is said to follow a Poisson distribution with parameter 
$\nu>0$ if the probability of observing exactly 
$y$ events is given by:
\begin{align}
    \emph{P}(Y=y\mid\nu)=\frac{\nu^{y}}{y!}e^{-\nu}\,.
\end{align}
The Poisson distribution has the following properties:
\begin{align}
    \mathbb{E}[Y]&=\nu\,,
    \\
    \mathbb{V}[Y]&=\nu\,,
\end{align}
i.e., the mean and variance are given by the parameter $\nu$, and for such $Y$ we write
\begin{align}
    Y&\sim \emph{Poisson}(\nu)\,.
\end{align}
\end{definition}

A natural question arises: Given that $Y\sim \text{Poisson}(\nu)$, what can be said about the parameter $\nu$ from the observation $y$? In this work, we will base our estimation on the maximum likelihood property:
\begin{definition}[Maximum Likelihood]
    For a random variable $Y\sim \emph{Poisson}(\nu)$, a maximum likelihood estimator (\textnormal{MLE}) for $\nu$, based on an observed count $y$, is the value $\hat{\nu}$ that maximizes the likelihood function
    \begin{align}
        \mathcal{L}(\nu)=\frac{\nu^{y}}{y!}e^{-\nu}\,,
    \end{align}
    satisfying the condition for any alternative $\tilde{\nu}\geq 0$
    \begin{align}
        P(Y=y\,|\,\hat{\nu})\geq P(Y=y\,|\,\tilde{\nu})\,.
    \end{align}
\end{definition}
In other words, an MLE $\hat{\nu}$ maximizes the probability to observe $y$, and a straightforward calculation yields that the MLE for $\nu$ is given by
\begin{align}
    \hat{\nu}=y\,,
\end{align}
which is an unbiased estimator, $\mathbb{E}[\hat{\nu}\,|\,\nu]=\nu$. This is known as \emph{inference for direct observations}.

However, in real experiments, the detector will distort the underlying process.
As a result, the observed data is a smeared representation of the
truth-level\footnote{The term \textit{truth-level} refers to the actual physical quantities or events before they are affected by detector resolution, efficiency, or other experimental effects.} counts. In such cases, we deal with \emph{inference for
indirect observations}, where---in statistical terms---an unbiased solution can result in an unacceptable
large variance~\cite{Cavalier2008NonparametricSI}. This complication
arises because the smearing transformation induced by the detector blur or mix the truth-level
counts, making direct inversion unstable. The inference process then becomes more complex, and a closed form solution of the maximum likelihood estimator typically does not exist.

In the following, we wish to generalize
from the discrete Poisson distribution to the Poisson point process in a
continuous setting. Practical limitations ultimately force us to consider a
discretized framework—for instance, by binning the energy spectrum into finite
intervals. Nonetheless, our aim is to abstract away from any specific
discretization so that the problem can be framed more generally.

\subsection{Poisson point processes}
We aim to model the detection of emitted gamma rays as a counting process, where the primary goal is to analyze how many events occur in specific energy regions. The appropriate mathematical tool for describing these types of processes is a point process. Essentially, a point process is a random mechanism that generates events within a continuous space.

Consider a state space $E\subseteq\mathbb{R}$, which represents the possible values for a physical observable of interest, e.g., energy. A point measure is a mathematical object
that counts the number of events  within a given set. For example, if we detect
gamma rays at specific energies, the point measure tells us how many gamma rays
are detected in a given energy range. To make this precise we need the following input:
\begin{itemize}
    \item On a set $E$, a $\sigma$-algebra is a nonempty collection of subsets of $E$ closed under complement, countable unions and countable intersections.
    \item A Borel $\sigma$-algebra on $E$, denoted $\mathcal{B}(E)$, is the smallest $\sigma$-algebra containing all open subsets of $E$. 
    \item For a Borel set $B\in\mathcal{B}(E)$ and $x\in E$, a Dirac measure $\delta_{x}$ is defined by
    \begin{align}
        \delta_x(B)=
      \begin{cases}
        1, & x\in B,\\
        0, & x\notin B\,.
      \end{cases}
    \end{align}
\end{itemize}
Then, we can define:
\begin{definition}[Point Measure]
     A point measure $\chi$ on $E$ is defined as
     % \footnote{A Borel set \( B \subseteq E \), where \( E \subseteq \mathbb{R} \), is a set that belongs to the Borel \( \sigma \)-algebra on \( E \). The Borel \( \sigma \)-algebra is the collection of all sets that can be formed from open subsets of \( E \) through countable operations, such as taking complements, countable unions, and countable intersections.
% }
    \begin{align}
        \chi(B)=\sum_{i\in I}\delta_{x_i}(B)\,,\hspace{0.2cm}B\in\mathcal{B}(E)\,,
    \end{align}
    where $I$ is a finite (or countable) index set and $\delta_{x_i}$ is the Dirac measure centered at the point $x_i\in E$, which counts whether the point $x_i$ is in the set $B$.
\end{definition}

A \emph{point process} $G$ is simply a random point measure, meaning that for each $B\in \mathcal{B}(E)$, the value $G(B)$ is a random integer that counts the number of points in $B$. For our purposes, the point process describes the random locations of gamma-ray detection events, and the distribution of these events can vary depending on the underlying physical process.

The Poisson point process is a special type of point process that arises naturally in many experimental settings. It has two key properties:
\begin{enumerate}
    \item For any Borel set $B$, the number of events $G(B)$ is a Poisson-distributed random variable with mean measure $\nu(B)$
    \begin{align}
        G(B)\sim\textnormal{Poisson}(\nu(B))\,.
    \end{align}
    \item The number of events in disjoint regions are independent.
\end{enumerate}
The mean measure $\nu(B)$ represents the expected number of points $\nu(B)=\mathbb{E}[G(B)]$ in the region
$B$, and is referred to as the mean measure of the process. This makes the Poisson process a good model for gamma-ray detection, where events occur independently, and the number of events in each region can be modeled using the Poisson distribution.

A key feature of the Poisson point process is the \emph{intensity function} $f(x)$, which describes the expected rate of events at each point $x \in E$. In a small region around $x$, the expected number of events is approximately $f(x)\,dx$. Thus, the intensity function plays a role analogous to the parameter $\nu$ in the standard Poisson distribution, but with the added flexibility of varying across space, allowing for the modeling of non-uniform event distributions. The total expected number of events in a bounded Borel (measurable) subset $B$ is given by
\begin{align}
    \nu(B) = \mathbb{E}[G(B)] = \int_{B} f(x)\,dx\,,
\end{align}
where $G(B)$ denotes the number of events occurring in $B$.

To illustrate this, consider a one-dimensional interval $[a,b] \subset \mathbb{R}$. If the intensity function is constant, $f(x) = \nu$, the expected number of events becomes \mbox{$\mathbb{E}[G([a,b])] = \nu(b-a)$}. In this case, $\nu$ represents the constant event rate per unit length, and the total expected number of events is simply the product of the rate and the length of the region. This is known as a \emph{homogeneous Poisson process}. When the intensity function varies with $x$, the process is referred to as an \emph{inhomogeneous Poisson process}.

\subsection{Poisson inverse problem}\label{sec:poisson inv prob}
In real experiments we do not observe the true intensity function directly. Instead, we observe a smeared version of the process due to detector imperfections, where the observed intensity $g$ is a blurred version of the truth-level intensity $f$. Performing inference in this scenario is referred to as a \emph{Poisson inverse problem}, and
generally falls into the class of \emph{statistical ill-posed problems} ~\cite{Antoniadis_2006}.

To model this, let us consider two Poisson processes $F$ and $G$. Let $F$
denote the truth-level spectrum of events with $D\subseteq\mathbb{R}$ as state space,
and let $G$ denote the smeared spectrum of events with $E\subseteq\mathbb{R}$ as
state space. We assume that both $E$ and $D$ are compact\footnote{Compact means bounded and closed, e.g., for $a,b\in\mathbb{R}$ the interval $[a,b]$
is compact. All closed sets in $\mathbb{R}$ are Borel sets because the Borel $\sigma$-algebra includes all closed (and open) sets. Thus, every compact interval is a Borel set.} intervals and we denote by $\mathcal{X}$ and $\mathcal{Y}$ spaces of
regular\footnote{Here, regular means sufficiently smooth functions,
i.e., a function that has derivatives of sufficient order at each point in its
domain.} functions on $D$ and $E$, respectively. Consider $f\in
\mathcal{X}$ and $g\in \mathcal{Y}$ intensity functions of $F$ and $G$. The
intensity functions can then be related by a bounded\footnote{The distinction between bounded and unbounded operators is theoretical. In practice, especially in numerical computations, whether or not an operator is theoretically bounded is less critical than ensuring that the numerical methods are stable and the model is well-posed.} linear operator
$\mathcal{R}:\mathcal{X}\rightarrow \mathcal{Y}$, giving the operator equation
\begin{align}\label{eq:operator equation}
    \mathcal{R}(f)=g\,.
\end{align}

Here, $\mathcal{R}$ is a linear operator whose \emph{integration
kernel} represents the response of a measuring device, e.g., the detector response. The detector response will act as  a smoothing operation on the truth-level spectrum, resulting in an observed smeared spectrum. This may be modeled by assuming that the intensity functions are related by a Fredholm equation of the first kind,
\begin{align}\label{eq:integral equation}
    g(y)=\big(\mathcal{R}f\big)(y)=\int_{D}R(y,x)f(x)dx\,,
\end{align}
where $R$ is a integration kernel\footnote{An integration kernel \(R(y,x)\) that makes Eq.~\eqref{eq:integral equation} meaningful for all \(f \in \mathcal{X}\) should be measurable in \((y,x)\) with integrability conditions ensuring \((\mathcal{R}f)(y)=\int_{D} R(y,x)\,f(x)\,dx\) is well-defined and measurable for each \(y \in E\) (for instance, this holds if \(R \in L^{2}(E\times D)\) with \(\mathcal{X}=\mathcal{Y}=L^{2}\)). These assumptions (together with boundedness and linearity of \(\mathcal{R}\)) specify a valid forward model \(g=\mathcal{R}f\); they do not, by themselves, imply that the inverse problem stated below is well-posed.
} such that the integral is meaningful and the forward model is well-defined.

% An idealized setting without smearing, $\mathcal{R}=I$, is what we have referred to as inference for direct observations. Making inferences for non-trivial $\mathcal{R}$ corresponds to inference for indirect observations. In practice, there is usually some uncertainty connected to the detector response, but in this work we treat it as known.
% \footnote{In this context, "well-defined" means that the integration kernel $R$ is measurable and satisfies the necessary conditions (such as continuity and integrability) to ensure that the integral in equation \eqref{eq:integral equation} exists for all functions $f \in \mathcal{X}$. This ensures that the operator $\mathcal{R}$ properly maps functions from $\mathcal{X}$ to $\mathcal{Y}$ without ambiguity.}
An inverse problem seeks to infer an unknown quantity from an indirect observation. This reconstruction task is often ill-posed; the stable solvability of a problem is given by:

% Recall, in an idealized setting without smearing, $\mathcal{R}=I$, we called this inference for direct observations. Making inferences for non-trivial $\mathcal{R}$ we called inference for indirect observations.
\begin{definition}[Hadamard Criteria]\label{def:hadamard} Let $f \in \mathcal{X}$ and $g\in\mathcal{Y}$. A problem is well-posed if the following conditions hold:
\begin{enumerate}[label=(\roman*)]
    \item Existence: $g$ is in the \emph{range} of the operator $\mathcal{R}$, i.e., there exists a solution $f$ to 
    $\mathcal{R}(f) = g$.
    \item Uniqueness: The solution $f$ is unique, implying that if $\mathcal{R}(f) = \mathcal{R}(f')$ then $f'=f$. 
    \item Stability: The solution $f$ is a continuous function of $g$. Essentially, this means that a small change in $\mathcal{Y}$ leads to small changes in $\mathcal{X}$.
\end{enumerate}
\end{definition}
If any of these conditions are not met, the problem is deemed ill-posed. Most practical applications are in fact ill-posed problems. 

Let us formulate this specifically for the problem at hand: The unfolding problem is to
make inferences about the truth-level intensity $f$ given a single observation\footnote{By single observation we mean that the experiment yields one spectrum.} of the Poisson
process $G$. In other words, the noise is modeled by regarding $g$ as a parameter of an underlying statistical Poisson model.
Given the stochastic nature of Poisson
processes, we do not have access to an exact $g$, but can only construct an estimate $\hat{g}$ based on the observed data and try to solve ~\cref{eq:operator equation} in an approximate sense. If we denote this approximate solution by $\hat{f}$, then it is possible that; $(i)$ $\hat{g}$ is not in the range of $\mathcal{R}$ and $\hat{f}$ is not an exact solution, $(ii)$ candidate solutions $\hat{f}$ map to the same $\hat{g}$, and $(iii)$ $\hat{f}$ is not a continuous function of $\hat{g}$. In addition,
even though the theoretical detector response is well-behaved, its practical implementation (e.g., discretization) does not necessarily carry this property. These practical limitations will most surely transform any well-behaved theoretical response to an ill-conditioned one, and with noise in the
model, small perturbations in data-space may yield drastically different
solutions in solution-space, and should therefore be regarded as unreliable.

In order to satisfy Hadamard's well-posedness conditions, one has to turn to regularization methods. Regularization methods---in the linear deterministic case---approach the ill-posedness in two different ways:
\begin{enumerate}[label=(\arabic*)]
        \item \textbf{Implicit Regularization:} This approach is based on so-called iterative methods, which start with an initial estimate of the solution and refine this estimate through a series of iterations. Each iteration aims to reduce the discrepancy between the observed data and the model's predictions, incorporating regularization implicitly through the iteration process itself and the application of a stopping criterion.
        \item \textbf{Explicit Regularization:} This method proceeds in two conceptual steps:
        \begin{enumerate}[label=(\roman*)]
            \item \textbf{Generalized Inverses:} For a non-injective detector response, the naive inverse is not well-defined. To address multiple solutions, one replace inversion by a selection rule among data-consistent solutions (fixed by geometry and possible constraints). A standard approach introduce a generalized inverse, $\mathcal{R}^{\dagger}$, and uses spectral/SVD projection to pick a canonical representative. For least squares, the Moore-Penrose pseudoinverse is one such selection (minimum norm in a Hilbert geometry) but is not stable in ill-posed settings.
            
            % for every $g\in\textnormal{ran}(\mathcal{R})$, $f^{\dagger}=\mathcal{R}^{\dagger}g$ is the unique mimimum-norm solution of the inverse problem $\mathcal{R}f=g;$ for $g$ outside the range one sets $\mathcal{R}^{\dagger}g=0$.
            
            \item \textbf{Parametric Regularization:} To address the problem of instability, a family of continuous operators $\{\mathcal{R}_{\alpha}\}$ is introduced, such that for a regularization parameter $\alpha > 0$ the operators $\mathcal{R}_{\alpha}$ converge (in an appropriate topology) to the generalized inverse in the limit $\alpha \rightarrow 0$.
        \end{enumerate}
\end{enumerate}
Ill-posed inverse problems are further compounded if the underlying model is probabilistic. This forces one to adapt the regularization methodology to a statistical setting, which may be classified as:
% In most cases of physical interest, this leads to a non-linear constrained optimization problem.
\begin{enumerate}[label=(\arabic*)]
    \item \textbf{Frequentist Unfolding:}
    Treats unfolding as a point-estimation problem: choose a solution that best fits the data (e.g., via maximum likelihood or loss minimization) under physical constraints, while stabilizing the inversion through regularization. Regularization may be \emph{explicit}---augmenting the objective with penalties that encode smoothness or shape---or \emph{implicit}---through early stopping of iterative schemes. The regularization strength or stopping rule should be selected by data-driven rules (e.g. discrepancy principle, L-curve), and uncertainties can be quantified with frequentist tools such as bootstrap.

    \item \textbf{Bayesian Unfolding:}
    
    This approach frames the unfolding problem in a purely probabilistic manner. Bayesian unfolding incorporates prior information about the solution and combines it with the observed data in a principled way:
    \begin{enumerate}[label=(\arabic*)]
    \item \textbf{Prior Knowledge:} To regularize the ill-posed problem, one introduces a prior probability distribution over the space of possible solutions. This prior encodes any a priori knowledge about the expected smoothness, shape, or other properties of the solution. Priors that favor smooth or constrained solutions regularize the problem by penalizing unlikely solutions, thus addressing both non-uniqueness and instability.
    
    \item \textbf{Posterior Distribution:} The observed data is modeled using a likelihood function, which accounts for the measurement process and the uncertainty in the observations. The posterior distribution is then obtained by applying Bayes' theorem,
    where the posterior reflects the updated belief about the solutions after observing the data.
    \end{enumerate}
\end{enumerate}
% \item (Regularization via Prior) The choice of the prior distribution plays the role of regularization. Priors that favor smooth or constrained solutions implicitly regularize the problem by penalizing unlikely solutions, thus addressing both non-uniqueness and instability.
    
    % \item (Marginalization or MAP Estimation) In practice, either the maximum a posteriori (MAP) estimate, which maximizes the posterior distribution, or the full posterior distribution may be used. In the case of MAP estimation, the problem becomes one of optimizing the posterior, which combines both the data fidelity (from the likelihood) and the regularization (from the prior).

In this work, the unfolding problem is approached using a frequentist
methodology, with an emphasis on explicit regularization to obtain acceptable
estimators. The most commonly used method for unfolding Oslo gamma-ray spectra,
FICS, relies on implicit regularization.\footnote{A detailed exposition of this
method is provided in~\cref{app:fics}.}
The case of Bayesian unfolding will be addressed
in a future work~\cite{mjos2025prep}.

As we will see in~\cref{sec:response}, trying to model the detector response as a pure smoothing operation will fail. That is, there are underlying nuclear processes that prevent the response from being a pure convolution operator, and standard deconvolution techniques are unsuitable.

% However, we can use this fact to our advantage in the regularization procedure. The specific choice of regularization in this work will be addressed in [\citations{ref to section on constructed regularized functional}].

Although the functional version of the unfolding problem is the most general, practical application necessitates a form that is amenable to computation. Therefore, we will next discretize the problem using histograms and discuss its ill-posedness in more detail.

\subsection{Discretization and ill-posedness}\label{sec:discretization}
The data in gamma-ray counting experiments are usually analyzed in binned form.\footnote{This is the most practical way to store the experimental data, but note that there is nothing fundamental about this choice. Also, the truth-level spectrum does not need to be restricted to histograms, e.g., another popular choice is splines. Nevertheless, for practical purposes, this is the choice we make.} Thus, Poisson problems are predominantly examined in a format where both the observable process and the unobservable process are made discrete using histograms.

Recall, we consider two distinct Poisson processes, $F$ and $G$, with intensity functions $f$ and $g$, respectively.
To discretize the smeared process $G$, we let $\{E_i\}_{i=1}^{n}$ be a partition of the smeared space $E$ into $n$ ordered intervals (energy bins). Similarly, for the truth-level process $F$, we let $\{D_i\}_{i=1}^{n}$ be a partition of the truth-level space $D$. Furthermore, we let $y_i$ denote the number of points falling within interval $E_i$, denoted $y_i = G(E_i)$, resulting in a random vector
\begin{align}
    \mathbf{y}=[G(E_1),\dots,G(E_n)]=[y_1,\dots,y_n]\,,
\end{align}
i.e., the $y_i$'s represent independent and Poisson distributed event counts. Then, we consider mean measures $\mu(D_i)$ and $\nu(E_i)$, such that we may construct mean vectors
\begin{align}
    \boldsymbol{\nu}&=[\nu(E_1),\dots,\nu(E_n)]=\Big[\int_{E_1}g(y)dy,\dots,\int_{E_n}g(y)dy\Big]
    \\
    \boldsymbol{\mu}&=[\mu(D_1),\dots,\mu(D_n)]=\Big[\int_{D_1}f(x)dx,\dots,\int_{D_n}f(x)dx\Big]\,,
\end{align}
where $\boldsymbol{\nu}\in\mathbb{R}_{+}^{n}$ represents the mean of the smeared histogram $\mathbf{y}$, and $\boldsymbol{\mu}\in\mathbb{R}_{+}^{n}$ represents the mean of some unobservable truth-level spectrum
\begin{align}
    \mathbf{x}=[F(D_1),\dots ,F(D_n)]\,.
\end{align}
We demand these vectors to belong to the non-negative orthant $\mathbb{R}_{+}=\{x\in\mathbb{R}\,|\,x\geq 0\}$ because negative counts are not physically possible.

% Also, for a constant rate, we observe that the mean vectors represents discrete approximations of the intensity functions.

The discretized unfolding problem then takes the following form: Given an observed smeared spectrum $\mathbf{y}$ and the model
\begin{align}
    \mathbf{y}\sim\text{Poisson } \Big(\mathbf{R}\boldsymbol{\mu}=\boldsymbol{\nu}\Big)\,,
\end{align}
what can be said about the means $\boldsymbol{\mu}$ of the truth-level spectrum $\mathbf{x}$? As indicated, the mean vectors are related by the matrix equation
\begin{align}
    \mathbf{R}\boldsymbol{\mu}=\boldsymbol{\nu}\,,
\end{align}
where the matrix $\mathbf{R}$ is the discretized version of the integral kernel
in~\cref{eq:integral equation}, representing \emph{migration
probabilities}.\footnote{Migration probabilities represent the probabilities that
events originating in specific bins of the truth-level spectrum are observed in
particular bins of the smeared spectrum due to detector effects such as
resolution limitations and inefficiencies. Each element $R_{ij}$ of the matrix
$\mathbf{R}$ corresponds to the probability that an event in the $j$-th
truth-level bin is reconstructed in the $i$-th observed bin. Thus, the matrix
$\mathbf{R}$ encodes the expected \emph{migration} of events between bins because of the
detector response.}
For an \emph{ill-conditioned}\footnote{This means that the matrix has a high
condition number, as defined in~\cref{def:conditionnumber}.} smearing matrix
$\mathbf{R}$, whose specific construction will be discussed in~\cref{sec:response}, an estimator $\hat{\boldsymbol{\mu}}$ is difficult to
obtain. Inferences for this model will be addressed in the next section.

It is important to note that for finite-dimensional systems a bijective
$\mathbf{R}$ yields necessarily well-posedness essentially as an application of
the Fredholm alternative.\footnote{In finite dimensions, the Fredholm alternative
simplifies to statements about the rank and nullity of $\mathbf{R}$ and is
closely related to the Rank-Nullity Theorem.}
In other words, such finite-dimensional linear problems are not ill-posed in the
sense that they violate the third condition of Hadamard's definition. The main
issue in such cases is that, due to an ill-conditioned $\mathbf{R}$, the solution
will overfit to noise.

% For example, we recall from (\citations{ref to MLE}) that the MLE for $\boldsymbol{\nu}$ is $\mathbf{y}$. However, this result no longer holds trivially in our current context.

To evaluate the scope of the issue, let us model the noise as the discrepancy between the data and the model prediction
\begin{align}
\label{eq:discreteproblem}
    \mathbf{y}=\mathbf{R}\boldsymbol{\mu}+\boldsymbol{\epsilon}\,.
\end{align}
Here, $\boldsymbol{\epsilon}$ represents the statistical noise in the observed data $\mathbf{y}$, accounting for discrepancies between the actual measurements and the model prediction $\mathbf{R}\boldsymbol{\mu}$. Each component $\epsilon_i$ of the noise vector corresponds to the deviation in the $i$-th bin, arising from inherent statistical fluctuations. Under the assumption of large number of counts, the Poisson distribution approaches a Gaussian distribution due to the Central Limit Theorem. In this regime, $\boldsymbol{\epsilon}$ can be modeled as a vector of independent Gaussian random variables with zero mean and variances equal to the expected counts, that is,
\begin{align}
\epsilon_i \sim \mathcal{N}\left(0, \nu_i\right), \quad \text{where} \quad \nu_i = \left(\mathbf{R}\boldsymbol{\mu}\right)_i\,.
\end{align}
It is important to note that while this additive noise model is an approximation—since Poisson noise is signal-dependent and not strictly additive—it captures the essential features of the ill-posedness we expect to encounter.
% While this noise model only serves as an example, i.e. Poisson noise is not strictly additive\footnote{Poisson noise is signal-dependent, with its variance equal to the signal level.}, in the case of large counts, the Poisson distribution approaches a Gaussian distribution due to the \emph{Central Limit Theorem}, making the additive approximation reasonable. In addition, \cref{eq:discreteproblem} is often the form used to take into account background signals $\mathbf{b}$, and the conclusions we find from this noise model directly translates to that case.

Since we cannot satisfy $\mathbf{y}=\mathbf{R}\boldsymbol{\mu}$ exactly, and there are possibly multiple $\boldsymbol{\mu}$, we seek a solution that approximates this equation. Let us handle this using the explicit regularization methodology: The first step is to ensure the existence of a solution, and if multiple solutions exist, we need to establish a notion of uniqueness. Once uniqueness is addressed, we then tackle potential instability by further refining the solution with supplementary information, thereby enhancing the robustness of the reconstruction.

A solution that always exists is the \emph{least squares} solution\footnote{Throughout, $\|\cdot\|$ denotes the Euclidean vector norm $\|\cdot\|_2$; for matrices, $\|A\|$ denotes the induced operator (spectral) norm $\|A\|_{2\to 2}$.
}: 
\begin{align}
    \hat{\boldsymbol{\mu}}_{\textnormal{LS}}=\underset{\boldsymbol{\mu} \in \mathbb{R}^n}{\arg\min}||\mathbf{R}\boldsymbol{\mu}-\mathbf{y}||^{2}\,,
\end{align}
but for rank deficient and ill-conditioned $\mathbf{R}$, this solution is not
unique and extremely unstable. The existence and non-uniqueness of the least
squares solution, and a possible bound on the reconstruction may be quantified
from the following theorem utilizing the Moore-Penrose pseudoinverse
(see ~\cref{app:mpinverse}).
\begin{theorem}\label{the:least squares}
    Let $\mathbf{R}\in\mathbb{R}^{n\times n}$ be a matrix with non-trivial null space $\textnormal{Ker}(\mathbf{R})$. Then, all the least squares solutions are of the form
    \begin{align}
        \boldsymbol{\mu}_{\textnormal{LS}}=\boldsymbol{\mu}_0+\boldsymbol{\lambda}\,,
    \end{align}
    which is a sum of a particular and a homogeneous solution
    \begin{align}
        \boldsymbol{\mu}_0&= \mathbf{R}^{\dagger}\mathbf{y}\in\mathbb{R}^n
        \\
        \boldsymbol{\lambda}&=(\mathbf{I}-\mathbf{R}^{\dagger}\mathbf{R})\mathbf{v}\in\textnormal{Ker}(\mathbf{R})\,,\hspace{0.2cm}\mathbf{v}\in\mathbb{R}^n\,,
    \end{align}
    where $\mathbf{R}^{\dagger}$ is the Moore-Penrose pseudoinverse and \mbox{$\mathbf{I}-\mathbf{R}^{\dagger}\mathbf{R}$} is the orthogonal projection onto $\textnormal{Ker}(\mathbf{R})$.
\end{theorem}

\emph{Proof:} The proof can be found in~\cref{app:proof_moorepenrose}.

\hfill $\square$

\medskip
For $\mathbf{R}$ of full column rank and square, the Moore-Penrose pseudoinverse reduces to the standard inverse, and the least squares estimator reduces to \mbox{$\hat{\boldsymbol{\mu}}_{\textnormal{LS}}=\mathbf{R}^{\dagger}\mathbf{y}=\mathbf{R} ^{-1}\mathbf{y}$}, which is unique:
\begin{corollary}\label{eq:corlstsq}
    The least squares solution $\hat{\boldsymbol{\mu}}_{LS}$ is unique if and only if the null space is empty, i.e., $\textnormal{Ker}(\mathbf{R})=\{\vecc{0}\}$.
\end{corollary}

\emph{Proof:} See~\cref{app:proof_cor_1}.

\hfill $\square$

Even if we obtain a unique solution, and \emph{if} the noise $\boldsymbol{\epsilon}\in\textnormal{Ran}(\mathbf{R})$,\footnote{A non-invertible $ \mathbf{R}$ implies that its range is a proper subspace of $\mathbb{R}^n$, making it highly probable for a randomly oriented noise vector $ \boldsymbol{\epsilon}$ to lie outside this range. That is, proper subspaces of $\mathbb{R}^n$ have Lebesgue measure zero in $\mathbb{R}^n$, so the probability of $\boldsymbol{\epsilon}$ being exactly within the range is zero if $ \boldsymbol{\epsilon}$ is drawn from a distribution with full support in $\mathbb{R}^n$.
}
 the naive inversion will---depending on how ill-conditioned $\mathbf{R}$ is---fit to noise and the reconstructed estimator $\hat{\boldsymbol{\mu}}_{\textnormal{LS}}$ may be far from the truth-level value $\boldsymbol{\mu}$. In most cases, however, the noise $\boldsymbol{\epsilon}\notin\textnormal{Ran}(\mathbf{R})$, which further compounds the problem.

% Furthermore, if $\mathbf{R}$ is not of full column rank (non-invertible) and simultaneously ill-conditioned, the instability of the solution is exacerbated. Thus, in the sense of Hadamard, the solution is unstable, but the primary issue lies in the ill-conditioning of $\mathbf{R}$.

Let us show this explicitly: the measure for the stable solvability of the problem is the \emph{condition number}, $\operatorname{Cond}(\mathbf{R})$:
\begin{definition}(Condition Number)
\label{def:conditionnumber}
    \begin{itemize}
        \item For a square matrix $\mathbf{R}\in\mathbb{R}^{n\times n}$ of full column rank, the condition number with respect to the matrix norm $||\cdot||$ is defined as:
        \begin{align}
            \operatorname{Cond}(\mathbf{R})=||\mathbf{R}||\cdot||\mathbf{R}^{-1}||\,.
        \end{align}
        \item For a square matrix $\mathbf{R}\in\mathbb{R}^{n\times n}$ with rank deficiency, the Moore-Penrose pseudoinverse may be used to define:
        \begin{align}
            \operatorname{Cond}(\mathbf{R})=||\mathbf{R}||\cdot||\mathbf{R}^{\dagger}||\,.
        \end{align}
    \end{itemize}
\end{definition}

% For simplicity, let us assume that $\mathbf{R}\in\mathbb{R}^{n\times n}$ is a symmetric positive definite matrix, i.e. $\mathbf{R}$ is of full column rank (invertible) and a square matrix\footnote{While the assumption of symmetry and positive definiteness simplifies the analysis, similar conclusions regarding stability hold for non-symmetric or rectangular matrices using the singular value decomposition (SVD).}
% . From the spectral theory of symmetric matrices we know that there exists eigenvalues $0< \lambda_1\leq\dots\leq\lambda_n$ and corresponding eigenvectors $\mathbf{r}_i$ such that
% \begin{align}
%     \mathbf{R}=\sum_{i=1}^{n}\lambda_i\,\mathbf{r}_i \mathbf{r}_{i}^{\textnormal{T}}\,.
% \end{align}

% The condition number is then the ratio of the largest and smallest eigenvalue
% \begin{align}
%     \textnormal{Cond}(\mathbf{R})=\frac{\lambda_n}{\lambda_1}\,,
% \end{align}
% implying that $\textnormal{Cond}(\mathbf{R})$ can become very large if $\lambda_1$ is very small.
Then, if we use the \emph{Singular Value Decomposition} (SVD), the condition number can be written as
\begin{align}
     \operatorname{Cond}(\mathbf{R})=\frac{\sigma_{\text{max}}}{\sigma_{\text{min}}}\,,
    \end{align}
where $\sigma_{\text{max}}$ is the largest singular value and $\sigma_{\text{min}}$ is the smallest non-zero singular value of $\mathbf{R}$.\footnote{For $\mathbf{R}$ of full column rank all singular values are non-zero, but not for rank-deficient matrices and therefore the definition only makes sense if we choose the smallest non-zero singular value.} If we (for simplicity) assume a normalization such that $||\mathbf{R}|| = \sigma_{\text{max}} = 1$, and that the noise in~\cref{eq:discreteproblem} satisfy $||\boldsymbol{\epsilon}||\leq \epsilon$,\footnote{Note that for a background vector $\mathbf{b}$ this would be $||\mathbf{b}||\leq b$.} we can for full column rank $\mathbf{R}$ estimate the difference in the reconstruction
\begin{align}
    ||\hat{\boldsymbol{\mu}}_{\textnormal{LS}}-\boldsymbol{\mu}||&=||\mathbf{R}^{-1}\boldsymbol{\epsilon}||\leq ||\mathbf{R}^{-1}||\cdot||\boldsymbol{\epsilon}||\,,
\end{align}
% \begin{align}
%     ||\hat{\boldsymbol{\mu}}_{\textnormal{LS}}-\boldsymbol{\mu}||&= \sum_{i=1}^{n}\lambda_{i}^{-2}|\mathbf{r}_{i}^{T}\boldsymbol{\epsilon}|^2\nonumber
%     \\
%     &\leq \lambda_{1}^{-2}||\boldsymbol{\epsilon}||^2
% \end{align}
giving that
\begin{align}\label{eq:epscond}
    ||\hat{\boldsymbol{\mu}}_{\textnormal{LS}}-\boldsymbol{\mu}||\leq \textnormal{Cond}(\mathbf{R})\epsilon\,.
\end{align}
Geometrically, this inequality defines a \emph{ball} around the true solution $\boldsymbol{\mu}$ within which $\hat{\boldsymbol{\mu}}$ must lie, and the radius of the ball is scaled by $\textnormal{Cond}(\mathbf{R})$. The implication is that a large condition number renders the reconstructed estimator far from the true value, 
\textit{even if the noise is small}. 

Furthermore, if $\mathbf{R}$ is not of full column rank (non-invertible) and simultaneously ill-conditioned, the instability of the solution is exacerbated. In this case, a similar calculation using the Moore-Penrose pseudoinverse yields that the difference in the reconstruction takes the form
\begin{align}
\label{eq:mucond}
    ||\hat{\boldsymbol{\mu}}_{\textnormal{LS}}-\boldsymbol{\mu}||\leq \textnormal{Cond}(\mathbf{R})\epsilon+||\boldsymbol{\lambda}||\,,
\end{align}
for some null-space vector $\boldsymbol{\lambda}\in\textnormal{Ker}(\mathbf{R})$. Geometrically, the solution is therefore part of a tube-like region extending in directions defined by $\textnormal{Ker}(\mathbf{R})$, allowing unbounded deviations.

In other words, for $\mathbf{R}$ of full column rank, we can always establish a bound on the difference that is proportional to $\textnormal{Cond}(\mathbf{R})$, ensuring that the estimated solution $\hat{\boldsymbol{\mu}}$ remains within a controlled vicinity of the true solution $\boldsymbol{\mu}$. However, when $\mathbf{R}$ is not of full column rank, no such bound can guarantee the proximity of $\hat{\boldsymbol{\mu}}$ to $\boldsymbol{\mu}$. In this scenario, the presence of a non-trivial null space allows solutions to deviate arbitrarily from the true value, thereby increasing the instability of the reconstruction.

Another challenge is the occurrence of negative components in the solution, which are physically impossible. That is, null space vectors can often contain components that are both positive and negative, and their amplification results in unphysical solutions:
\begin{proposition}
\label{prop:nullvectors}
Let $\mathbf{R} \in \mathbb{R}^{n \times n}$ be a non-negative matrix, $R_{ij} \geq0$ for all $i,j$, with no zero rows or columns, and suppose that $\mathbf{R}$ has a non-trivial null space. Then any non-zero vector $\boldsymbol{\lambda} \in \textnormal{Ker}(\mathbf{R})$ must have both positive and negative components.
\end{proposition} 

\emph{Proof:} See ~\cref{proofprop:negativecomponents}

\hfill $\square$

This result implies that any reconstruction method involving matrices $\mathbf{R}$ with
the stated properties, and that fails to mitigate the influence of null space
vectors, will inevitably produce solutions with unphysical features.

Another challenge is the proper treatment of the background spectrum. Simply subtracting the background can amplify the background error, in direct analogy with the additive noise model~\cref{eq:discreteproblem}.  
In the presence of a background \(\mathbf{b} = \boldsymbol{\beta} + \boldsymbol{\epsilon}_\beta\) with additive noise \(\boldsymbol{\epsilon}_\beta\),~\cref{eq:discreteproblem} becomes
\begin{equation}
    \mathbf{y} = \mathbf{R}\boldsymbol{\mu} + \boldsymbol{\epsilon} + \boldsymbol{\beta} + \boldsymbol{\epsilon}_\beta\,.
\end{equation}
Since \(\mathbf{b}\) is not directly observable, the naive approach is to subtract an estimate \(\hat{\mathbf{b}}\) from \(\mathbf{y}\), yielding
\begin{equation}
    \tilde{\mathbf{y}} = \mathbf{R}\boldsymbol{\mu} + \boldsymbol{\epsilon} + \boldsymbol{\beta} + \boldsymbol{\epsilon}_\beta - \hat{\mathbf{b}}\,.
\end{equation}
The resulting background error,
\begin{equation}
\label{eq:condmuback}
\|\mathbf{b} - \hat{\mathbf{b}}\| = \|\boldsymbol{\beta} + \boldsymbol{\epsilon}_\beta - \hat{\mathbf{b}}\|\,,
\end{equation}
acts as an additional source of additive noise. As in ~\cref{eq:epscond} and ~\cref{eq:mucond}, this error is amplified by the condition number.
In general, any attempt to subtract away unwanted structure without accounting for its uncertainty will destabilize the reconstruction, by analogous argument.

In conclusion, a large condition number indicates an ill-conditioned problem
with a high potential for noise amplification. In finite-dimensional settings,
noise may be present in the data, but it is the ill-conditioning that makes the
reconstruction highly sensitive to such noise, potentially causing the solution
to fit noise rather than signal. Importantly, a large condition number reflects
fundamental sensitivity to \emph{any} perturbation—whether from measurement
error, modeling inaccuracies, or numerical instability. Ill-conditioned problems
therefore require careful handling, typically through regularization. Moreover,
if the response matrix is rank-deficient, the problem also lacks uniqueness. In
such cases, additional constraints or regularization are essential to avoid
inaccurate or unphysical solutions.

\subsection{Tikhonov regularization}
\label{sec:explicitregularization}
Having established the challenges posed by ill-conditioning and potential rank
deficiency in the discretized unfolding problem, we now turn to regularization techniques
designed to yield stable and physically meaningful solutions. One prominent
strategy involves incorporating prior knowledge or desired
properties of the solution (e.g., non-negativity) are incorporated directly into the problem
formulation, typically by adding a penalty term to an objective function like
the least squares criterion.

%In most cases, this can only be properly handled by carefully adding information into the problem, e.g. via a regularization methodology together with other relevant physical constraints (e.g. non-negativity). In this work, we only consider regularization for frequentist unfolding, which consists of either an iterative methodology by early stopping or by explicitly adding a regularization term to the minimization problem. For completeness, as we propose improvements on a previous iterative approach, we will discuss strengths and weaknesses of both approaches. 

%To finish the explicit regularization methodology, let us continue with the least squares problem, where the impact of a regularization term is easy to track.
To get around the non-uniqueness of the least squares solution, we observe that the set of all solutions in~\cref{the:least squares}
\begin{align}
    \mathcal{N}=\{\boldsymbol{\mu}_{\textnormal{LS}}=\mathbf{R}^{\dagger}\mathbf{y}+\boldsymbol{\lambda}\,|\,\boldsymbol{\lambda}\in\textnormal{Ker}(\mathbf{R})\}\,,
\end{align}
is a closed convex set.\footnote{The solution set is convex because any convex combination of solutions remains a solution. The solution set is an affine subspace, and in finite-dimensional settings, affine subspaces are closed. This means they contain all their limit points, ensuring that the solution set is topologically closed.} Hence, we may use the \emph{closest-point theorem}, which states that there exists a unique point in $\mathcal{N}$ corresponding to the minimum norm solution, i.e.,
\begin{align}
    \hat{\boldsymbol{\mu}}_{\textnormal{MNLS}}=\underset{\boldsymbol{\mu}_{\textnormal{LS}}\in \mathcal{N}}{\arg\min}||\boldsymbol{\mu}_{\textnormal{LS}}||=\mathbf{R}^{\dagger}\mathbf{y}\,,
\end{align}
is called the \emph{minimum norm least squares estimator} and is unique, but again, even though a unique solution exists, noise amplification might still be a problem. To remedy this, we seek a smooth cut-off of the singular values such that the estimates $||\hat{\boldsymbol{\mu}}-\boldsymbol{\mu}||$ are dampened. As we have already obtained a notion of a unique solution, we can achieve such a smooth cut-off by considering a solution that continuously converge to the minimum norm least squares solution.
\begin{theorem}\label{theorem:Tikhonov reg}
    Let $\alpha >0$ be a constant. For the least squares problem $\mathbf{y}=\mathbf{R}\boldsymbol{\mu}$, where $\mathbf{R}\in\mathbb{R}^{n\times n}$, consider the regularized solution defined by:
    \begin{align}\label{eq:Tikhonov}
        \hat{\boldsymbol{\mu}}_{\alpha}=\underset{\boldsymbol{\mu}}{\arg\min}\big(||\mathbf{y}-\mathbf{R}\boldsymbol{\mu}||^2 +\alpha||\boldsymbol{\mu}||^2\big)\,.
    \end{align}
    Then:
    \begin{enumerate}
        \item \emph{Uniqueness:} The solution
        \begin{align}
            \hat{\boldsymbol{\mu}}_{\alpha}=(\mathbf{R}^{\textnormal{T}}\mathbf{R}+\alpha\mathbf{I})^{-1}\mathbf{R}^{\textnormal{T}}\mathbf{y}\,,
        \end{align}
        exists and is unique for all $\alpha >0$.
        \item \emph{Limiting Case:} As $\alpha\rightarrow 0$, the regularized solution $\hat{\boldsymbol{\mu}}_{\alpha}$ continuously converges to the minimum norm least squares estimator
        \begin{align}
            \lim_{\alpha\rightarrow 0}||\hat{\boldsymbol{\mu}}_{\alpha}-\hat{\boldsymbol{\mu}}_{\textnormal{MNLS}}||=0\,.
        \end{align}
        \item \emph{Noise Dampening:} The regularization introduces a smooth cut-off of the singular values of $\mathbf{R}$, dampening the amplification of noise in the estimates. Specifically, using the Singular Value Decomposition
        \begin{align}
            \mathbf{R}=\sum_{i=1}^{\operatorname{rank}(\mathbf{R})}\sigma_i \mathbf{u}_i \mathbf{v}_{i}^{\textnormal{T}}\,,
        \end{align}
        the regularized expression can be expressed as:
        \begin{align}
            \hat{\boldsymbol{\mu}}_{\alpha}=\sum_{i=1}^{\textnormal{rank}(\mathbf{R})}\frac{\sigma_i}{\sigma_{i}^{2}+\alpha} \big(\mathbf{u}_{i}^{\textnormal{T}}\mathbf{y}\big) \mathbf{v}_{i}\,,
        \end{align}
        showing that each component of the solution is weighted by a factor that for $\alpha > 0$ dampens the impact of small singular values, mitigating noise amplification.
    \end{enumerate}
\end{theorem}

\medskip
\emph{Proof:} See~\cref{prooftheorem:Tikhonov reg}

\hfill $\square$

% Effectively, the modification~\cref{eq:Tikhonov} alters the original problem by
% adjusting the response matrix $\matt{R}$ through \emph{normal equations}, making
% the method sensitive to loss of critical physical information that $\mathbf{R}$
% represent.

The main insight from this result is that $\hat{\boldsymbol{\mu}}_{\alpha}$ is obtained by reconciling fitting to the data, and finding a solution with small norm. This type of regularized solution is most often called \emph{Tikhonov regularization}, and it can be further generalized by adding any convex term $||\mathbf{L}\boldsymbol{\mu}||^{2}$, where $\mathbf{L}$ is referred to as a regularization matrix. Typically, generalized Tikhonov regularization employs $\mathbf{L}$ as a discretized version of the first or second order derivative operator, such as $\nabla$ or $\nabla^2$. This form of regularization is particularly well-suited for smooth spectra, where the underlying signal is expected to vary gradually without abrupt changes.

In contrast, for spectra characterized by sparsity, Tikhonov regularization may fail to capture these sharp features adequately. In such cases, a regularization term that promotes sparsity is preferred. For example, incorporating an $\ell_1$ norm penalty encourages many coefficients to be exactly zero, making it ideal for sparse solutions where only a few significant components are present. However, if the underlying signal is not only sparse but also contains sharp peaks, pure $\ell_1$ regularization may inadvertently penalize these important features, leading to a loss of critical information. To address this, more sophisticated regularization techniques that balance sparsity with the preservation of sharp transitions must be employed.

With this demonstration, we have shown that an explicit regularization method can ensure stable and unique solutions. However, we anticipate several challenges if this method is naively applied to the relevant spectra in this study. One significant issue is that achieving a unique and stable solution required modifying the original problem through \emph{normal equations}, which substantially increases the condition number. As a consequence it is extremely difficult to find an appropriate regularization and the reconstruction does not retain important physical structure. Additionally, the Tikhonov regularization applied to the least squares problem does not enforce the non-negativity constraint. The consequence of this is that while standard Tikhonov allows for a closed-form solution via normal equations, the addition of non-negativity constraints necessitates optimization techniques;\footnote{This is often called \emph{regularized non-negative least squares} in the context of least squares estimation.} closed-form solutions are generally unattainable. While this is not a major problem in itself, as there exists customized optimization methods to handle this, the correct handling of the appropriate noise and background of the physical system is in general infeasible with this simple setup.

To obtain stable and accurate solutions, alternative strategies are necessary---specifically, incorporating explicit regularization techniques within an appropriate statistical framework that effectively handles relevant noise and background. That is, noise and background are stochastic and need to be treated as such. Therefore, we turn our attention towards \emph{regularized maximum likelihood estimation} where these problems have an innate implementation.

\subsection{Regularized maximum likelihood estimation}
\label{sec:regularizedunfolding}
Since we deal with Poisson distributed data, we are interested in performing maximum likelihood estimation for the mean $\boldsymbol{\mu}$. In the case of direct observation, the MLE had a closed form solution, but this is no longer the case for indirect observations.

Given the unfolding problem and data model
\begin{align}\label{eq:discretizedPoisson}
    \mathbf{y}\sim\operatorname{Poisson} \Big(\mathbf{R}\boldsymbol{\mu}=\boldsymbol{\nu}\Big)\,,
\end{align}
we construct the likelihood of the true mean histogram $\boldsymbol{\mu}$ as:
\begin{align}\label{eq:discretizedNlog}
    \mathcal{L}(\boldsymbol{\mu}|\mathbf{y})=\prod_{i}\frac{\Big(\sum_{j}R_{ij}\mu_j\Big)^{y_i}}{y_{i}!}\exp{\Big(-\sum_{j}R_{ij}\mu_j\Big)}\,,
\end{align}
which can equivalently be expressed in terms of the smeared histogram $\boldsymbol{\nu}$, through the relation $\mathbf{R}\boldsymbol{\mu}=\boldsymbol{\nu}$. 

However, a challenge in applying maximum likelihood estimation here lies in the nonlinearity of the likelihood function in the constrained $\boldsymbol{\mu}\in\mathbb{R}_{+}^{n}$. This leads to a nonlinear constrained optimization problem that lacks a closed-form solution. This means that the MLE must be solved using numerical optimization methods, and due to $\mathbf{R}$ being rank-deficient and ill-conditioned, the existence and stability of the MLE must be carefully examined.
 
Instead of directly maximizing the likelihood function, we minimize the negative log-likelihood. The  negative log-likelihood retains the same information as the likelihood through a monotonic transformation, which preserves the location of the optimal solution. Additionally, the transformation simplifies the optimization process by converting products into sums, making the problem more tractable, while also ensuring that the resulting problem is framed as a minimization. 

For the model~\cref{eq:discretizedNlog}, the negative log-likelihood takes the form:
\begin{align}
\label{eq:loglikedef}
\ell(\boldsymbol{\mu}|\mathbf{y}) = \sum_{i} &\left[ \sum_{j} R_{ij} \mu_j - y_i \log\left(\sum_{j} R_{ij} \mu_j \right) \right]\nonumber
\\
&+ \sum_{i} \log(y_i!)\,,
\end{align}
where the last term is independent of $\boldsymbol{\mu}$ and does not affect the optimization process. The model requires  $\boldsymbol{\nu}=\mathbf{R}\boldsymbol{\mu}\geq 0$, and for physical interpretation we must have that $\mathbf{R}\geq 0$, $\boldsymbol{\mu}\geq 0$. Given this, the task is to find the estimator
\begin{align}
\hat{\boldsymbol{\mu}}_{\text{MLE}} = \underset{\boldsymbol{\mu} \in\mathbb{R}_{+}^{n}}{\arg\min} \, \ell(\boldsymbol{\mu}|\mathbf{y})\,.
\end{align}
The negative log-likelihood has---under certain conditions---some very useful properties. First, however, it is useful to differentiate between two notions of indeterminacy in this model:
\begin{itemize}
     \item \emph{Identifiability (model-level)}: Different parameter values induce different distributions. This is a property of the statistical model, and independent of the data.
     
     % The parameters $\boldsymbol{\nu}$ are identifiable if $\forall\,\boldsymbol{\nu},\boldsymbol{\nu}'\in\mathbb{R}_{+}^{\,n},\quad$
    % \begin{align}
    %     \boldsymbol{\nu}\neq\boldsymbol{\nu}' \;\Rightarrow\;
    %     \exists\,\mathbf y\in\mathbb{N}_{0}^{\,n}\ :\
    %     p(\mathbf y \mid \boldsymbol{\nu}) \neq p(\mathbf y \mid \boldsymbol{\nu}')\,,
    % \end{align}

    % where $p(\cdot|\cdot)$ is the probability mass function.
    \item \emph{Uniqueness (estimator-level)}: Given data and a fitting rule, there exists only one solution to the optimization problem.
\end{itemize}
In the Poisson model~\cref{eq:discretizedPoisson}, the smeared means $\boldsymbol{\nu}$ are identifiable, but the truth-level means become identifiable
only when the matrix $\mathbf{R}$ has full column rank; by the identifiability of $\boldsymbol{\nu}$, it holds that $\mathbf{R}(\boldsymbol{\mu}-\boldsymbol{\mu'})=0$, implying that the following set maps to the same $\boldsymbol{\nu}$
\begin{align}
    \{\boldsymbol{\mu}=\boldsymbol{\mu}_0+\boldsymbol{\lambda}\,|\,\boldsymbol{\lambda}\in \textnormal{Ker}(\mathbf{R})\}\,,
\end{align}
which in general is a unphysical and unbounded set. The truth-level means $\boldsymbol{\mu}$ are therefore identifiable if and only if $\textnormal{Ker}(\mathbf{R})=\{0\}$. However, it is not necessarily true that identifiability and uniqueness coincide. 

% In other words, for rank-deficient $\mathbf{R}$, the truth-level means are elements of the set
% \begin{align}
%     \mathcal{N}=\{\boldsymbol{\mu}=\boldsymbol{\mu}_0+\boldsymbol{\lambda}\,|\,\boldsymbol{\lambda}\in \textnormal{Ker}(\mathbf{R})\}\cap\mathbb{R}_{+}^{n}\,,
% \end{align}

% \begin{proposition}\label{prop:identifiabilityPoisson}
%     For the Poisson model~\cref{eq:discretizedPoisson} with likelihood~\cref{eq:discretizedNlog}, the smeared means $\boldsymbol{\nu}$ are identifiable, and all truth-level means are identifiable if and only if $\mathbf{R}$ are of the form:
%     \begin{align}
%         \boldsymbol{\mu}_{\text{MLE}}=\boldsymbol{\mu}_0+\boldsymbol{\lambda}\,,
%     \end{align}
%     where $\boldsymbol{\mu}_0$ is a particular solution and $\boldsymbol{\lambda}\in\textnormal{Ker}(\mathbf{R})$ is a homogeneous solution.
% \end{proposition}

% \emph{Proof:} See ~\cref{proofprop:identifiability}

% \hfill $\square$

% without direct access to the spectral machinery of linear operator theory,

To address uniqueness, we note that the non-negativity requirement changes the feasible set to a bounded one
\begin{align}
    \{\boldsymbol{\mu}=\boldsymbol{\mu}_0+\boldsymbol{\lambda}\,|\,\boldsymbol{\lambda}\in \textnormal{Ker}(\mathbf{R})\}\cap\mathbb{R}_{+}^{n}\,,
\end{align}
preserving physicality (see ~\cref{app:feasible set} for a demonstration), but it is difficult to further determine the impact of the non-negativity constraint beyond boundedness. However, 
if we assume that $\mathbf{R}\geq0$ and that there are no zero rows in $\mathbf{R}$ for indices with $y_i>0$\footnote{If some row $i$ of $\mathbf{R}$ were identically zero while $y_i > 0$, the Poisson likelihood would be infinite for all $\boldsymbol{\mu} \ge 0$, and the problem would be infeasible.
}, the following properties hold:
\begin{itemize}
    \item The function $\ell(\boldsymbol{\mu}|\mathbf{y})$ attains a minimum over the feasible set of $\boldsymbol{\mu}$.
    \item The function $\ell(\boldsymbol{\mu}|\mathbf{y})$ is convex with respect to $\boldsymbol{\mu}$; thus, all local minima are global minima.
    \item If $\mathbf{R}$ has full column rank, $\ell(\boldsymbol{\mu}|\mathbf{y})$ is strictly convex and the solution is unique.
\end{itemize}
The existence of a minimum follows from the continuity and coercivity of $\ell(\boldsymbol{\mu}|\mathbf{y})$ over the closed domain $\mathbb{R}_{+}^{n}$. The convexity of the $\ell(\boldsymbol{\mu}|\mathbf{y})$ follows from the positive semi-definiteness of its Hessian matrix, and if $\mathbf{R}$ is of full column rank, the Hessian is positive definite and $\ell(\boldsymbol{\mu}|\mathbf{y})$ is strictly convex. In essence, the practical problem of indeterminacy has been reduced to possible rank-deficiency of the detector repsonse.
% In practice, we often have that $R_{ij}\geq 0$ rather than $R_{ij}>0$, which means some entries of $\mathbf{R}$ could be zero. This can weaken the strict convexity of the negative log-likelihood function and, consequently, affect the uniqueness of the solution. Despite this, the negative log-likelihood remains convex, and optimization methods designed for convex problems can still be applied to find a global minimum.

In summary: From the given properties of $\ell$, we know that a solution exists, but it is not necessarily unique if $\mathbf{R}$ lacks full column rank. In optimization terms; the negative log-likelihood is convex but not strictly convex when $\mathbf{R}$ is rank-deficient. Additionally, if $\mathbf{R}$ is ill-conditioned, the solution is highly sensitive to noise. As we have established, this will cause a multitude of problems in the optimization. In contrast to the least squares method, the non-negative constraint is inherent when using likelihoods, since the likelihood is only defined for $\boldsymbol{\nu} \geq 0$. From this, physical requirements sets non-negativity constraint on $\mathbf{R}$ and $\boldsymbol{\mu}$. While these constraints reduces the feasible set of solutions and narrows the search space, it does not eliminate the occurrence of multiple solutions.

By adding a regularization term, the number of possible solutions and noise amplification in the feasible set is expected to be further reduced and dampened. In general, the objective is therefore to find the regularized estimator
\begin{align}
    \hat{\boldsymbol{\mu}}_{\text{RMLE}} = \underset{\boldsymbol{\mu} \in\mathbb{R}_{+}^{n}}{\arg\min}\big(\ell(\boldsymbol{\mu}|\mathbf{y})+\Omega(\boldsymbol{\mu};\boldsymbol{\alpha})\big)\,,
\end{align}
where $\Omega(\boldsymbol{\mu};\boldsymbol{\alpha})$ is the appropriate regularization term(s) for a given spectrum with regularization parameter(s) $\boldsymbol{\alpha}$.

% \lasse{Still not finished with this section, but only minor details. Maybe include discussion of link functions (or maybe appendix?).}

Having established the ill-posed nature of the unfolding problem and the role of
the detector response \(\mathbf{R}\) in smearing the truth-level spectrum,
the next step is to examine this matrix in detail. Section~3 describes the
physical basis of the detector response and explains how particle interactions
and detector effects give rise to the response matrix \(\mathbf{R}\).

\section{The detector response}
\label{sec:response}

The experimental data analyzed in this work is structured as two-dimensional
histograms, giving rise to a matrix structure $\mathbf{Y}\in\mathbb{R}^{m\times
n}$ for the data (see~\cref{app:exp} for the experimental setup). To model this,
we recall from Section \ref{sec:poisson inv prob} that $F$ and $G$ are two distinct Poisson processes with corresponding
intensity functions $f$ and $g$. $F$ models the truth-level process and
$G$ models the smeared process. The rows $E$ are partitioned into $m$ ordered
intervals $\{E_i\}_{i=1}^{m}$ and the columns $E'$ into $n$ ordered intervals
$\{E'_j\}_{j=1}^{n}$, where each pair $(E_i, E'_{j})$ defines a grid cell $E_i
\times E'_{j}$. For each grid cell $E_i \times E'_{j}$, we let $Y_{ij}$ denote
the number of events falling into that cell, $Y_{ij}=G(E_i \times E'_j)$,
resulting in a stochastic matrix
\begin{equation}
\mathbf{Y} = \begin{bmatrix}
Y_{11} & Y_{12} & \dots & Y_{1n} \\
Y_{21} & Y_{22} & \dots & Y_{2n} \\
\vdots & \vdots & \ddots & \vdots \\
Y_{m1} & Y_{m2} & \dots & Y_{mn} \\
\end{bmatrix} \in \mathbb{R}_{+}^{m\times n}.
\end{equation}
From the mean measures
\begin{align}
    \mu_{ij}=\mu(D_i \times D'_{j})=\int_{D_i \times D'_{j}}f(x,z)\,dxdz
    \\
    \nu_{ij}=\nu(E_i \times E'_{j})=\int_{E_i \times E'_{j}}g(y,w)\,dydw\,,
\end{align}
we construct the mean matrices
\begin{equation}
\boldsymbol{\mu} = \begin{bmatrix}
\mu_{11} & \mu_{12} & \dots & \mu_{1n} \\
\mu_{21} & \mu_{22} & \dots & \mu_{2n} \\
\vdots & \vdots & \ddots & \vdots \\
\mu_{m1} & \mu_{m2} & \dots & \mu_{mn} \\
\end{bmatrix}\in \mathbb{R}_{+}^{m\times n}.
\end{equation}
\begin{equation}
\boldsymbol{\nu} = \begin{bmatrix}
\nu_{11} & \nu_{12} & \dots & \nu_{1n} \\
\nu_{21} & \nu_{22} & \dots & \nu_{2n} \\
\vdots & \vdots & \ddots & \vdots \\
\nu_{m1} & \nu_{m2} & \dots & \nu_{mn} \\
\end{bmatrix}\in \mathbb{R}_{+}^{m\times n}.
\end{equation}
To describe the smearing process, we must separate the smearing across the rows and the smearing across the columns. This can be represented by a double-sided matrix equation
\begin{align}\label{eq:doublesided}
    \mathbf{R}_1\boldsymbol{\mu}\mathbf{R}_2=\boldsymbol{\nu}\,,
\end{align}
where $\mathbf{R}_1\in\mathbb{R}^{m\times m}$ describes the smearing along the row dimension and $\mathbf{R}_2\in\mathbb{R}^{n\times n}$ describes the smearing along the column dimension, giving rise to the matrix model
\begin{align}
    \mathbf{Y}\sim \text{Poisson}(\mathbf{R}_1\boldsymbol{\mu}\mathbf{R}_2=\boldsymbol{\nu})\,.
\end{align}

% The response to individual photons and charged particles is additive, and interactions within
% the detectors do not introduce nonlinearities under normal operating conditions.

The detector response consists of two distinct transformations: row-wise
smearing $\mathbf{R}_1$ modeling gamma-ray detection, and column-wise smearing
$\mathbf{R}_2$ modeling particle detection. A property of the OCL
detector setup is that these transformations are independent --- the gamma-ray
detection (OSCAR) and particle detection (SiRi) are uncorrelated measurements.
This independence can be seen in the experimental data as symmetric Gaussian peaks
rather than rotated ellipses that would indicate correlation.
%\begin{figure}
%    \centering
%    \includegraphics{figures/28Si76Ge.pdf}
%    \caption{Prompt spectra of \ce{^{28}Si} (top) and \ce{^{76}Ge} (bottom) 
%    bla bla. The axes of the Gaussian peaks for\ce{^{28}Si} are vertical and horizontal: 
%    there is no off-diagonals in the covariance matrix. In contrast, the axes}
%    \label{fig:enter-label}
%\end{figure}

The particle detector response $\mathbf{R}_2 = \mathbf{G}_\text{in}$ is modeled
as a Gaussian smoothing matrix with constant resolution $\sigma_\text{in}$. The
specific resolution depends on the experimental setup and particle type --- for
instance, the instrumental energy resolution when detecting protons (at forward
angles) in SiRi~\cite{Siri} is approximately \SI{150}{keV} (full width at half
maximum, FWHM).

The gamma-ray detector response $\mathbf{R}_1$ can be factored as $\mathbf{R}_1 =
\mathbf{G}_\gamma\mathbf{D}$, where $\mathbf{D}$ is a discrete response matrix
capturing the fundamental detector physics, and $\mathbf{G}_\gamma$ is a
smoothing matrix accounting for finite detector resolution. This factorization
separates the distinct physical processes affecting gamma-ray detection into
components that can be handled separately in the unfolding process.

The discrete response matrix $\mathbf{D}$ can be determined either through
GEANT4 simulations~\cite{geant4} or experimental measurements. For the OSCAR
array, we use GEANT4-simulated responses~\cite{oscarfabio} as they provide
better resolution than direct experimental measurements. Since the simulated
response energies typically don't align with experimental spectrum binning,
interpolation is required (see~\cite{GUTTORMSEN1996371}).

%After obtaining the discrete response, the spectrum is smoothed using Gaussian
%kernels to account for the resolution of the gamma-detector resolution. This smoothing process is represented
%by $\mathbf{G}_\gamma$. The resolution needs to be calibrated to each specific experimental spectrum.

% \erlend{Adjusting the text below to fit with the notation we have used above.}
% The detector responses are modeled as linear operations using matrices.
% The physical motivation for this double sided structure is that the response to individual photons and charged particles is additive, and interactions within
% the detectors do not introduce nonlinearities under normal operating conditions.

% For a $m\times n$ experimental matrix $\matt{Y}$, the folding of the spectrum 

% \begin{equation}
% \vnu = \underbrace{\Gg{}\matt{D}}_{\matt{R}}\vmu\Gin
% \end{equation}
% under the assumption that
% \begin{equation}
%  \matt{Y} \sim\operatorname{Poisson}(\vnu)
% \end{equation}
% Here, the matrix $\Gg\matt{D}$ models the various processes by which a
% $\gamma$-detector distorts the true $\gamma$-spectrum.
% The $\gamma$-response can be factored into the
% discrete response $\matt{D}$, which captures
% the $\gamma$-detector physics, and the smoothing matrix $\Gg$.
% $\Gin$ is the smoothing matrix describing the
% smearing along the $\Ex$ axis due to the charged particle detector.
% Both smoothing matrices, $\Gg$ and $\Gin$, are square matrices  of
% dimensions $m\times m$ and $n \times n$, respectively. 

\begin{figure}
    \centering
    \includegraphics{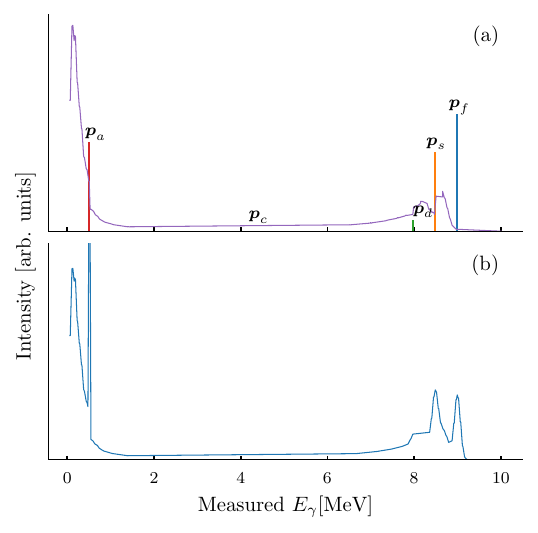}
    \caption{(a) Each component of the discrete response for true $\Eg=\SI{9}{MeV}$.
    The peaks have been scaled down to make them visually comparable.
    (b) A single sharp peak at \SI{9}{MeV} folded with $\Gg\matt{D}$.}
    \label{fig:discretecomponents}
\end{figure}

The gamma-detector physics is modeled by
four discrete structures: the full energy peak ($\mathbf{p}_{\text{f}}$), which corresponds to the
true energy of the incoming gamma-ray; the single escape peak ($\mathbf{p}_{\text{s}}$); the double
escape peak ($\mathbf{p}_{\text{d}}$); and the annihilation peak ($\mathbf{p}_{\text{a}}$). 
In addition, there is a
continuum composed of Compton scattering, backscatter, and low-energy processes.
This continuum is traditionally referred to as the \textit{Compton background}
($\mathbf{P}_{c}$). The top panel of~\cref{fig:discretecomponents} illustrates the components
for a single true gamma energy, as a function of measured gamma energy. 

The discrete response $\matt{D}$ is the matrix sum
of these components. By construction, $\matt{D}$ is row-stochastic, with each
row normalized to unity:
\begin{equation}
    \sum_{j=1}^{m} D_{ij} = 1 \quad \forall i =1,2,\dots,m\,.
\end{equation}
The last element of the $i^{\textnormal{th}}$ row of $\matt{D}$ is the element
$(p_{\text{f}})_{i}$ of $\mathbf{p}_{\text{f}}$, located at the diagonal entry $\elemof{D}{ii}$, which
makes $\matt{D}$ a lower triangular matrix. The other vector components are thus placed below the main diagonal: $\mathbf{p}_{\text{s}}$ is placed along the subdiagonal corresponding to $511$ keV below the full energy peak, $\mathbf{p}_{\text{d}}$ placed along the subdiagonal corresponding to $2\cdot 511$ keV below the full energy peak and $\mathbf{p}_{\text{a}}$ is placed along the column corresponding to $511$ keV. Given such a structure, the vector components may be embedded into a matrix as:

\begin{align}
    \mathbf{P}=\sum_{k=0}^{m-1}\text{diag}_{k}(\mathbf{p}_k)+\sum_{j=1}^{m}\text{col}_{j}(\mathbf{q}_j)\,,
\end{align}
where the first sum places vectors along the $k$-th diagonal and the second sum places vectors into the $j$-th column. To preserve the lower triangular structure: for each column $j$, the entries $q_{j,i}$, corresponding to rows $i<j$ must be zero.

%%%% Just for summary, to be edited:
% \begin{align}
%     \mfe &= \diag{\vfe}\\
%     \mse &= \subdiag{k}{\vse}\\
%     \mde &= \subdiag{k}{\vde}\\
%     \map &= \operatorname{col}_{m_e}\left({\vap}\right)
% \end{align}
%%%%
These diagonal and column components together with the Compton component yields the discrete matrix
\begin{align}
\label{eq:Dcomponents}
    \mathbf{D}=\mathbf{P}_{\text{f}}+\mathbf{P}_{\text{s}}+\mathbf{P}_{\text{d}}+\mathbf{P}_{\text{a}}+\mathbf{P}_{\text{c}}\,.
\end{align}

\begin{figure}
    \centering
    \includegraphics[]{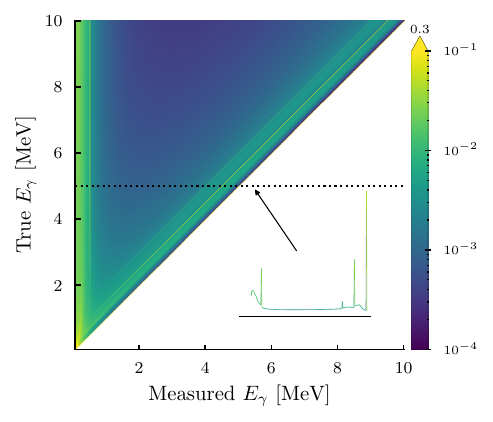}
    \caption{The discrete OSCAR response $\matt{D}$ for a $1000\times 1000$ matrix.
    The discrete peaks $\vfe, \vse$ and $\vde$ are along the diagonal and at offsets
    \SI{511}{keV} and \SI{1024}{keV}, respectively.
    The $\vap$ is a sharp vertical structure at measured $\Eg = \SI{511}{keV}$.
    The remaining bulk is the $\mcb$ component.
    The inset axes shows an example of a single row at true $\Eg = \SI{5}{MeV}$.
    The color scale is logarithmic, scaled to prevent outlying bins from
    affecting the color. The numbers on the colorbar indicate 
    $0.3\%$ of the bins lie above the range. 
    }
    \label{fig:responsematrix}
\end{figure}

The response matrix $\matt{D}$ for OSCAR is shown in~\cref{fig:responsematrix}.\footnote{While the matrix is mathematically lower
triangular, it is displayed with reversed rows to show increasing energy on the
y-axis, making it appear upper triangular in the figure.}

\begin{figure}
    \centering
    \includegraphics{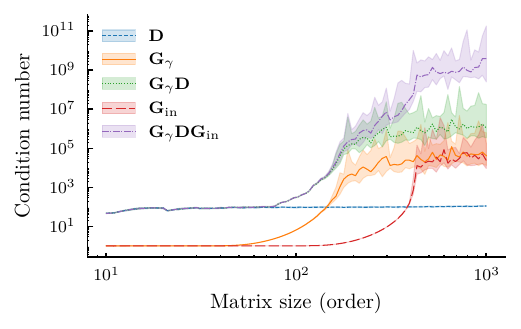}
    \caption{
Condition numbers versus matrix order for matrices $\matt{D}$, $\Gg{}$, $\Gg{}\matt{D}$,
$\Gin$ and $\Gg\matt{D}\Gin$. The smoothing operators $\Gg{}$ and $\Gin$ exhibit substantially higher
condition number growth with increasing order than $\matt{D}$. Shaded bands represent the range of condition numbers 
across 100 instances with small perturbations, simulating numerical fluctuations in matrix construction. The resolution is held constant at $\sigma_{\text{in}} = \SI{40}{keV}$
and $\sigma_\gamma(\SI{1330}{keV}) = \SI{40}{keV}$. The energy range is from $0$ to \SI{10}{MeV},
with $\Delta E$ determined by the order.}
    \label{fig:condition_number}
\end{figure}
% \begin{equation}
%     \matt{D} = \mfe + \mse + \mde + \map  + \mcb.
% \end{equation}
% By construction, $\matt{D}$ is row-stochastic, with each row normalized to unity:
% \begin{equation}
%     \sum_j D_{ij} = 1 \quad \forall i
% \end{equation}
% The last element of the $i^{\textnormal{th}}$ row of $\matt{D}$ is
% $\elemof{\left(\vfe{}\right)}{i}$, located at the diagonal entry $\elemof{D}{ii}$, which
% makes $\matt{D}$ an lower triangular matrix.
As $\matt{D}$ is a lower triangular matrix with all non-zero diagonal elements,
it is theoretically full rank and invertible. In practice, however, $\matt{D}$ is not effectively full rank. 
Its condition number (\cref{def:conditionnumber})
as function of matrix order is shown in~\cref{fig:condition_number}. The value is near constant, but
too large to be considered well-conditioned. Therefore, we must effectively treat the discrete
response $\mathbf{D}$ as being rank-deficient and ill-conditioned.

\begin{figure}
    \centering
    \includegraphics{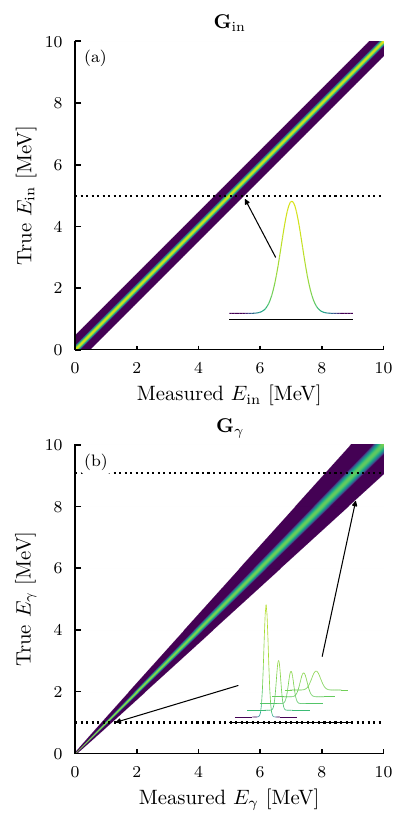}
    \caption{Examples of smoothing matrices for the initial-excitation-energy axis,
$\Gin$ (a), and gamma-ray-energy axis, $\Gg$ (b).
The inset axes show examples of the Gaussians at specific true energies marked with dotted lines.
The color scale (not shown) is logarithmic
with $0$ mapped to white.}
    \label{fig:smoothingmatrices}
\end{figure}

To model the smearing, we use a Gaussian smoothing matrix $\matt{G}\in\mathbb{R}^{n\times n}$, defined element-wise as
\begin{equation}\label{eq:gaussiancenters}
G_{ij} = \frac{1}{Z_i}\exp{\left[-\frac{\left(\elemof{E}{j} - \elemof{E}{i}\right)^2}{2\sigma(\elemof{E}{i})^2}\right]}\,,
\end{equation}
with the normalization\footnote{
This normalization is valid only when unfolding the full spectrum. If only a subregion is unfolded, the normalization must account for smearing into unobserved lower-energy regions. In practice, the most straightforward approach is to construct the full matrix $G_{ij}$ as described and then extract the relevant submatrix.}
\begin{equation}
    Z_i = \sum_{k=1}^{n}\exp{\left[-\frac{\left(\elemof{E}{j} - \elemof{E}{i}\right)^2}{2\sigma(\elemof{E}{i})^2}\right]}\,.
\end{equation}

Here $\mathbf{E}=\{E_1,E_2,\dots, E_n\}$ represents the energy bins, and $\sigma(E_i)$ denotes the resolution. Each row
$\mathbf{G}_i$ of $\matt{G}$ corresponds to a Gaussian centered at
$\elemof{E}{i}$. The normalization constant $Z_i$ ensures that the sum of elements in each row, $\sum_{j=1}^{n}G_{ij} =1$, thereby making $\mathbf{G}$ row-stochastic. The $\matt{G}$ matrices for OSCAR and SiRi are given in~\cref{fig:smoothingmatrices}.

%Because the energy values in $\mathbf{E}$ are distinct, the rows
%of $\matt{G}$ are unique and linearly independent; hence $\mathbf{G}$ is of full rank and invertible.
% \begin{theorem}\label{the:gaussian matrix}
%     Let $\mathbf{G}$ be a Gaussian matrix with entries given by \cref{eq:gaussiancenters}, where $\{E_1,E_2,\dots,E_n\}$ are distinct center points and $\sigma >0$ is a fixed constant. Then, the matrix $\mathbf{G}$ is of full rank and hence invertible.
% \end{theorem}

% \emph{Proof:} See (\citations{Appendix})

% \hfill $\square$

As the bin width $\Delta E$ of $\mathbf{E}$ narrows, the rows become increasingly similar, compromising their linear independence: 
\begin{enumerate}
        \item As $\Delta E$ becomes much smaller than $\sigma$, the Gaussian functions centered at adjacent energy bins overlap significantly. Consequently, the rows of $\mathbf{G}$ becomes nearly identical, leading to effective linear dependence. $\mathbf{G}$ loses its full rank and becomes non-invertible.
        \item When $\Delta E$ is much larger than $\sigma$, the Gaussian functions centered at each energy bin have minimal overlap. The rows of $\mathbf{G}$ remain distinct and linearly independent. As a result, $\mathbf{G}$ maintains full rank and remains invertible.
    \end{enumerate}
% \begin{corollary}
%     Let $\mathbf{G}$ be as in \cref{the:gaussian matrix}, and let
%     \begin{equation}
%         \Delta E=\underset{1\leq i\leq n}{\min}(E_{i+1}-E_i)\,.
%     \end{equation}
%     Then:
%     \begin{enumerate}
%         \item As $\Delta E$ becomes much smaller than $\sigma$, the Gaussian functions centered at adjacent energy bins overlap significantly. Consequently, the rows of $\mathbf{G}$ becomes nearly identical, leading to effective linear dependence. $\mathbf{G}$ loses its full rank and becomes non-invertible.
%         \item When $\Delta E$ is much larger than $\sigma$, the Gaussian functions centered at each energy bin have minimal overlap. The rows of $\mathbf{G}$ remain distinct and linearly independent. As a result, $\mathbf{G}$ maintains full rank and remains invertible.
%     \end{enumerate}
% \end{corollary}

% \emph{Proof:} See (\citations{Appendix})

% \hfill $\square$
% The Euclidean distance between adjacent rows follows
% \begin{equation}
%     || \vecc{G}_{i+1} - \vecc{G}_i ||^2 = 2\left(1 - \exp\left[-\frac{\Delta E^2}{4\sigma(E_i)^2} \right] \right).
% \end{equation}

\begin{figure}
    \centering
    \includegraphics[]{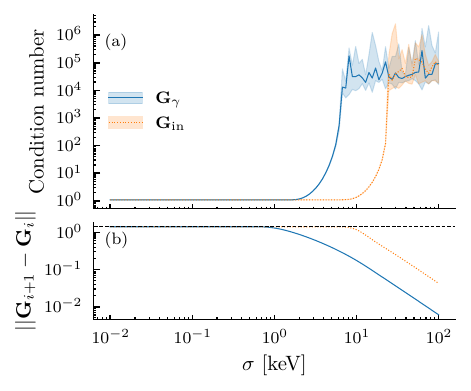}
    \caption{(a) Condition number and (b) mean row distance of Gaussian smoothing matrix $\matt{G}$ versus resolution parameter $\sigma$, with fixed bin width $\Delta E = \SI{10}{keV}$. 
    The dashed line in the bottom panel marks $\sqrt{2}$.
    The shaded bands represent the range of condition numbers across 100 instances with small perturbations, 
    simulating numerical fluctuations in matrix construction.
    The transition from constant to linear distance scaling occurs when $\sigma \approx \Delta E$, coinciding with rapid growth in condition number. For $\Gg$, the $\sigma_\gamma(\Eg)$
    is calibrated so that the mean $\sigma_\gamma$ over $\Eg$ equals $\sigma$.}
    \label{fig:condition_number_sigma}
\end{figure}

In the
regime where $\Delta E \gg \sigma$, the Gaussian distributions centered at $E_i$
and $E_{i+1}$ have negligible overlap. Each row $\mathbf{G}_{i}$ and
$\mathbf{G}_{i+1}$ is sharply peaked around $E_i$ and $E_{i+1}$, respectively.
The difference between adjacent rows resembles the difference
between two orthogonal unit vectors, and the distance saturates to
\begin{equation}
    || \mathbf{G}_{i+1} - \mathbf{G}_i || \approx \sqrt{2},
\end{equation}

while for $\Delta E \ll \sigma$, adjacent rows significantly overlap, and by converting discrete sums to integrals, the distance obey the scaling behavior
\begin{equation}
    || \mathbf{G}_{i+1} - \mathbf{G}_i || \approx \left(\frac{1}{4\sqrt{\pi}}\frac{\Delta E ^3}{\sigma^3}\right)^{1/2}.
\end{equation}
When the row distance becomes sufficiently small, floating point errors
accumulate, leading to a rapid increase in the condition number of $\matt{G}$ and
effective rank deficiency. \Cref{fig:condition_number_sigma} demonstrates
the ill-conditioning of $\mathbf{G}_{\gamma}$ and $\mathbf{G}_{\text{in}}$: at
low $\sigma$, the row distance remains constant at $\sqrt{2}$. Once
$\sigma$ becomes comparable to $\Delta E$, the distance decays linearly on
a log-log scale, accompanied by a sharp increase in the condition number.

\begin{figure}
    \centering
    \includegraphics[]{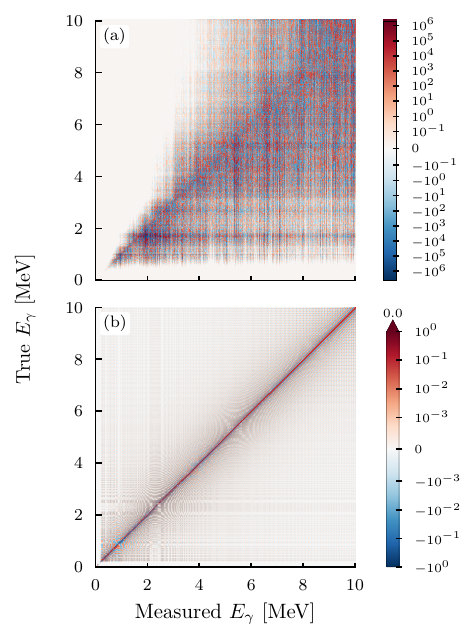}
    \caption{(a) The product $\Gg\Gg^{-1}$.
    (b) The product $\Gg\Gg^{\dagger}$, where $\Gg^{\dagger}$ is the
    Moore-Penrose pseudoinverse of $\Gg$. The latter product is six orders
    of magnitude
    closer the identity than the former. The colors are linear between 
    $\pm 10^{-1}$
    and $\pm 10^{-3}$ for the top and bottom plot, respectively.
    }
    \label{fig:Ginv}
\end{figure}

A direct consequence is that the inverse $\matt{G}^{-1}$ is not well defined. 
\Cref{fig:Ginv} demonstrates the effective rank deficiency of
$\mathbf{G}_{\gamma}$ in the form of
$\mathbf{G}_{\gamma}\mathbf{G}_{\gamma}^{-1}\neq \mathbf{1}$. Instead of an identity matrix,
the product fluctuates twelve orders of magnitude. The Moore-Penrose
pseudoinverse \mbox{$\mathbf{G}_{\gamma}\mathbf{G}_{\gamma}^{\dagger} = \matt{P}_\matt{G} \approx \mathbf{1}$}
is expected to be more well-behaved, which the lower panel confirms.

For the matrix $\Gin$, we set a constant width $\sigma_{\text{in}}$, which results in symmetric elements \mbox{$\elemof{\left(\Gin\right)}{ij} = \elemof{\left(\Gin\right)}{ji}$}. This symmetry of $\Gin$ is visible in the top panel of~\cref{fig:smoothingmatrices}.
If the values of $\vEx$
are equally spaced, $\Gin$ forms a symmetric Toeplitz matrix.
Furthermore, when $\sigma_{\text{in}}$ is sufficiently small, the elements
rapidly decay to zero, resulting in an approximately banded symmetric Toeplitz matrix.
These conditions are usually met in practice. 
Notably, all banded symmetric Toeplitz matrices commute,
a desirable property when solving inverse problems.

In contrast, the resolution of the {gamma-ray} energy axis  
is modeled by a square root quadratic function~\cite{oscarfabio}:

\begin{equation}
\sigma_{\gamma}:=\sigma\left(\Eg \right) = \sqrt{a_0 +a_1 \Eg{} + a_2 \Eg{}^2}.
\end{equation}
This energy dependency disrupts both the symmetry and the Toeplitz structure\footnote{
Convolutions are represented by (circular) Toeplitz matrices. Since our matrices do not have this structure, unfolding gamma-spectra should
not be called \textit{deconvolution}.
}
of the resulting matrix
$\Gg{}$, and $\Gg$-like matrices no longer commute.
An example of a $\Gg{}$ matrix is given in the bottom panel of~\cref{fig:smoothingmatrices}.

These properties---non-commutativity of the response matrices and the degeneracies they introduce---can undermine the reliability of standard methods. In the next section, we demonstrate how these factors can lead to significant inaccuracies when conventional techniques are applied and present a framework to address these challenges effectively.
%The coefficients $a_0, a_1$ and $a_2$ are determined by fitting to
%experimentally measured full energy peaks of known energy (see \cite{oscarfabio}).
%In practice,  each experiment requires its own minor calibration, specific to
%the measured spectrum.

%%%%%%%%%%%%%%%%%%%%%%%%%%%%%%%%%%%%%%%%%%%%%%%%%%%%%%%%%%

\section{Mitigating ill-posedness}
\label{sec:unfoldingspaces}

Here we derive algebraic limitations of gamma-spectrum unfolding and outline how to construct workable solutions. We begin by showing why the system’s ill-posedness cannot be resolved by using sharper Gaussian matrices. Accepting the experimental resolution, we map the unfolding problem to a smoothed space, which lifts much of the solution degeneracy. We then introduce a reparameterization that enforces nonnegativity and removes part of the degeneracy caused by the null space. Finally, we identify the remaining sources of variation and explain how regularization and uncertainty quantification fit into the overall workflow.

\subsection{Algebraic limitations of generalized transformations}

Having established that the structure of the double-sided matrix equation~\cref{eq:doublesided},
\begin{align}
    \boldsymbol{\nu} = \mathbf{G}_{\gamma} \mathbf{D} \boldsymbol{\mu} \mathbf{G}_{\text{in}}\,,
\end{align}
leads to a highly ill-posed problem due to the numerical properties of the involved matrices, we now turn to strategies for mitigating this ill-posedness.
In particular, we consider a transformation that later will prove instrumental in addressing the limitations of more explicit constraint- and regularization-based approaches.

\begin{figure}
    \centering
\begin{tikzpicture}[
  >=stealth,                    % nice arrow tips
  node distance=2.5cm,          % default separation
  circle node/.style={          % style for all circles
    draw, circle, minimum size=1.5cm, inner sep=1ex, 
    font=\small, align=center
  },
  arrow/.style={->}      % style for all arrows
]

  % 1) Define the five nodes
  \node[circle node] (Top)    at (0,  3) {$\vtau\in\rtau$};
  \node[circle node] (Left)   at (-3, 0) {$\vmu\in\rmu$};
  \node[circle node] (Center) at (0,  0) {$\retak$};
  \node[circle node] (Right)  at (3,  0) {$\veta\in\reta$};
  \node[circle node] (Bot)    at (0, -3) {$\vnu\in\rnu$};

  % 2) Three curved arrows out of Left
  \draw[arrow] (Top)  to[out=210, in=60]   node[midway, above left] {$\Psi(\vtau)$} (Left);
  \draw[arrow] (Left) to[out=60, in=120]   node[midway, below, yshift=-7pt] {$\Gl\vmu\Gk$} (Center);
  \draw[arrow, looseness=0.8] (Left) to[out=60,  in=120]   node[midway, above] {$\Gg\vmu\Gin$} (Right);

  % 3) Three straight arrows into Bottom
  \draw[arrow,dotted] (Center) -- node[right] {$\Rlk$} (Bot);
  \draw[arrow] (Left)   -- node[left]  {$\Gg\matt{D}\vmu\Gin$} (Bot);
  \draw[arrow,dotted,] (Right)  -- node[right,xshift=5pt] {$\matt{R}_\sigma$} (Bot);

\end{tikzpicture}
\caption{The relation between the optimization space $\rtau$, the peaked space $\rmu$, the smoothed space $\reta$, the folded space $\rnu$, and the intermediate ``sharper" space $\retak$. Each arrow represents a map from one space to another in the direction of the arrow. The dashed arrows for $\Rlk$ and $\matt{R}_\sigma$ indicate that these maps are only well defined in special cases.}
     \label{fig:diagram_muetanu}
\end{figure}
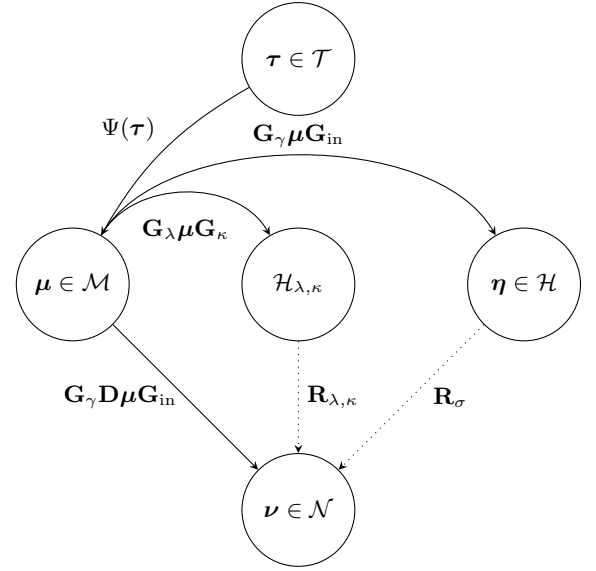

As shown in~\cref{fig:condition_number}, the matrices $\Gg$ and $\Gin$ are the primary contributors to the high condition number of the system and thus the main drivers of degeneracy in the solution space. In contrast, the smearing introduced by the discrete matrix $\matt{D}$ is far more pronounced than that of $\Gg$ and $\Gin$. This observation suggests a tradeoff: if one is willing to accept the Gaussian smearing imposed by $\Gg$ and $\Gin$, it is possible to reduce the ill-posedness of the inversion. Accordingly, rather than unfolding directly to the sharply peaked spectrum $\vmu \in \rmu$, we instead unfold to a smoothed representation $\veta \in \reta$, where
\begin{equation}
    \veta = \Gg\vmu\Gin\,.
\end{equation}
The relations are illustrated in the diagram in~\cref{fig:diagram_muetanu}.
We denote by $\matt{R}_\sigma$ the map from $\reta$ to $\rnu$, if such a map exists.

Before proceeding, one might question whether the specific choice of $\Gg$ and $\Gin$, corresponding to the experimental resolution, is overly restrictive. In practice, some approaches advocate using sharper Gaussian kernels to recover more detailed spectral features.\footnote{%
See, e.g., Ref.~\cite{sukosd_spectrum_1995} which appears to be the earliest work proposing this approach in the context of gamma-spectrum unfolding. Although the method is not presented in formal terms, it effectively corresponds to using Gaussian kernels narrower than the experimental resolution. Specifically, the authors advocate unfolding with artificially increased resolution and propose a heuristic rule for doing so, albeit without a theoretical justification. As demonstrated here, no universal rule of this kind can exist in general, though the underlying intuition is noteworthy.
}

To examine the validity of this idea, we temporarily generalize the
transformation $\rmu \to \reta$ in~\cref{fig:diagram_muetanu} by introducing
Gaussian smearing matrices $\Gl$ and $\Gk$. These matrices have Gaussian width parameters $\lambda$ and
$\kappa$, respectively, and play the roles of $\Gg$ and $\Gin$ \emph{only} in
the mapping from $\rmu$ to a new image space $\retak$. This
construction allows us to test whether alternative smoothing scales is possible. The analysis that follows shows that the
structure of the problem ultimately forces a return to the original choice
$\Gl=\Gg$ and $\Gk=\Gin$.

The mapping from $\retak$ to the observable space $\rnu$ is denoted by $\Rlk$. One might imagine unfolding to $\retak$ via $\Rlk$ instead of using the full operators $\Gg\matt{D}$ and $\Gin$. Yet the diagram imposes strict algebraic constraints, which turn out to preclude the existence of a well-defined $\Rlk$.

To obtain an expression for $\Rlk$, we require the diagram~\cref{fig:diagram_muetanu} to commute, i.e.\ the path $\rmu \!\to\! \retak \!\to\! \rnu$ must equal the direct path $\rmu \!\to\! \rnu$:
\begin{subequations}\label{eq:Rlk}
\begin{align}
    \overbrace{\Rlk\,\Gl\vmu\Gk}^{\rmu\to\retak\to\rnu} &= \overbrace{\Gg\,\matt{D}\vmu\Gin}^{\rmu\to\rnu}\\
    \Rlk\,\Gl\vmu\Gk\Gk^{-1} &= \Gg\,\matt{D}\vmu\Gin\Gk^{-1}\\
    \Rlk\,\Gl\vmu &= \Gg\,\matt{D}\vmu\Gin\Gk^{-1}. 
\end{align}
\end{subequations}
To obtain a bounded operator $\Rlk$ that is independent of $\vmu$, the right-hand factor must cancel, 
requiring $\Gin\Gk^{-1}=\mathbf{1}$ and thus forcing $\Gk=\Gin$.
Although we write $\Gk^{-1}$ and use $\Gk\Gk^{-1}=\mathbf{1}$ formally, this identity holds only in exact arithmetic; in practice, the Gaussian matrices are too ill-conditioned to be inverted.

Because~\cref{eq:Rlk} holds for all $\vmu\in\rmu$, we focus solely on the operators. Multiplying by $\Gl^{-1}$ on the right
yields

\begin{equation}
    \Rlk = \Gg\matt{D}\Gl^{-1} \;=\; \matt{D}\Gg\Gl^{-1} \;-\; \bigl[\matt{D},\Gg\bigr]\Gl^{-1}.
\end{equation}
Here the commutator \([\matt{D},\Gg]\) is taken in the matrix algebra acting on the gamma-energy bins. Since both operators are \(n \times n\) matrices in the same space, the bracket is well-defined and quantifies their failure to commute.

There are essentially two limiting regimes to consider.  
Although an intermediate range exists in which $\lambda$ is non-negligible yet still too small to drive the condition number of $\Gl$ to extreme values, this window is practically irrelevant because the condition number of $\Gl$ already exceeds $10^{5}$ before the resolution reaches $\sigma=\SI{10}{keV}$ (well below the typical OSCAR resolution), so its behavior is indistinguishable from the near-identity case.

In the limit $\lambda \to 0$, $\Gl$ approaches the identity, and $\retak$ collapses to the trivial space $\rmu$.  
As $\lambda$ increases, the condition number of $\Gl$ grows rapidly (see~\cref{fig:condition_number_sigma}), making $\Rlk$ increasingly ill-defined.  
For $\Rlk$ to remain well defined, every term must stay bounded.  
A necessary, though not sufficient, requirement is that the product $\Gg\Gl^{-1}$ in the first term equals the identity, which forces $\Gl = \Gg$.

With both $\Gk = \Gin$ and $\Gl = \Gg$, the space $\retak$ coincides with $\reta$, and the map $\Rlk$ reduces to
\begin{equation}
  \label{eq:commutatorerror}
  \matt{R}_\sigma \;=\; \matt{D} \;-\; \bigl[\matt{D},\Gg\bigr]\Gg^{-1}.
\end{equation}

\begin{figure}
  \centering
  \includegraphics[]{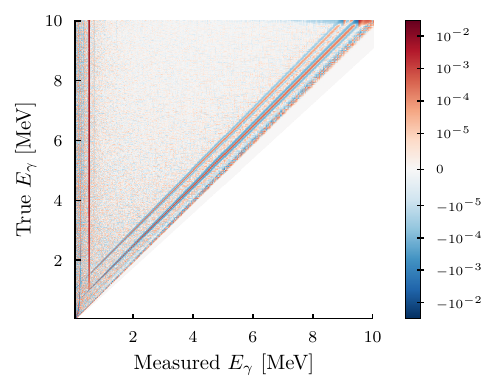}
  \caption{The commutator $[\matt{D},\Gg]$.  
  The largest feature is the vertical peak at \SI{511}{keV} caused by the annihilation peak.  
  Other discrete structures also fail to commute but are several orders of magnitude weaker.  
  The Compton background’s contribution is indistinguishable from numerical noise, except for its low-energy peak.}
  \label{fig:commutator}
\end{figure}

Whether $\matt{R}_\sigma$ is well defined depends on the second term.  
The previous section showed that neither $\matt{D}$ nor $\Gg$ has a structure that allows them to commute, and numerical tests confirm this result, see~\cref{fig:commutator}.  
The dominant feature of the commutator is a vertical peak at \SI{511}{keV}, which arises from the annihilation peak’s energy-independent position.  
All other features are oscillations that are several orders of magnitude smaller.

The nonzero commutator $[\matt{D},\Gg]$ shows that smearing and Gaussian smearing are order dependent, so this residue cannot be removed by any redefinition of $\matt{R}_\sigma$.  
Moreover, the inverse $\Gg^{-1}$ rarely exists in practice because its condition number is extremely large.  
Consequently, $\matt{R}_\sigma$ cannot be well defined in practical computations.

The identifications $\Gl = \Gg$ and $\Gk = \Gin$ are also motivated by practical considerations. Using a resolution sharper than the experimental resolution leaves residual degeneracy in the solution space that must otherwise be removed, a near impossible task in practice.

While methods such as those in Refs.~\cite{sukosd_spectrum_1995,GUTTORMSEN1996371} do not explicitly define a mapping
of the form $\Rlk$, the operational structure of their algorithms suggests that such a transformation is assumed.
Our derivation shows that this assumption leads to unavoidable structural error. Fortunately, a direct unfolding via $\Rlk$ is unnecessary due to a transformation trick, which we now describe.

\subsection{Constructing physically constrained solutions}
\label{subsec:constructingconstrained}

The previous analysis established that unfolding directly to the smoothed space
$\reta$ is only rarely well defined. Rather than unfolding to $\reta$ explicitly via the
ill-defined map $\matt{R}_\sigma$, we instead formulate the optimization in the degenerate
space $\rmu$ and recover physically meaningful quantities through the mapping to
$\reta$, lifting the degeneracy.

To enforce non-negativity, the optimizer operates in an unrestricted space
$\mathcal{T}\subseteq\mathbb{R}^{N\times M}$, producing a candidate solution
${\htau} \in \mathcal{T}$. This candidate is mapped to
$\rmu\subseteq\mathbb{R}^{N\times M}_+$ by a reparameterization $\Psi$
(discussed in the next subsection), and then to $\reta$ and $\rnu$ through the
known operators. Specifically, we construct $\heta = \Gg \Psi({\htau}) \Gin$ and
$\hnu = \Gg \matt{D} \Psi({\htau}) \Gin$. The comparison to data is performed in
$\rnu$, while regularization terms can be applied in either $\rmu$ or $\reta$,
depending on the nature of the constraint.

\begin{figure}
    \centering
    \includegraphics[]{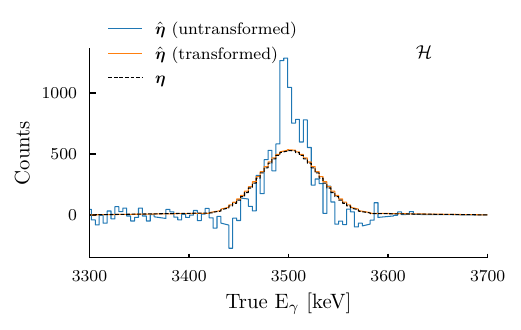}
      \caption{Unfolding without using the transformation nor reparameterization yields an $\heta$ that lacks the correct bin-to-bin correlations and non-negativity constraints. By contrast, unfolding with the transformations produces a solution that correctly preserves the expected correlations.
      The label $\mathcal{H}$ is included to clarify that the vectors shown reside in the space $\mathcal{H}$. The $\heta$s were
      found by RMLE as described in~\cref{sec:modelest}.}
    \label{fig:correlations}
\end{figure}

This setup allows us to construct spectra in $\reta$ through a mathematically justified transformation, rather than attempting to solve the inverse problem directly in that space. Importantly, the expected correlations between bins in $\heta$—which are difficult to enforce through regularization alone—are inherently preserved by the transformation. An example is shown in~\cref{fig:correlations}.

Obtaining $\hnu$ from $\heta$ is not possible, as the operator $\matt{R}_\sigma$ does not exist. In practice, however, this is rarely a limitation. Instead, the solution can be stored as either $\htau$ or $\hmu$, from which $\heta$ and $\hnu$ can be recovered through the known operators. It is important to note that neither $\htau$ nor $\hmu$ has a direct physical interpretation due to the degeneracy of the inverse problem that we have accepted in the tradeoff.

A substantial benefit of this transformation is that it introduces global correlations across the spectrum, governed by the expected experimental resolution. The implications of this are examined in~\cref{subsec:gin}.

%\placeholderfig{unfoldingdiagram}{Diagram of the unfolding process. A candidate
%solution $\htau$ is first proposed by the optimizer in the space $\rtau$. This
%solution is then mapped to $\rmu$ using the link $\Psi$, which ensures that all
%bin values are non-negative. From there, $\hmu$ is mapped into two different
%spaces. It is mapped into $\hnu$ in the space $\rnu$, where it is compared to
%the data $\vn$ using the likelihood function. Simultaneously, $\hmu$ is mapped
%into $\heta$ in the space $\reta$, which will be given to the user when the
%optimization is complete. The penalty terms $\Omega(\hmu)$ and $\Omega(\heta)$
%are applied separately in their respective spaces. Because $\heta$ is
%constructed explicitly from $\Gg$, the correct correlations between
%the bins in $\heta$ are automatically preserved. When
%the map $\matt{R}_\sigma'$ is well-defined, $\heta$ can be reliably mapped to
%$\hnu$. }

\subsection{Null space degeneracy}
\label{subsec:paramregular}
\begin{figure}
    \centering
    \includegraphics[]{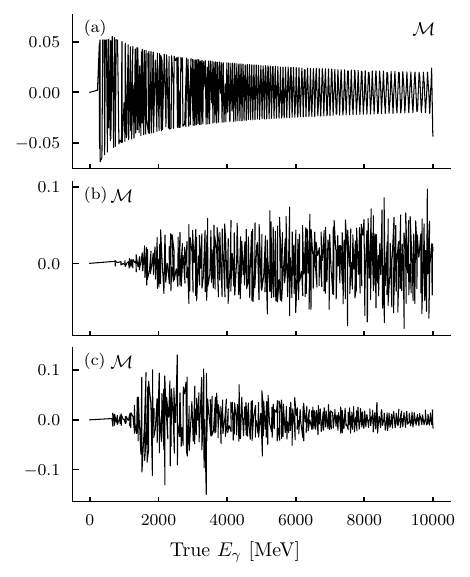}
    \caption{
Three null vectors of \(\Gg\matt{D}\), computed with a relative condition number
\(\varepsilon = 10^{-5}\). These vectors oscillate between negative and positive
values, but fold to $\vecc{0}$ in observable space $\rnu$.}
    \label{fig:nullvectors}
\end{figure}

Null vectors present a fundamental challenge by creating degeneracy in the
solution space. The effective null space of $\Gg$ is substantial: for a
$2000\times 2000$ matrix with a relative condition number\footnote{While the
matrices are theoretically full rank, the null space issues of $\matt{G}$-like
matrices are \textit{numerical} in nature. The relative condition number
$\varepsilon$ determines which vectors are effectively treated as zero.
Specifically, for the largest singular value $\sigma_{\text{max}}$, any singular
value less than $\varepsilon \cdot \sigma_{\text{max}}$ is considered zero.} of
$10^{-5}$, the null space dimension exceeds $1500$. Three examples of elements of
the null space are shown in~\cref{fig:nullvectors}. They are all oscillatory about
zero, as expected
by~\cref{prop:nullvectors}.

\begin{figure}
    \centering
    \includegraphics[]{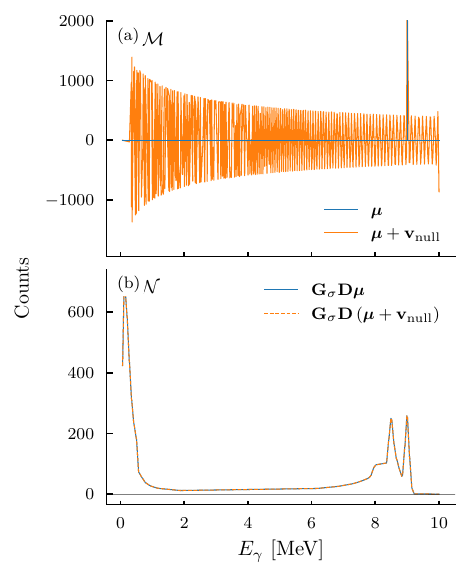}
    \caption{
Comparison between a sharp peak \(\vmu\) and \(\vmu + \vecc{v}_{\text{null}}\) for a $\vecc{v}_{\text{null}}$(a),
alongside their folded counterparts \(\Gg\matt{D}\vmu\) and \(\Gg\matt{D}(\vmu +
\vecc{v}_{\text{null}})\) (b). While the addition of $\vecc{v}_{\text{null}}$ gives large oscillations in
$\rmu$,  the folded vectors are identical (up to a bin-difference $\propto 10^{-3}$), showing the
issue caused by null-space vectors. The y-axis of the upper plot as been truncated.}
    \label{fig:nullfolding}
\end{figure}

\begin{figure}
    \centering
    \includegraphics[]{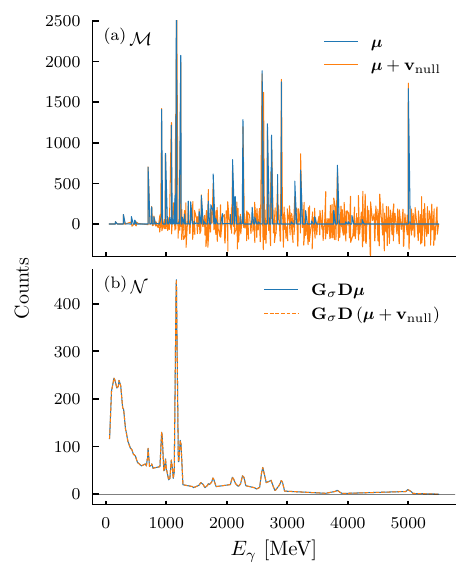}
    \caption{Similar comparison as~\cref{fig:nullfolding} but for simulated \ce{^{120}Sn}, showing the impact of null-space vectors on a realistic spectrum. The identical folded spectra (b) demonstrate that the null-space degeneracy persists even in complex, physically-motivated cases. The y-axis in (a) has been truncated.}
    \label{fig:nullfolding2}
\end{figure}

\begin{figure}
    \centering
    \includegraphics[]{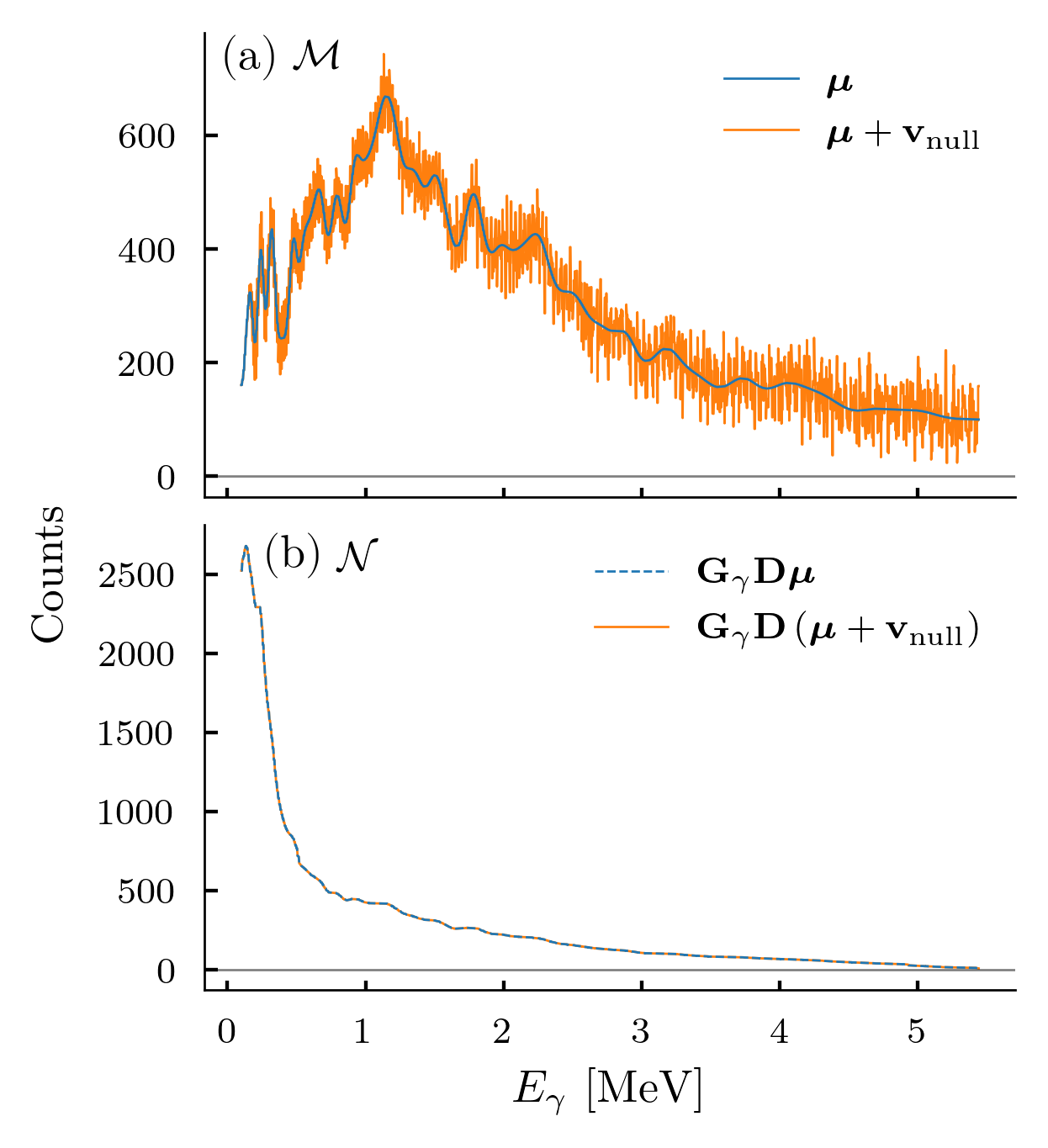}
    \caption{Demonstration that null-space vectors can affect even strictly positive solutions. Despite both solutions being non-negative in (a), their difference is a null vector, resulting in identical folded spectra (b).}
    \label{fig:nullfolding3}
\end{figure}

A non-empty null space renders the unfolding problem \textit{non-identifiable} (see~\cref{sec:regularizedunfolding}).
Given any data vector $\my$, recovering the true $\vmu$ becomes impossible
without additional constraints or information. This non-identifiability is illustrated
in~\cref{fig:nullfolding,fig:nullfolding2,fig:nullfolding3}, which compare
$\vmu$ solutions with and without null vector components. Despite substantial
differences in the $\vmu$ vectors when null vectors are added, their folded
counterparts in observable space remain identical. This demonstrates a critical
limitation: optimizing or evaluating solutions solely in $\rnu$ will almost
inevitably lead to unphysical solutions distorted by null vector components.

The degeneracy of the solution space can be partially addressed by exploiting
the oscillatory nature of the null vectors. While enforcing non-negativity
constraints does not completely exclude null vectors from the solution space (as
long as solution components remain sufficiently close to zero), it can help
regularize the problem. This is implemented by using a reparameterization to
transform the solution space to be non-negative.

As the data follows a Poisson distribution, an exponential reparameterization
$\Psi(\vtau) = \exp(\vtau)$ is a natural choice, but it leads to optimization
instability. Instead, we employ a quadratic reparameterization $\Psi(\vtau) =
\vtau^2$, which provides better numerical stability.%\footnote{The quadratic
%function also simplifies the transformation of covariance matrices, with
%$\text{cov}(\vmu) = \text{cov}(\vtau)^2$, due to the moment expansion of
%$\Psi$.}
See~\cref{app:math} for more discussion on reparameterizations.

Optimizing over $\vtau$ allows the optimizer (see the next section for details on the optimizer) to traverse the parameter space
without encountering hard boundaries that could cause convergence issues, or
soft constraints that might permit small negative values. This transformation
restricts the search space by eliminating unphysical negative components
associated with null-space vectors. Although the exact reduction in search space
dimensionality is difficult to quantify, this approach improves convergence in
cases where unconstrained optimizations fail. However, as shown
in~\cref{fig:nullfolding3}, the presence of null vectors is not completely
eliminated—a solution can remain strictly positive while still containing null
vector components.

\subsection{Sources of variation}
\label{sec:fluctuations}
In previous works on gamma-ray unfolding (e.g., \cite{sukosd_spectrum_1995,GUTTORMSEN1996371}) the term \enquote{fluctuations} has been used ambiguously to describe several distinct phenomena. This ambiguity has led to confusion in interpreting results and comparing methodologies. We propose a more precise terminology to distinguish between six sources of variation:
\begin{enumerate}
    \item \textit{Spectral complexity}, which is the intrinsic shape of $\vmu$.

    \item \textit{Stochastic variation}, arising from the Poisson nature of the prompt data \mbox{$\Ps \sim \text{Poisson}(\vnu+\vbeta)$}, where $\vbeta$ is the background spectrum.
    
    \item \textit{Background error}, which captures the systematic bias introduced when using 
    an observed background $\B_i$ in place of the true (unobservable) background $\B$.
    We discuss the background in~\cref{sec:bg}.
    
    \item \textit{Model degeneracy}, which describes the fundamental non-uniqueness in solutions $\hmu$ due to the non-empty null space of $\matt{R}$. 
    
    \item \textit{Monte Carlo variance}, which appears when using Monte Carlo (MC) simulations to construct confidence intervals.
    The MC method is presented in~\cref{sec:uncquant}.

    \item \textit{Variance in $\matt{R}$}. Model uncertainty in $\matt{R}$ will affect the solution space. This lies outside the scope of this work, but should be modeled. %, but is discussed briefly in~\cref{sec:responsediscrepancy}.
\end{enumerate}

Spectral complexity refers to variations intrinsic to the underlying signal
structure -- $\vmu$, $\veta$, or $\vnu$ --
distinct from stochastic noise introduced during measurement or data processing.
This complexity can be quantified by analyzing the properties of $\veta \in
\reta$, see~\cref{app:smoothness}.

\begin{figure}
    \centering
    \includegraphics{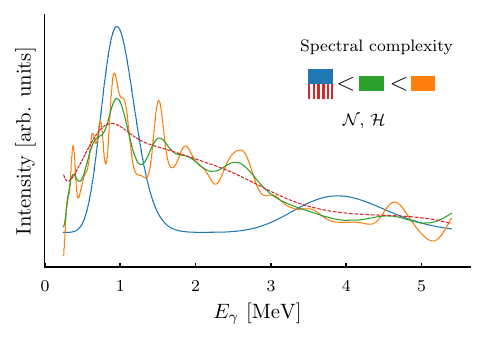}
    
\caption{%
Illustration of spectral complexity categories. The orange curve exemplifies a non-smooth spectrum %
%with high $s_{\text{max/med}}$ and high $s_{\text{std}}$,
characterized by non-isolated peaks with varying amplitudes. The blue curve represents a pseudo-smooth spectrum %
%(high $s_{\text{max/med}}$, low $s_{\text{std}}$)
where sharper, isolated peaks are present within a generally smooth profile. The red curve demonstrates a smooth spectrum %
%(low $s_{\text{max/med}}$, low $s_{\text{std}}$)
with low first and second derivatives, indicating a lack of sharp features. The green curve exhibits an intermediate level of complexity between the orange and red examples. Notably, all curves lack high-frequency bin-to-bin fluctuations, as their shape is determined by the underlying function space of $\veta$.}
    \label{fig:complexity}
\end{figure}

\begin{figure}
    \centering
    \includegraphics{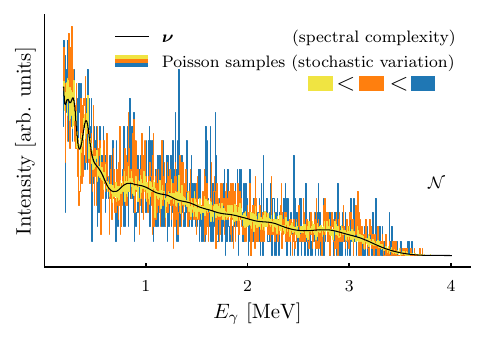}
    \caption{The spectral complexity compared to three different
    stochastic variations. Stochastic variation is a property of the 
    experiment, while the spectral complexity is independent of the experiment. The
    stochastic variations are Poisson samples of the
    same spectrum at different number of counts, normalized
    to be visually comparable.}
    \label{fig:stochastic variation}
\end{figure}

The spectral complexity is the most important feature when selecting a regularization procedure. We study this extensively in~\cref{subsec:low_spectral_complexity}--\ref{sec:regularization_systematics}.
The fluctuations in the observed experimental data $\mathbf{Y}$ can be decomposed into two distinct components: spectral complexity, represented by variations in the true expected spectrum $(\boldsymbol{\nu} + \boldsymbol{\beta})$, and stochastic variation, arising from Poisson sampling noise:
\begin{align}
\my &= \underbrace{(\vnu + \vbeta)}_{\text{Spectral complexity}} + \underbrace{(\my - (\vnu + \vbeta))}_{\text{Stochastic variation}}.
\end{align}
\Cref{fig:complexity,fig:stochastic variation} illustrate the difference.

Without any prior model on $\vnu+\vbeta$, these two components cannot be separated.\footnote{In fact, \emph{denoising} $\my$ is an equally hard problem as unfolding to $\reta$.} 
As a direct consequence, the bin-to-bin fluctuations give almost no information on how to regularize $\heta$.
On the contrary, it is instead the \emph{prior} assumptions of Poisson distribution and modeling of the background that 
allow for the separation of the spectral complexity and stochastic variation.

The estimated unfolded solution does not admit a simple additive variance decomposition. However, for illustrative purposes, we can approximate its variance as follows:

\begin{align} 
\text{Var}(\hmu) &\approx \underbrace{\text{Var}(\hmu|\vnu,\vbeta)}_{\text{Stochastic variation}} + \underbrace{\text{Var}(\hmu|\B_i \neq \B)}_{\text{Background error}}\notag\\
&\quad + \underbrace{\text{Var}[\hmu|\operatorname{Ker}(\Gg\matt{D}\Gin)]}_{\text{Model degeneracy}} + \underbrace{\text{Var}{\text{MC.}}}_{\text{MC variation}} 
\end{align}
While these contributions are not statistically independent and may exhibit substantial covariance, we do not attempt to quantify those cross-terms here. This decomposition is meant to provide a rough indication of the dominant sources of variance.
Similar decompositions apply to $\hnu$ and $\heta$.

Notably, the spectral complexity does not appear explicitly in the decomposition of $\operatorname{Var}(\hmu)$. This is because the spectral shape is intrinsic and unchanging for a given $\vmu$. When these sources of variation are properly accounted for, they influence the geometry of the solution space rather than any individual point within it. A specific point—such as $\heta$—will always reflect the spectral complexity corresponding to its location in the space. However, the collection of possible solutions inherits this complexity structure, which is implicitly shaped by the nature of the variations listed above.

\begin{figure}
    \centering
    \includegraphics{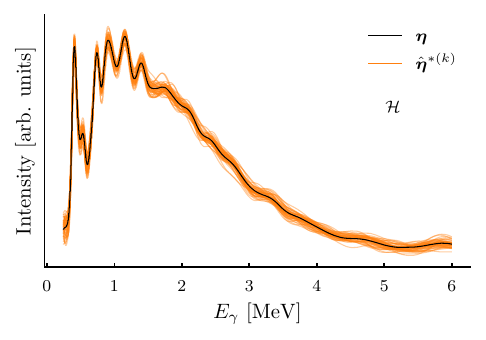}

\caption{Spectral complexity compared to $\operatorname{Var}(\heta)$. 
Each Monte Carlo sample $\heta^{*(k)}$ represents a draw from the allowed solution space, with each sample exhibiting the same spectral complexity. The ensemble variance $\operatorname{Var}(\heta)$ reflects the spread of these solutions. A properly constrained solution space should only permit spectral complexities consistent with that of the true solution.}
    \label{fig:mcvariation}
\end{figure}

In \cref{fig:mcvariation} the true $\veta$ is shown together with an ensemble of MC $\heta^*$. While each ensemble member exhibits similar spectral complexity, the ensemble as a whole expresses the variance of the solution space. It is this ensemble-level spread—not the fluctuations of individual solutions—that reflects the full extent of uncertainty.

Unfolding methods that fail to account for these variations produce solutions that are explicitly affected by them. For instance, Richardson’s iteration (see \ref{sec:richardson}) tends to overfit to statistical noise, causing stochastic variation to appear directly in the solution. Backgrounds subtracted naively allow the background error to propagate
(see~\cref{eq:condmuback} and~\cref{sec:bg}). Finally, the model degeneracy would not be accounted for, causing oscillatory features of the null space to appear, as illustrated in \cref{fig:nullfolding,fig:nullfolding2,fig:nullfolding3}.

Among the variance components, the MC variation is the most straightforward to
control, as it decreases with the number of samples. Both stochastic
variation and background error scale with the Poisson uncertainty in each bin,
roughly $\sqrt{Y_{ij}}$, which decreases as data counts increase.
Regularization suppresses variance contributions from stochastic fluctuations, background mismodeling, and degeneracy, but at the cost of introducing bias.
In the absence of regularization, stochastic variation and background error are greatly amplified by the condition number.
Model degeneracy, in
contrast, is unbounded, as demonstrated in \cref{sec:discretization}, particularly by
\cref{eq:condmuback}.

%%%%%%%%%%%%%%%%%%%%%%%%%%%%%%%%%%%%%%%%%%%%%%%%%%%%%%%%%%%%%%%%%%%%%%%%%%%%
\section{Unfolding method}
\label{sec:modelest}

\noindent\begin{minipage}{\columnwidth}
\NotationBlock{
    \notation{\Gg}{Smearing matrix along gamma-energy axis.}
    \notation{\Gin}{Smearing matrix along the initial-excitation-energy axis.}
    \notation{\matt{D}}{Discrete response.}
    \notation{\vtau}{Optimization variable}
    \notation{\vmu}{The expectation value of the unfolded spectrum.}
    \notation{\veta}{The expectation value of the unfolded spectrum refolded by $\Gg$ and $\Gin$.}
    \notation{\vnu}{The expectation value of the folded spectrum.}
    \notation{\matt{Y}}{The observed data \mbox{$\matt{Y}\poi{\vnu=\Gg\matt{D}\vmu\Gin}$}.}
    \notation{\rmu,\reta\\ \rnu,\rtau}{The spaces which $\vmu, \veta, \vnu$ and $\vtau$ belong,
    respectively.}
    \notation{\Psi}{Non-negativity constraint $\vmu = \Psi(\vtau)$.}
}
\end{minipage}

\subsection{Regularized maximum likelihood estimation}

\begin{algorithm}[H]
\caption{Maximum Likelihood Unfolding with Regularization}
\label{alg:mlunfolding}
\newcommand{\algbf}[1]{\texttt{#1}}
\begin{algorithmic}[1]
\Require 
    \Statex \quad Observed data matrix $\my$
    \Statex \quad Discrete response matrix $\mathbf{D}$
    \Statex \quad $\Eg{}$ smoothing matrix $\Gg$
    \Statex \quad $\Ex$ smoothing matrix $\Gin$
    \Statex \quad Regularization parameters $\vecc{\theta}$
    \Statex \quad Detector efficiency $\varepsilon(\Eg)$
\Ensure 
    \Statex \quad Optimized unfolded spectrum $\heta$

\State \algbf{Initialize} 
    \Statex \quad Find $\sigma_\text{max}$ from $\operatorname{SVD}(\Gg\matt{D})$
    \Statex \quad Set step size $d\tau < \frac{1}{2\sigma^2_{\text{max}}}$
    \Statex \quad Compute bound $\mu_{\text{max}} = \frac{||Y||_1}{||\Gg||_1 ||\matt{D}||_1 ||\Gin||_1}$
    \Statex \quad Set initial guess $\tau_i \leftarrow \operatorname{Uniform}(10^{-1}, \mu_{\text{max}})$
    \Statex \quad Compute initial $\hmu \leftarrow \Psi(\vtau)$

\While{not converged}
    \State \algbf{Fold}
    \Statex \qquad  $\hnu \leftarrow \Gg\matt{D}\hmu\Gin$
    \State \algbf{Compute Loss Function} $L(\hmu, \vn; \vecc{\theta})$
    \Statex \qquad $L \leftarrow \sum_{ij} \nu_{ij} -
    Y_{ij}\log\left(\nu_{ij}\right) +
    \Omega({\hmu ; \vecc{\theta}})$
    
    \State \algbf{Calculate Gradients (JAX)}
    \Statex \qquad $\nabla_{\vtau} L$%$ = \frac{\partial L}{\partial \elemof{\tau}{i}}$
    
    \State \algbf{Update Parameters (NAdam)}
    \Statex \qquad $\vtau \leftarrow \text{Optimizer}(\vtau, d\tau, \nabla_{\vtau}L)$
    \Statex \qquad $\hmu \leftarrow \Psi(\vtau)$

    \State \algbf{Check Convergence}
    \If{Change in $L$ below threshold\\
     \qquad\textbf{or} maximum iterations reached}
        \State \algbf{Break}
    \EndIf
\EndWhile

\State \algbf{Correct for detector efficiency}
\Statex \quad $\hmu \leftarrow {\hmu}/{\vecc{\varepsilon}}$
\State \textbf{Return}
\Statex \quad $\hmu$
\Statex \quad \mbox{$\heta \leftarrow \Gg\hmu\Gin$}
\end{algorithmic}
\end{algorithm}

% ML
Having established the theoretical and practical challenges of the inverse
problem, we now present the regularized maximum
likelihood method for unfolding gamma-ray spectra.
A summary of the unfolding method is given in~\cref{alg:mlunfolding}.

The unfolded spectrum is obtained by minimizing a loss function, $L(\hmu, \my)$, which balances the fit
to the data and expected physical characteristics of the solution using gradient descent.
The loss consists of the Poisson log-likelihood, $\ell(\vmu)$ (see \cref{eq:loglikedef}),
along with an added penalty term, $\Omega\left(\hmu;\vecc{\theta} \right)$, to enforce the physicality:
\begin{align}
    L(\hmu, \my ;\vecc{\theta} ) &= \ell(\hmu | \my) + \Omega\left(\hmu; \vecc{\theta}\right),
    \label{eq:lossdef}
    \intertext{where}
    \ell\left(\hmu | \my\right) &=  \Gg\matt{D}\hmu\Gin - \my\log\left(\Gg\matt{D}\hmu\Gin\right),
\end{align}
and $\vecc{\theta}$ are the penalization parameters.\footnote{Both the likelihood and different
forms of regularizations contain $\log$ and 
divisions by matrices. Both are susceptible to numerical instability, and must be stabilized by the
addition of small numerical constants. These are not shown in the equations to avoid visual clutter.}
Some convergence requirements on the optimization step size are derived in~\cref{sec:convergence}.

The choice of regularization strength should be guided by empirical expectations regarding the spectrum’s structure.
Discrete regions are expected to exhibit sharp peaks,
continuous regions should show smooth transitions, and the quasi-continuum
combines both smoothness and structure. At low initial excitation energies, the level
density---and consequently the number of possible \mbox{gamma transitions}---is low,
leading to high sparsity. In contrast, the continuum is characterized by a lack
of sparsity, while the intermediate quasi-continuum presents a challenge, as its
sparsity decreases with increasing initial excitation energy. To account for these
variations in the spectrum, the strength of the regularization parameter can be
adjusted based on initial excitation energy using simulated representative spectra.

The penalty term used for regularization should be chosen according to the characteristics of the spectrum. 
For a discrete spectrum, where sparsity is expected, a sparsity‑promoting cost
encourages fewer non‑zero elements in the $\hmu$‑spectrum. 
Although an $\ell_1$ penalty promotes sparsity, it
can also attenuate important peaks. A smoother alternative that retains salient
peaks while still favoring sparsity is a sigmoid‑shaped penalty, for example,
based on the arctangent,
\begin{equation}
  \Omega(\hmu;\vecc{\theta}) =
  \theta_0 \sum_{ij}
  \frac{1}{2}\!\left(
    1 + \frac{2}{\pi} \arctan\!\left(
      \frac{\mu_{ij} - \theta_1}{\theta_1/\theta_2}
    \right)
  \right),
\end{equation}
where $\theta_0$ controls the overall regularization strength, $\theta_1$ sets the lower threshold, and $\theta_2$ determines the width of the sigmoid. 

To promote smoothness in the estimated spectrum \(\heta\), a generalized Sobolev penalty can be applied to penalize high-frequency variations. The regularization term is given by
\[
  \Omega(\hat\mu;\boldsymbol{\theta}) \;=\;
  \sum_{k=1}^{K}\theta_{k}
  \sum_{i,j}
    \bigl|\bigl(\nabla^{k}\,G_{\gamma}\,\hat\mu\,G_{\mathrm{in}}\bigr)_{ij}\bigr|^{2},
\]
where \(\nabla^{k}\) denotes the \(k^{\text{th}}\) discrete derivative operator.
This formulation allows for control over which frequency components are
penalized, with higher \(k\) targeting increasingly rapid oscillations in
\(\heta\).

A complementary strategy is to minimize or maximize the \textit{entropy},
\begin{equation}
  \Omega(\hmu;\theta) =
  -\theta \sum_{ij}
  \Bigl(G_\gamma \hat\mu G_{\text{in}}
        \log\bigl[G_\gamma \hat\mu G_{\text{in}}\bigr]\Bigr)_{ij},
\end{equation}
where the sign of $\theta$ selects the desired behavior: minimizing the entropy
($\theta>0$) sharpens the distribution, effectively mimicking the sigmoid
sparsity penalty, whereas maximizing it ($\theta<0$) encourages a broader,
smoother spectrum similar to Sobolev regularization.

While it is in principle possible to combine penalty terms to accommodate spectra with both discrete and continuous features, selecting regularization strengths that yield consistently good results remains challenging in practice. For data of the type measured in Oslo experiments, this remains an open problem, as discussed in~\cref{sec:sims}.
A comment on how to select regularization method and regularization strength is 
given in~\cref{sec:regularization_systematics}.

% where the singular values $\sigma_{\text{min}}$ and $\sigma_{\text{max}}$ 
% are found by singular value decomposition.
% \erlend{this is not true for regularized since the loss landscape is changed}

% Regularization
%**Enforcing Non-Negativity to Avoid Null-Space Solutions**
% \erlend{rewrite to flow with theory}
% In solving inverse problems for counting experiments, it is important to ensure
% that the solution vector \(\hat{\vmu}\) has no negative elements, as negative
% counts are physically meaningless. The response matrix
% \(\matt{R}\) has a non-empty null space, which can introduce solutions
% that differ only by a null-space component, making them indistinguishable when
% folded back into the observable space.

% For any solution \(\hat{\vmu}\), adding a null-space vector
% \(\vecc{v}_{\text{null}}\) gives another indistinguishable solution:

% \begin{equation}
% \matt{R}\left(\hat{\vmu} + \vecc{v}_{\text{null}}\right) = \matt{R}\hmu
% + \matt{R}\hmu_{\text{null}} = \hat{\vnu} + \vecc{0}. 
% \end{equation}

% Many null-space vectors oscillate between negative and positive values, causing
% spurious solutions with non-physical negative components.
% \cref{fig:nullvectors} demonstrates these oscillations, and
% \cref{fig:nullfolding} shows how such solutions lead to indistinguishable
% folded vectors, underscoring the need to exclude null-space components.

% To avoid negative elements and null-space solutions, we transform the search
% space by introducing an unconstrained variable \(\vtau \in \rtau \) and a link function
% \(\Psi\) such that $\vmu = \Psi(\vtau)$.

\subsection{Effect of $\Gin{}$}
\label{subsec:gin}

Standard  analyses of Oslo gamma-spectra have addressed $\Gin$ smearing
separately from the main unfolding framework. This approach typically manages
count sparsity through $\Ex$-axis rebinning, increasing statistics per row at
the cost of introducing discretization artifacts. While the effects of $\Gin$
may appear less pronounced than those of $\Gg\matt{D}$, excluding them from the
unfolding process overlooks important aspects of the measurement structure.

The forward model $\vnu = \Gg\matt{D}\vmu\Gin$ shows that $\Gin$ is an integral
part of the spectrum formation. While $\Gg$ induces smearing and correlations along the
$\Eg$ axis, $\Gin$ plays an equivalent role along the $\Ex$ axis. A complete
inversion of this process must therefore account for both transformations.

\begin{figure*}
    \centering
    \includegraphics[]{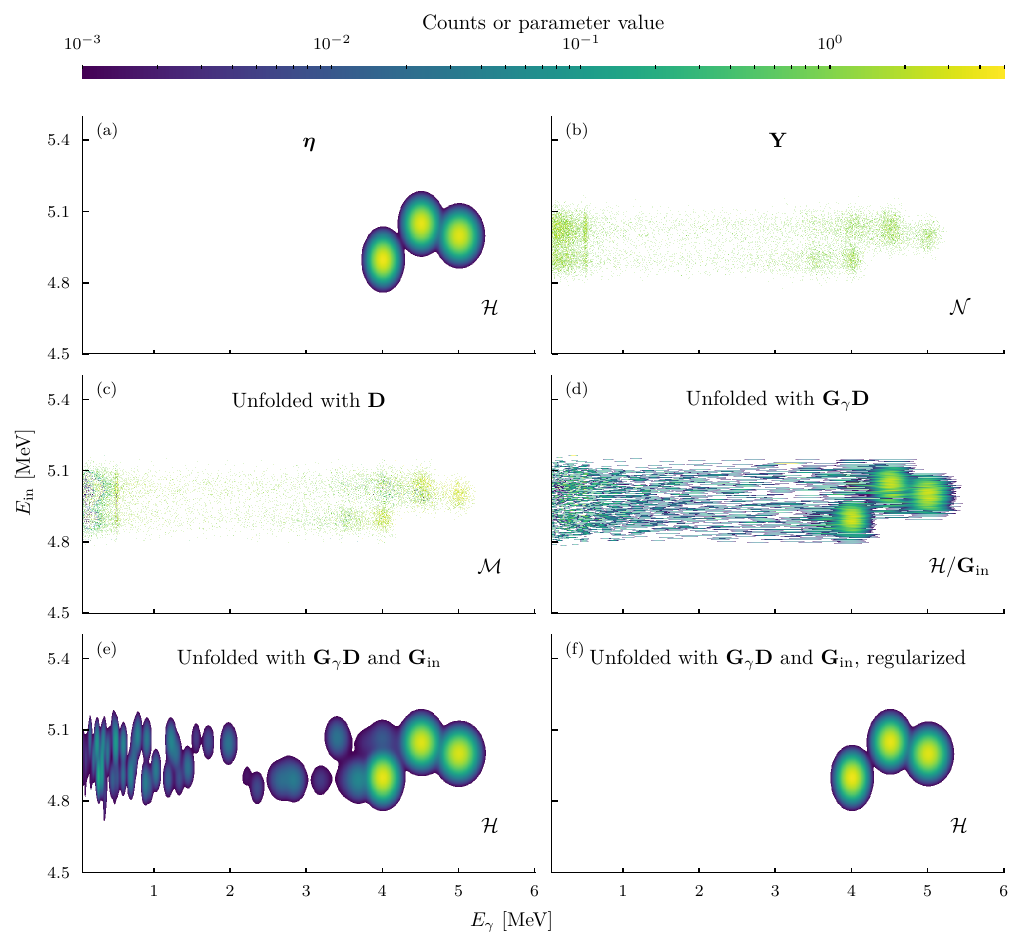}
    \caption{%
Comparison of unfolding methods applied to a high-resolution $1000\times 1000$ spectrum containing three sharp peaks, each with $10^4$ counts. 
Panel (a) shows the true $\veta$ spectrum, while panel (b) displays its Poisson-sampled observation $\matt{Y}$, illustrating the challenge posed by sparse data, particularly in the $\SIrange{4.8}{5.2}{MeV}$ region, which contains only 73 counts on average per $\Ex$ vector. 
The subsequent panels present unfolding results with progressively more prior information: 
(c) $\matt{D}$ alone (negligible improvement), 
(d) $\Gg\matt{D}$ (partial improvement but dominated by noise), 
(e) $\Gg\matt{D}\Gin$ (sharper reconstruction but with spurious peaks), and 
(f) $\Gg\matt{D}\Gin$ with sparsity regularization (accurate recovery of the original $\veta$). 
The incorporation of correlations from both $\Gg$ and $\Gin$, along with regularization, yields the most effective reconstruction under sparse conditions.%
}
    \label{fig:unfoldinggin}
\end{figure*}

Numerical tests show the advantages of this unified approach. In
\cref{fig:unfoldinggin}, we compare reconstructions with progressively more
complete model specifications. Unfolding with $\matt{D}$ alone completely fails
to recover the underlying structure due to the count sparsity.
Including $\Gg\matt{D}$ begins to resolve
the peaks but produces noisy results with poorly defined boundaries. The
complete $\Gg\matt{D}\Gin$ model achieves a more faithful reconstruction of
$\veta$, though still containing noise artifacts. Since the unfolding problem
remains ill-posed, even this complete model cannot uniquely determine the true
$\veta$ without additional constraints.
However, the addition of sparsity regularization proves sufficient for the
recovery of $\veta$. This unified treatment achieves full resolution without
rebinning to increase statistics, and incurs negligible additional computational cost. 
Importantly, the incorporation of $\Gg$ and $\Gin$ with the remapping to $\reta$ (as described in~\cref{subsec:constructingconstrained}) allows the optimizer to exploit the correlations of
neighboring bins while not being affected by their induced degeneracy.

\FloatBarrier

\subsection{Background modeling}
\label{sec:bg}
Experimental spectra always contain background signals that must be accounted for in the analysis.
We model both the background and prompt spectra as independent Poisson processes with means 
$\vbeta$ and $\vpi$, respectively:
\begin{align}
    \B &\sim \text{Poisson}(\vbeta) \\
    \Ps &\sim \text{Poisson}(\vpi)\,.
\end{align}
A naive approach to extract the data spectrum would be to subtract the background
spectrum $\B$ from the prompt spectrum $\Ps$, yielding $\tilde{\matt{Y}} = \Ps - \B$.
However, in addition to the problems shown in~\cref{sec:discretization}, this difference $\tilde{\matt{Y}}$ does not follow a Poisson distribution. While the sum of
two Poisson-distributed variables yields another Poisson distribution with mean
$\vbeta + \vpi$, their difference follows a \textit{Skellam}
distribution:
\begin{align}
    \Ps - \B \sim\operatorname{Skellam}(\vpi, \vbeta)\,.
\end{align}
The Skellam distribution poses computational challenges. Its likelihood evaluation 
is computationally demanding, and relaxing the non-negativity constraints on
$\matt{Y}$, $\veta$, and related variables introduces additional degeneracy that 
destabilizes the optimization.
To avoid these challenges, we model $\matt{Y}$ as a latent, unobserved variable:
\begin{align}
    \matt{Y} &\sim \text{Poisson}(\vnu) \\
    \Ps &=  \matt{Y} + \B, \quad \Ps \sim \text{Poisson}(\vnu + \vbeta)\,.
\end{align}
The likelihoods of all parameters are now Poisson and easy to handle.

An additional complexity arises because, while $\Ps$ is directly
observable, the actual background $\B$ of $\Ps$ is not.
However, we can observe $N$ non-prompt peaks, which can be modeled as independent
samples from the same background distribution parameterized by \(\vbeta\):
\begin{equation}
\label{eq:background}
    \B_i \poi\vbeta\quad\text{for } i = 1, \ldots, N.
\end{equation} 
To unfold with this model, we jointly optimize the log-likelihood of $\vnu$ and $\vbeta$,
combining their contributions along with appropriate regularization terms:
\begin{align}
    L(\hmu, \hbeta, \Ps, \B_i; \vecc{\theta}_\mu, \vecc{\theta}_\beta) &= \ell\left(\hmu\mid \Ps\right) 
    + \sum_i^N \ell\left(\hbeta\mid\B_i\right)\notag\\ 
    &\quad+\Omega_\mu\left(\hmu; \vecc{\theta}_\mu\right) +
    \Omega_\beta\left(\hbeta; \vecc{\theta}_\beta \right),
\end{align}
where
\begin{align}
    \ell\left(\hmu\mid\Ps\right) &= \Gg\matt{D}\hmu\Gin + \hbeta\notag\\
    &\quad -\Ps\log\left( \Gg\matt{D}\hmu\Gin+\hbeta\right),\\
    \intertext{and}
\ell\left(\hbeta\mid\B_i\right) &= \hbeta - \B_i\log\left(\hbeta\right),
\end{align}
represent the log-likelihood terms for the prompt and background spectra, respectively. 

While the background parameter $\hbeta$ can be regularized through a penalty
term $\Omega_\beta\left(\hbeta; \vecc{\theta}_\beta\right)$, analogous to the
signal, this approach faces similar challenges, particularly in selecting an
appropriate regularization model. The background consists of several distinct
physical processes, which makes it difficult to construct a meaningful prior. In
practice, a Sobolev regularization with order $k \geq 5$ is recommended to suppress
spurious bin-to-bin fluctuations in the absence of more specific structural
assumptions.

\begin{figure*}[th]
\centering
\includegraphics[]{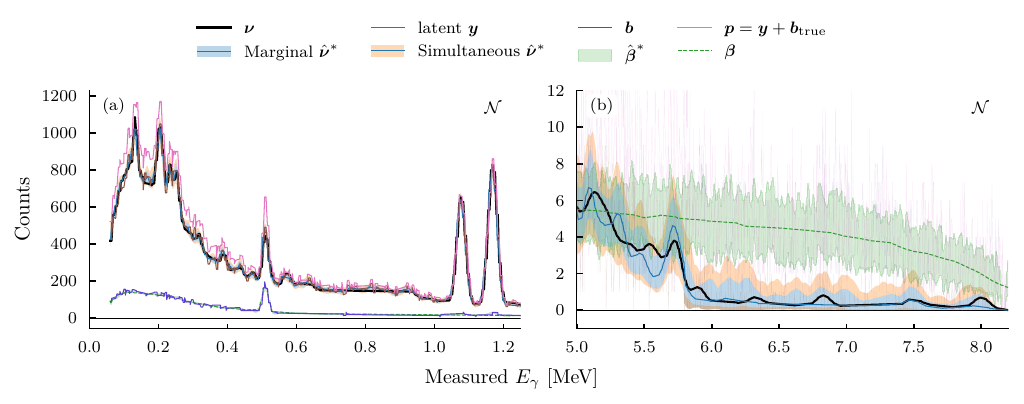}
\caption{Demonstration of background unfolding in the $\rnu$ space.
The observed prompt spectrum $\vecc{p}$ represents the sum of two unobserved
components: the signal spectrum $\vecc{y}\poi{\vnu}$ and the background spectrum
$\vecc{b}\poi{\vbeta}$, with
their relative contributions varying across the energy range. The left panel (a)
shows the low-energy region where signal dominates, while the right panel (b) shows
the high-energy region where background becomes dominant.
The $95\%$ confidence
intervals around $\hnu$ successfully contain the true $\vnu$ across the entire
spectrum, demonstrating robust recovery even in regions of low signal-to-noise
ratio. }
\label{fig:bgnu}
\end{figure*}

\begin{figure*}[th]
\centering
\includegraphics[]{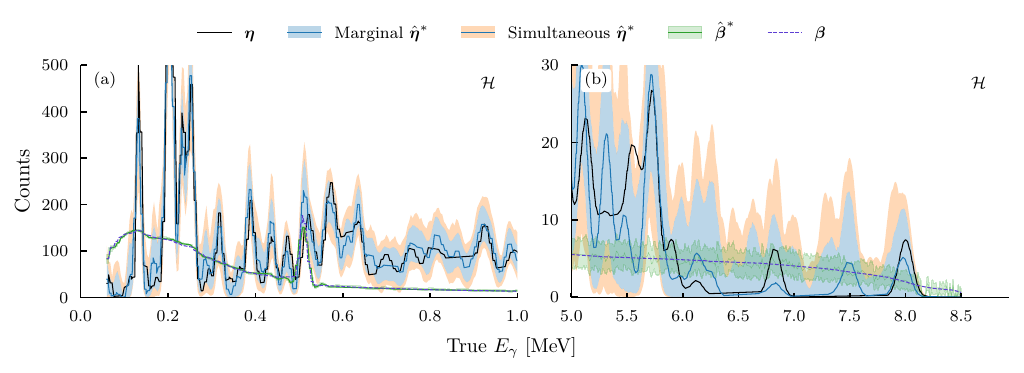}
\caption{
Results of background unfolding shown in the $\reta$ space in
a low-energy region (a) and a high-energy region (b). The $95\%$ confidence intervals around both the unfolded spectrum $\heta$ and the background spectrum $\hbeta$ largely capture their true values ($\veta$ and $\vbeta$, respectively). The contrasting smoothness between signal and background highlights the impact of regularization: $\heta$ exhibits smooth behavior due to implicit regularization, while the unregularized $\hbeta$ shows greater spectral variation, resulting in wider and less smooth confidence intervals.}
\label{fig:bgeta}
\end{figure*}

We demonstrate the combined unfolding approach with background in~\cref{fig:bgnu,fig:bgeta}
using synthetic \ce{^{120}Sn}-like spectrum at $\Ex=\SI{8}{MeV}$, with uncertainty quantification 
as explained later in~\cref{sec:uncquant}.  
The $95\%$ confidence intervals largely
encompass the true expectations $\veta$ and $\vbeta$ for both the signal and the background. 
Notably, the method successfully recovers and separates signal and background
components even in regions where the signal-to-noise ratio is low.
This capability stems from the signal's smeared
distribution across the entire spectrum, allowing information from regions
with better signal-to-noise ratios to inform the recovery in the noisier regions.
The increased uncertainty in these regions is reflected in wider confidence intervals
consistent with $0$.

\subsection{Contaminant modeling}

Gamma spectra from Oslo-type experiments often contain contaminant peaks from unwanted background
sources. Typical contaminants are transitions from \ce{^{16}O} and \ce{^{12}C}.
Contaminants may have different response functions from the
main spectrum due to Doppler broadening. While standard unfolding methods have
difficulties in handling contaminants, the flexibility of RMLE
handles them effectively.

Unlike a background spectrum, which can be measured separately, contaminants 
overlap with the prompt signal, making temporal separation impossible.
An arbitrary contamination is practically
impossible to separate from the signal unless it can be accurately modeled.
Fortunately, many contaminants in Oslo-type spectra are 
characterized by their simplicity --- typically just a handful of 
well-identified transitions. This characteristic allows us 
to place strong constraints on both their peak locations and intensities 
during the unfolding process. By incorporating these constraints into the 
RMLE optimization, we can effectively separate the contaminant contributions 
from the prompt spectrum.

For $M$ well-identified contaminant peaks $\vcons_{i=1}^{M}$, we can construct 
individual response functions $\left\{\mathbf{G}_{\gamma,i}\mathbf{D}_i\mathbf{G}_{\text{in},i}\right\}_{i=1}^M$ 
for each contaminant component $\vcon_i$. The observed spectrum $\my$ can then 
be modeled as a sum of the main spectrum and these contaminant contributions (for notational simplicity we ignore the background model of~\cref{sec:bg}, but they are easily combined additively):
\begin{equation}
\my \sim \text{Poisson}\left(\mathbf{G}_\gamma\mathbf{D}\vmu\mathbf{G}_\text{in} + \sum_{i=1}^{M} 
\mathbf{G}_{\gamma,i}\mathbf{D}_i\vcon_i\mathbf{G}_{\text{in},i}\right).
\end{equation}
The unfolding is performed by 
optimizing the log-likelihood with appropriate regularization terms: 
\begin{subequations}
\begin{align}
L\left(\hmu, \hcon_i, \mathbf{Y}; \vecc{\theta}, \vecc{\theta}_{\xi,i}\right) &= \ell\left(\hmu, \hcons_{i=1}^M|\mathbf{Y}\right) \notag \\
& +\Omega\left(\hmu, \hcons_{i=0}^M; \vecc{\theta}, \vecc{\theta}_{\xi,i}\right) \\
&= \hnu + \hxi_\nu - \mathbf{Y}\log\left(\hnu + \hxi_\nu\right)\notag\\
& + \Omega_\mu(\hmu; \vecc{\theta}_\mu) + \sum_{i=1}^{M}\Omega_i(\hcon_i;\vecc{\theta}_{\xi,i}),
\end{align}
\end{subequations}
with response terms
\begin{align}
\hnu &= \mathbf{G}_\gamma\mathbf{D}\hmu\Gin\\
\hxi_\nu &= \sum_{i=1}^{M} \mathbf{G}_{\gamma,i}\mathbf{D}_i\hcon_i\mathbf{G}_{\text{in},i}.
\end{align}
The regularization terms $\left\{\Omega_i(\hcon_i; \vecc{\theta}_{\xi,i})\right\}_{i=0}^M$ enforce sparsity and other user-specified constraints on the 
contaminant components, constraining their contributions to the 
known peak regions. 

\begin{figure}
    \centering
    \includegraphics[]{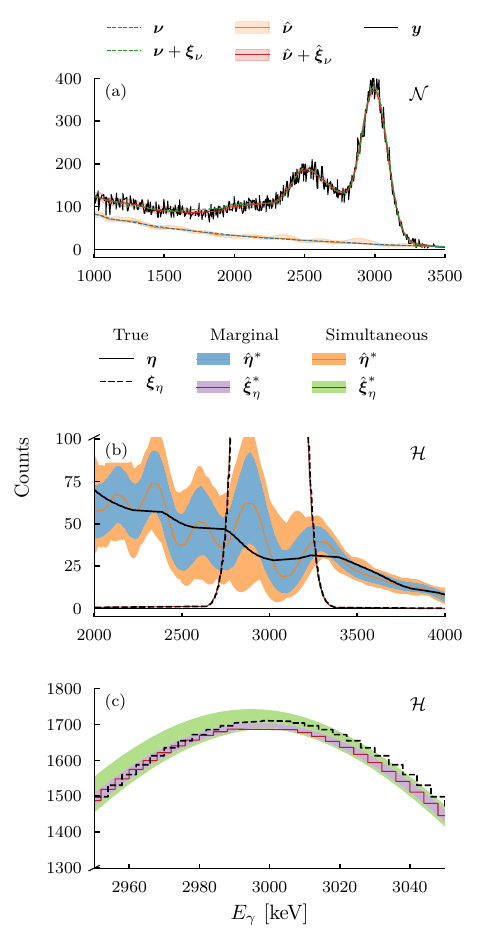}
\caption{
Unfolding of a simulated \ce{^166Ho}-like spectrum with a dominant contaminant peak at \SI{3}{MeV}. 
With sufficient model constraints on both the peak and the overlap region, the signal and contaminant can be effectively separated. 
(a) The raw spectrum $\vn$, a Poisson sample of the combined signal $\vnu$ and contaminant $\boldsymbol\xi$. The refolded solutions accurately recover both $\vnu$ and $\vnu + \boldsymbol\xi$. The uncertainty bands are marginal.
(b) Zoomed-in comparison of the true spectrum $\veta$ and the unfolded Monte Carlo solution $\heta^*$. The confidence interval (CI) widens near the contaminant peak, fully encompassing $\veta$.
(c) Zoomed-in comparison of the true contaminant $\boldsymbol\xi$ and its unfolded estimate $\hat{\boldsymbol\xi}^*$. The CI here is much narrower than that of $\heta^*$ due to stronger constraints in the contaminant model.}
    \label{fig:contaminant}
\end{figure}

To illustrate the method, a simulated \ce{^166Ho}-like spectrum at $\Ex=\SI{4}{MeV}$
was contaminated with a single peak at \SI{3}{MeV} and unfolded
(see~\cref{fig:contaminant}. The uncertainty bands are explained later in~\cref{sec:uncquant}). 
A prior Gaussian fit of the contaminant peak
provided initial estimates and constraints for the central value, standard deviation and $\nu$-amplitude.
%($\hat E_0$, $\hat\Gamma$, $\hat A_\nu$).
%, leading to
%model boundaries of $\hat E_0 \in [\SI{2.8}{MeV}, \SI{3.2}{MeV}]$, $\hat \Gamma
%= \SI{41}{keV}$, and $\hat A_\nu \in [380, 390]$.
These constraints alone
resulted in overfitting by the unfolding algorithm, producing a
spurious peak in $\heta$. To address this, a weighted Tikhonov regularization was applied
between \SI{2.6}{MeV} and \SI{3.4}{MeV} to penalize excessive counts in this
region. The penalization weight was determined through trial and error,
balancing the avoidance of dips in the unfolded spectrum ($\heta$) with the
suppression of excessive counts.

This underscores a fundamental
limitation: the necessity for strong, user-defined constraints on the
contaminant model, which introduces the potential for bias through overly strict
or lenient specifications. Lacking a model that sufficiently constrains the solution space, it is best to construct a model that
allows for the most variance in order to reduce the bias. 

The RMLE method provides a stable and physically constrained point estimate of the unfolded spectrum. However, to enable statistically sound interpretation of the results, uncertainty quantification is necessary. In the next section, we describe how to construct calibrated confidence intervals for each bin of the unfolded spectrum through Monte Carlo resampling.

%END SECTION
%%%%%%%%%%%%%%%%%%%%%%%%%%%%%%%%%%%%%%%%%%%%%%%%%%%%%%%%%%%%%%%%%%

%%%%%%%%%%%%%%%%%%%%%%%%%%%%%%%%%%%%%%%%%%%%%%%%%%%%%%%%%%%%%%%%%%
%START SECTION
\section{Uncertainty quantification}
\label{sec:uncquant}

\subsection{Monte Carlo resampling}
\label{subsec:mcresampling}

\begin{algorithm}[H]
\caption{Constructing confidence intervals using Monte Carlo ensemble.} 
\newcommand{\algbf}[1]{\texttt{#1}}
\label{alg:confidence_intervals}
\begin{algorithmic}[1]
\Require 
    \Statex \quad Observed data $\my$ or $\Ps$
    \Statex \quad Observed background(s) $\B_i$
    \Statex \quad Models for contaminants $\left\{\hcon_{\text{sol},i} \right\}_{i=1}^M$\Comment{If present}
    \Statex \quad Refolded solution $\snu$\Comment{Recommended}
    \Statex \quad Background parameter $\hbeta_\text{sol}$
    \Statex \quad Number of Monte Carlo samples $N$
\Ensure 
    \Statex \quad Confidence intervals for $\seta, \snu, \hbeta_\text{sol}, \left\{\hcon_{\text{sol},i}\right\}_{i=1}^M$

\State \algbf{Generate Monte Carlo Ensemble}
    \For{$k = 1$ to $N$}
        \State Sample new observation $\mc{Y_{ij}} \sim \text{Poisson}(\lambda_{ij})$, 
        \Statex\qquad where $\lambda_{ij} = \left(\hat{\nu}_\text{sol}\right)_{ij}$ or 
        $\left(\hat{\nu}_\text{sol} + \sum_{l=0}^M\hat{\con}_{\text{sol},l} \right)_{ij}$
        or $Y_{ij}$
        \State Sample new observation $\mc{B_{ij}} \sim \text{Poisson}(\kappa_{ij})$, 
        \Statex\qquad where $\kappa_{ij} = \left(\sbeta\right)_{ij}$ or $B_{ij}$
        \State Unfold $\mc{\my}$ with $\mc{\B}$ to get $\mc{\heta}$, $\mc{\hbeta}$ and
        $\left\{\mc{\hcon_i} \right\}_{i=0}^M$
    \EndFor
    \State Collect the ensembles \[\mcbox{\heta} = \left\{ \mc[1]{\heta}, \mc[2]{\heta}, \dots, \mc[N]{\heta} \right\}\]
    \[\mcbox{\hbeta} = \left\{ \mc[1]{\hbeta}, \mc[2]{\hbeta}, \dots, \mc[N]{\hbeta} \right\}\]
    and
    \[\left[\left\{\hcon_i^{*(k)}\right\}\right]_{i=0}^M = \left[\hcon_i^{*(1)},
    \hcon_i^{*(2)},\ldots, \hcon_i^{*(N)}
    \right]_{i=0}^M \]

\State \algbf{Construct Confidence Intervals}
    \State\quad Use the desired method to compute confidence intervals on the ensemble of each parameter.
    %\State\quad Estimate the acceleration parameter using Jackknife resampling

\State \algbf{Output} 
    \Statex \quad Confidence intervals for $\seta$, $\snu$, $\hbeta_\text{sol}$ and
    $\left\{\hcon_{\text{sol},i}\right\}_{i=1}^M$
    based on the constructed ensemble.

\end{algorithmic}
\end{algorithm}

A Monte Carlo ensemble method is used to construct confidence intervals on
the unfolded solution, similar to Midtb{\o} \textit{et al.}~\cite{MIDTBO2021107795}.
The algorithm is summarized in~\cref{alg:confidence_intervals}.

We generate simulated observations by sampling from either the raw data $\my$ or
the refolded solution $\hnu$.
Under the assumption that these are representative
of the true parameter $\vnu$ and that the data are Poisson-distributed, we can
create an ensemble $\mcbox\my$ by sampling:
\begin{equation}
    \mc{Y_{ij}}\poi{\lambda_{ij}}\quad \text{with } \lambda_{ij} = Y_{ij}\text{ or }\elemof{\hat\nu}{ij}.
\end{equation}
This ensemble is unfolded as before, resulting in an ensemble of solutions $\mcbox{\heta}$
from which we can construct a distribution. The choice of $\lambda_{ij}$ affects the variance of
the ensemble, especially when the counts are low. Using $\hat{\nu}_{ij}$ as a mean is preferred 
because it incorporates information from the unfolding process, providing a more stable
estimate than the raw counts $Y_{ij}$. There is a risk of introducing bias if $\hnu$ is a
poor solution, but in practice this is not a concern since the uncertainty intervals
would be appropriately widened.

When background is present, we can either resample the prompt and background spectra separately:
\begin{align}
    \mc{B_{ij}}&\poi{B_{ij}},\\
    \mc{P_{ij}}&\poi{P_{ij}},
\end{align}
or sample the estimated parameters and construct the prompt spectrum hierarchically:
\begin{align}
    \mc{B_{ij}}&\poi{\left(\hat\beta\right)_{ij}},\\
    \mc{Y_{ij}}&\poi{\left(\hat\nu\right)_{ij}},\\
    \mc{P_{ij}}&=\mc{Y_{ij}} + \mc{B_{ij}}.
\end{align}
Again the latter approach is preferred as it provides more stable samples by incorporating more information.
Contaminants $\left\{\hcon_{i}\right\}_{i=1}^M$ are treated identically.

\begin{figure}
    \centering
    \includegraphics[]{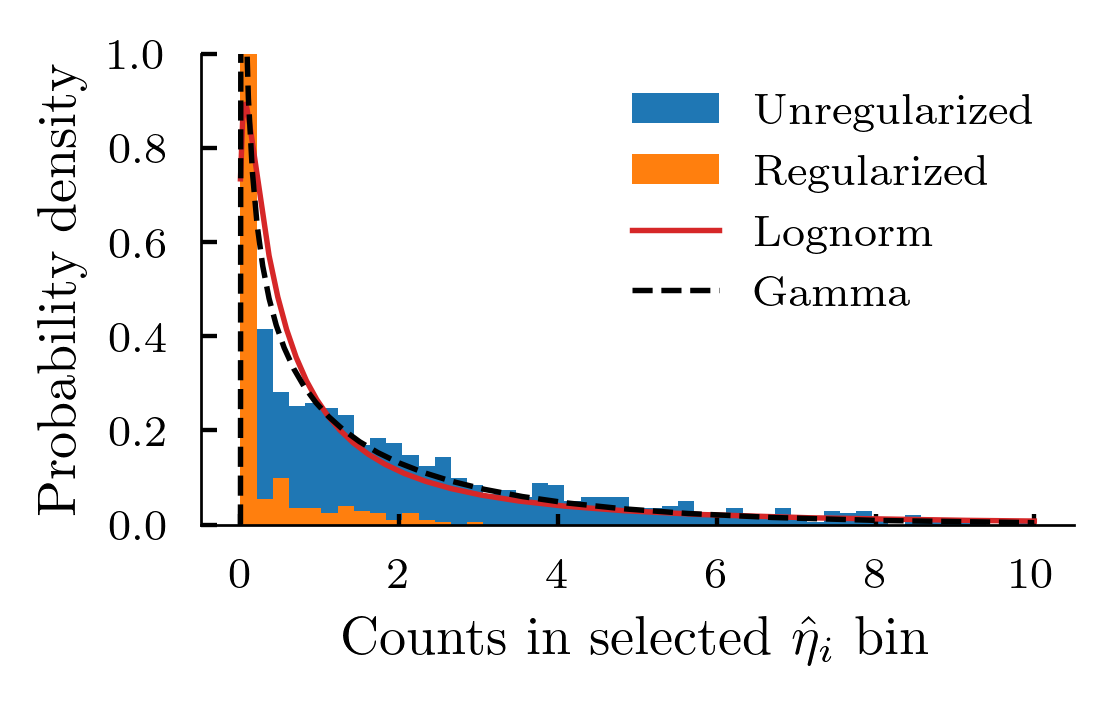}
    \caption{
Distribution of ensemble estimates $\mcbox{\hat{\eta}_{ij}}$ for a single bin $ij$ with true value $\eta_{ij} = 0$, comparing unregularized MLE and regularized MLE with sparsity penalty. The regularized distribution shows stronger concentration at zero with reduced tail thickness. Fitted gamma and log-normal distributions (shown for unregularized case) capture the general shape but fail Kolmogorov-Smirnov goodness-of-fit tests. The y-axis is truncated at 1.0.
}
    \label{fig:bindist}
\end{figure}

The ensemble distributions tend to be non-normal. \Cref{fig:bindist} illustrates
this by showing the distribution of $\mcbox{\hat{\eta}_{ij}}$ for a bin where
the true value is $\eta_{ij} = 0$, comparing unregularized and regularized MLE
solutions. The distributions display pronounced left-skewness with extended
right tails. In the regularized case, there is a sharp concentration at zero,
consistent with the sparsity-promoting regularization that drives the bin to zero.

The distributions are clearly non-normal.
Although gamma and lognormal
distributions capture the general shape of the unregularized histogram, they
fail Kolmogorov-Smirnov goodness-of-fit tests at conventional significance
levels. As the number of counts $\eta_{ij}$ increases, the distributions become
more normal, but rarely reach significance in a Shapiro-Wilks test for normality. 

The departure from normality can be quantified through higher-order moments. 
%\cref{fig:skewkurtosis} shows the skewness and excess kurtosis for bins with
%mode less than 5 counts. \anders{To aid physicist readers who are not that into statistics, 
%should we maybe add: ``For normally distributed data we would expect skewness and 
%excess kurtosis to be close to zero.''} 
Both unregularized and regularized solutions exhibit
substantial skewness (observed range $1-10$) and excess kurtosis (observed range $1-10$), with the regularized case showing
markedly larger moments. This behavior aligns with expectations: the
regularization enhances concentration around true values, while the
non-negativity constraint introduces asymmetry near $\eta=0$. The high excess
kurtosis reflects the presence of significant outliers in the distribution
tails.

The non-normality of the ensemble $\mcbox{\heta}$ is not an artifact of the
Monte Carlo simulations but fundamental to how the unfolding process
transforms the distribution of the observed data.
Using standard percentile intervals for confidence intervals is unsuitable in
this context because they do not account for the non-normality of the bin
distributions of $\mc{\heta}$. Instead, we draw on the bootstrap literature to
construct bias-corrected and accelerated (BCa) confidence intervals
\cite{efron1987}, which adjust for bias, variance, and skewness in the sampling
distribution. The acceleration parameter required for the BCa method is computed
using Jackknife.

Because the data consist of spectra with highly correlated bins, the confidence
intervals must account for these correlations to support valid spectra-wise
inference. This is achieved using \textit{simultaneous} confidence intervals,
which are discussed further in~\cref{app:margsimci}. All simultaneous intervals
shown here use the Bonferroni correction. For reference, we also include the
more familiar \textit{marginal} confidence intervals in each plot.

Alternative approaches to quantify the uncertainty were explored but proved
unsuccessful. Methods involving the inverse of the Fisher information matrix
fail because they assume that higher-order moments beyond the second are
negligible, which is not valid in our context. Additionally, inverting the
Hessian matrix is ill-posed due to its near-singular nature, leading to
unreliable variance estimates. Likelihood profiles are also unsuitable as
they require asymptotic normality, which we do not have.

Methods relying on local curvature, such as those using the second derivative of
the likelihood function, also fail for the same reasons. They assume that the
log-likelihood surface can be well-approximated by a quadratic form near the maximum
likelihood estimate, which is not valid in our case due to the ill-conditioned
nature of the problem and the significance of higher-order terms.

Given these significant limitations, the Monte Carlo ensemble method emerges as
the most viable approach for accurately quantifying uncertainty in our unfolding
process, as it handles non-normal distributions and avoids the
computational and numerical challenges associated with alternative methods.

%END SECTION
%%%%%%%%%%%%%%%%%%%%%%%%%%%%%%%%%%%%%%%%%%%%%%%%%%%%%%%%%%%%%%%%%%

%%%%%%%%%%%%%%%%%%%%%%%%%%%%%%%%%%%%%%%%%%%%%%%%%%%%%%%%%%%%%%%%%%
%START SECTION
\section{Systematics}
\label{sec:sims}

\NotationBlock{
    \notation{\Gg}{Smearing matrix along the gamma energy axis.}
    \notation{\sigma_\gamma(E_\gamma)}{The resolution along the gamma energy axis.}
    %\notation{\Gin}{Smearing matrix along the initial excitation energy axis.}
    \notation{\matt{D}}{Discrete response.}
    \notation{\vmu}{The expectation value of the unfolded spectrum.}
    \notation{\veta}{The expectation value of the unfolded spectrum refolded by $\Gg$.}
    \notation{\vnu}{The expectation value of the folded spectrum.}
    \notation{\vn}{The observed data \mbox{$\vn\poi{\vnu=\Gg\matt{D}\vmu}$}.}
    \notation{\rmu,\reta,\rnu}{The spaces which $\vmu, \veta $ and $ \vnu$ belong,
    respectively.}
    \notation{\heta}{RMLE estimate of $\veta$.}
    \notation{\heta^*}{Monte Carlo estimate or ensemble, depending on context.}
}

The performance of RMLE is mostly determined by structural features of the input data and the unfolding setup. In
this section, we analyze how of spectral complexity influence the reliability of
the reconstructed spectrum and the coverage of the associated confidence
intervals.

\subsection{Low spectral complexity}
\label{subsec:low_spectral_complexity}
Spectra exhibiting low spectral complexity are generally more amenable to unfolding. As outlined in \cref{sec:fluctuations}, spectral complexity relates to the presence of fluctuations and features in the underlying signal $\veta$. When this signal has a relatively simple structure, the task of applying and tuning appropriate regularization methods becomes more straightforward, mitigating spurious peaks and oscillations in the unfolded result $\heta$. We can broadly categorize these simpler cases into discrete and smooth:

\begin{description}
    \item[Discrete spectra] This category includes spectra composed of a small
    number of distinct, well-separated peaks. It is their separability that simplifies the
    problem. Regularization techniques that promote sparsity are particularly
    effective in such cases, as they align with the inherently sparse nature of
    the true signal $\veta$, such as sparsity cost
    functions or minimum entropy criteria (see \cref{subsec:paramregular}). 
    An instance of this category is the single-peak scenario, which will be discussed in \cref{sec:compfics}.
    Although the
    discrete peak region in Oslo Method data is not typically used for direct
    extraction of the nuclear level density or gamma-strength function,
    its accurate unfolding is important for recovering the first-generation
    spectrum, see~\cref{app:exp}.
    
    \item[Smooth spectra] The second category involves smooth spectra,
    often arising in scenarios with high nuclear level density, where individual
    energy levels are unresolved and merge into a (quasi-)continuum, \emph{and} where 
    the gamma cascades don't strongly feed through some low lying states. These spectra
    lack dominant, isolated structures. For such cases, regularization
    methods designed to enforce smoothness are preferred. Techniques like
    Tikhonov regularization, minimizing a Sobolev norm, or employing maximum
    entropy principles penalize high-frequency oscillations and favor the
    expected smooth behavior of $\veta$.
\end{description}

\begin{figure}[h]
    \centering
    \includegraphics[]{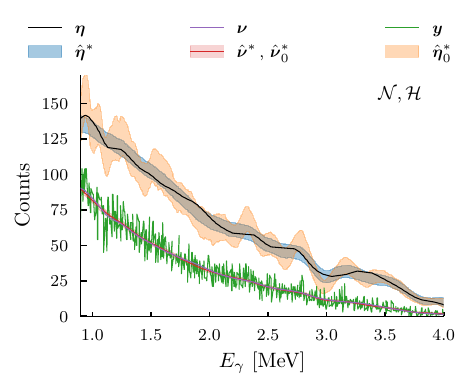}
    \caption{Unfolding of a simulated \ce{^{166}Ho}-like spectrum at $\Ex = \SI{4}{MeV}$. Shown are the unregularized solution $\heta^*_0$ and the regularized solution $\heta^*$ (using a Sobolev norm), both with marginal uncertainty bands. The true underlying spectrum $\veta$ is shown as solid black line. Although the corresponding refolded spectra, $\hnu^*_0$ and $\hnu^*$, both closely match the measured spectrum $\vnu$—with narrow uncertainty bands not discernible in the plot and therefore plotted together—the unregularized $\heta^*_0$ exhibits substantial oscillations compared to the smoother $\heta^*$. Simultaneous CI are not shown.}
    \label{fig:ho166}
\end{figure}

The spectrum of a \ce{^{166}Ho}-like spectrum serves as a practical example of a smooth case,
owing to its high level density. \Cref{fig:ho166} presents results from
unfolding a simulated \ce{^{166}Ho} spectrum at \SI{4}{MeV}. It compares an
unregularized solution $\heta^*_0$ with a solution regularized using a Sobolev
norm $\heta^*$. While the refolded versions of both solutions ($\hnu^*_0$ and
$\hnu^*$) provide excellent fits to the ideal folded data $\vnu$ (with
uncertainty bands significantly smaller than the Poisson noise), their unfolded
counterparts $\heta$ differ markedly. The unregularized solution $\heta^*_0$
exhibits excessive higher-frequency oscillations. In contrast, the regularized
solution $\heta^*$ is smooth, consistent with the expected nature of a continuum
spectrum. Consequently, the uncertainty bands associated with the unregularized
$\heta^*_0$ are wider and less smooth than that of the regularized $\heta^*$. 
%As we will see in~\cref{sec:coverage},
%this results in poorer coverage of the true $\veta$ compared to the narrower,
%smoother bands of the appropriately regularized solution $\heta^*$.
%\anders{Småplukk: Begrepet ``poorer coverage'' virker litt uklart for meg. Er
%coverage feil? Og i så fall, i hvilken retning? Undercoverage eller
%overcoverage?}\erlend{Godt poeng, gjorde en liten reformulering, men trengs
%kanskje mer}

\subsection{High spectral complexity}
\label{sec:highspectralcomplexity}

\begin{figure}[h]
    \centering
    \includegraphics{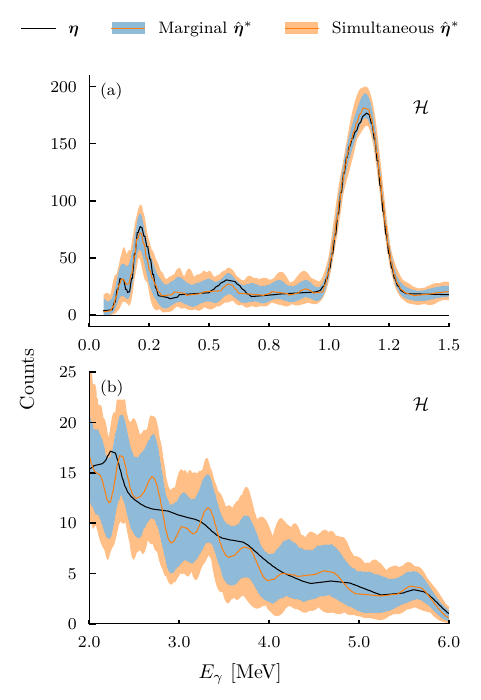}
    \caption{Unfolding of a complex simulated \ce{^{120}Sn}-like spectrum ($\Ex = \SI{6}{MeV}$). (a) Low energy region with a large peak in a otherwise smooth spectrum. (b) Higher energy region with a smoother spectrum. The spectrum's complex structure poses a challenge for regularization, resulting in an unfolded solution with wide, fluctuating uncertainty bands. Both marginal and simultaneous confidence intervals (CI) are displayed. The simultaneous CI always contains the true solution in this particular example, while the marginal CI does not (e.g., $\Eg=\SI{2.2}{MeV}$).}
    \label{fig:120sn}
\end{figure}

A significant challenge in analyzing Oslo spectra arises from their inherent
high spectral complexity, which manifests as overlapping distinct peaks with
varying amplitudes and widths, and a smoother underlying component. These
structures are exemplified in the simulated \ce{^{120}Sn}-like spectrum shown
in~\cref{fig:120sn}. The presence of such structures makes the application of
regularization techniques  challenging. 
No single regularization method has proven consistently effective for accurately unfolding these complex spectra.
Consequently, unfolding high complexity spectra results in wide and
fluctuating uncertainty bands. In this specific instance, the simultaneous
confidence interval contains the true $\veta$, whereas not all marginal CIs do.
However, the simultaneous CIs are not guaranteed to always contain the true
solution.

Further analysis of both regularization type and strength, as they relate to spectra of varying spectral complexity, is presented in~\cref{sec:regularization_systematics}.

%\clearpage

%\FloatBarrier
\subsection{Coverage probability}
\label{sec:coverage_tendency}

Having mapped how different regularization schemes influence point estimates
we now turn to
the associated uncertainty estimates. Specifically, we assess whether the
confidence intervals produced by the RMLE procedure attain their nominal
coverage probability. The definition and equations for coverage are given in~\cref{sec:appcoverage}. 

Ideally we would assess the coverage behavior of RMLE for every possible gamma-ray
spectrum, but this is impractical. A given element \(\eta_i\) will correspond to
different physical features in different spectra, and the confidence-region
width varies with both spectral complexity and the chosen regularization.
Coverage must therefore be evaluated within a defined class of spectra and
regularization scheme. Taking a pragmatic approach, we here test RMLE on three
representative cases spanning the complexity range studied earlier: a discrete
single-peak spectrum, a smooth \ce{^{166}Ho}-like spectrum, and a
high-complexity \ce{^{120}{Sn}}-like spectrum.

\emph{Spectrum coverage} is used to denote
the set of empirical coverage probabilities $\hat p (\veta)$ across all bins for a given spectrum:
\begin{equation}
  \hat p(\boldsymbol{\eta})
  \;=\;
  \bigl\{\hat p (\eta_i) : i=1,\dots,n\bigr\}.
  \label{eq:cov_tendency_vector}
\end{equation}
This set is summarized by the mean and the median.
%Since 
%neighboring bins $\heta_i$ are strongly correlated due to the resolution, a weighted mean
%is also used in which each bin is weighted by \(1/\sigma_\gamma(E_\gamma)\),
%as a rough correction for correlations.\erlend{think we can drop the weighting correction. I don't think people will notice.} \anders{Happy with whatever you prefer.}

\begin{figure}
  \centering
  \includegraphics{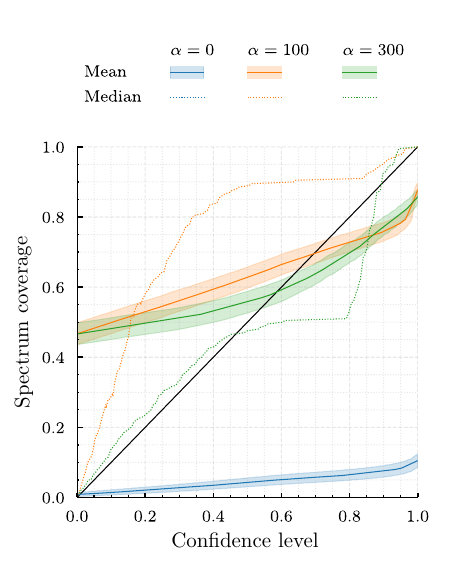}
  \caption{%
    Spectrum coverage for a peaked delta-spectrum under different regularization strengths $\alpha$.
    The colored bands indicate the standard error
    of the Monte Carlo estimate~\eqref{eq:coverage_rate}.
    Regularization strength \(\alpha=100\) minimizes the Wasserstein
    distance \(W_1(\hat{\boldsymbol{\eta}},\boldsymbol{\eta})\)
    (see~\cref{sec:regularization_systematics}).
    The coverage of simultaneous intervals is not shown as they are indistinguishable from $0$.
    The black diagonal represents perfect coverage.%
  }
  \label{fig:coverage_rate_peak}
\end{figure}

\Cref{fig:coverage_rate_peak} illustrates this notion for a sharply peaked spectrum,
using three different regularization strengths \(\alpha\). When no regularization is used
(\(\alpha=0\)), the unfolding introduces spurious peaks that draw counts away from the true signal.
This results in systematic undercoverage across all confidence levels. These intervals
fail to capture the signal, despite the fact that the overall reconstruction \(\hat{\boldsymbol{\eta}}\)
may be close to \(\boldsymbol{\eta}\) in terms of global metrics like \(D_{\text{KL}}\) and
Wasserstein distance \(W_1\). As \(\alpha\) increases, these spurious features are suppressed,
and spectrum coverage improves. The spectrum coverage becomes conservative at low confidence 
levels, but still undercover at high confidence levels.

Even with an appropriate regularization strength, the distribution of spectrum coverage
is very wide. Bins far from
the signal are often numerically zero, and their intervals trivially cover the true value,
leading to overcoverage. In contrast, bins on the lower tails of peaks may consistently
miss the target due to narrow ensemble spread, resulting in near-zero coverage. These effects
lead to the flat behavior of the spectrum coverage curves, e.g., the dotted green curve ($\alpha = 300$) in~\cref{fig:coverage_rate_peak}, which stays roughly constant from 0.5 to 0.8. 
By the same reason, simultaneous intervals achieve zero coverage, and the associated curves are therefore omitted from the figure.
%intervals. \anders{If I understand it correctly, it is ``stuck bins'' that explain the near-flat parts of the spectrum coverage vs confidence level graphs, e.g.\ the green dotted graph in Fig.\ 24? If so I think we should point the reader to one of these explicit examples in the plots, to help the reader see the connection between the explanation and the plots. For instance just insert a sentence along the lines of ``The dotted green curve in Fig.\ 24 shows an example of this.''}
%\erlend{Would it be easier to just not mention `stuck bins' at all? In the grand scheme it is just an annoying detail.}
%\anders{It's an annoying detail, but I think it's a nice way to show that we actually understand these results in quite some detail. And if I was a referee and there wasn't any comment/discussion about those ``flat regions'' in the coverage graphs, I suspect I would ask for it. So I vote for keeping it.}

\begin{figure}
    \centering
    \includegraphics{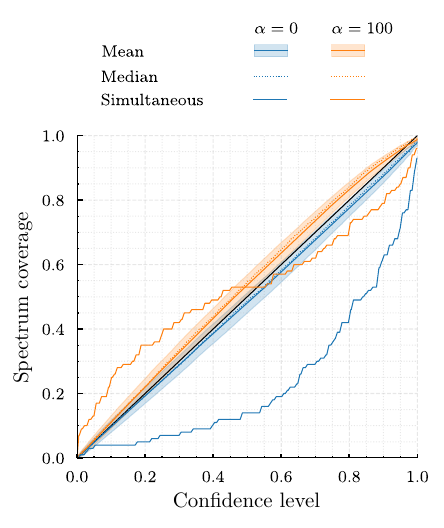}
    \caption{Spectrum coverage for a \ce{^166{Ho}}-like spectrum.
    Without regularization ($\alpha=0$), the (marginal) spectrum coverage is mostly calibrated,
    but with systematic undercoverage. Sobolev regularization with strength that
    minimizes the $W_1$ cost ($\alpha=100$) gives minor systematic overcoverage.
    The spectrum coverage of studentized supremum simultaneous interval greatly improves with regularization.}
    \label{fig:coverage_rate_ho}
\end{figure}

In contrast, a \ce{^{166}Ho}-like spectrum does not have these problems, as shown in~\cref{fig:coverage_rate_ho}.
The MC ensemble is much narrower for both the unregularized and regularized case, leading to the means and median
coinciding.
Without regularization the spectral complexity is too high, but on average the MC ensemble $\mc{\heta}$
will generally cover the true $\veta$. The marginal spectrum coverage is nearly calibrated, but with
minor undercoverage for all confidence levels. Sobolev regularization models the spectral complexity correctly,
resulting in overcoverage for nearly all confidence levels. The simultaneous spectrum coverage has severe
undercoverage without regularization, but has much better calibration with regularization, as desired.

\begin{figure}
  \centering
  \includegraphics{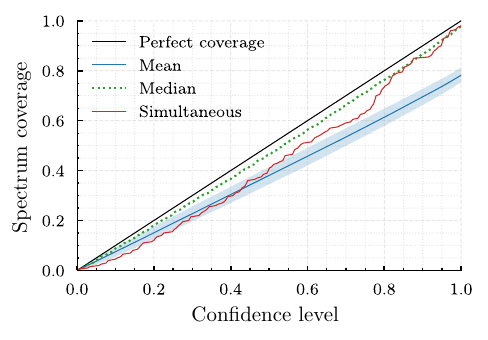}
  \caption{%
    Spectrum coverage for a complex \ce{^{120}Sn} spectrum.
    The Monte Carlo spread is much narrower than in
    Fig.~\ref{fig:coverage_rate_peak}, and the mean and median nearly coincide.%
  }
  \label{fig:coverage_rate_sn}
\end{figure}

\Cref{fig:coverage_rate_sn} shows the result for a \ce{^{120}Sn}-like
spectrum. Despite the higher spectral complexity and lack of regularization,
the spectrum coverage here behaves more predictably. While some systematic
undercoverage remains due to the mismatched spectral complexity, it is less severe and more evenly distributed than
the peaked spectrum, but worse than \ce{^{166}Ho}.
The spectral mismatches are
stochastic and do not lead to large systematic over- or undercoverage, which makes the simultaneous coverage behavior
nearly calibrated, but with some undercoverage.

In general, the RMLE confidence intervals exhibit reasonably good calibration, particularly when an appropriate regularization scheme is used. However, confidence levels should never be interpreted naively; their validity depends on the spectral context and the underlying assumptions of the unfolding process, which the practitioner must always take into account.

\section{Comparison to FICS}
\label{sec:compfics}

FICS (Folding Iteration with Compton Subtraction) is the standard unfolding
algorithm used in the Oslo Method and serves as a natural benchmark for comparison. Examples of its applications include~\cite{%
PhysRevC.73.064301,PhysRevC.90.044311,GUTTORMSEN2021136206,PhysRevC.106.034322,PhysRevC.111.015803}).
As a variant of Richardson's iterative method
(see~\cref{sec:richardson}), FICS inherits its core operational principle and
limitations. Iterative methods like FICS function by minimizing the
residuals $|\vn - \hnu|$ between the refolded estimate $\hnu$ and the measured
data $\vn$, effectively performing a stepwise partial inversion of the
detector response.

However, a fundamental issue arises because the minimization target is the noisy
experimental data $\vn$, rather than the underlying true folded distribution
$\vnu$. Consequently, the iterative process inevitably incorporates and
amplifies noise (see discussion in~\cref{sec:fluctuations}) 
present in $\vn$, leading to
overfitting. The severity of this overfitting is influenced by the condition
number of the response matrix as shown in~\cref{sec:discretization}. The primary technique
used to mitigate this in FICS and similar methods is early stopping, halting the
iteration before convergence becomes excessive.
FICS employs a cost function that combines a weighted chi-square term with a fluctuation penalty based on the residuals between the unfolded estimate $\heta$ and its smoothed counterpart. For a detailed theoretical discussion, including methodological challenges in directly comparing FICS and RMLE, see \cref{app:fics}.

\begin{figure*}[t]
    \centering
    \includegraphics{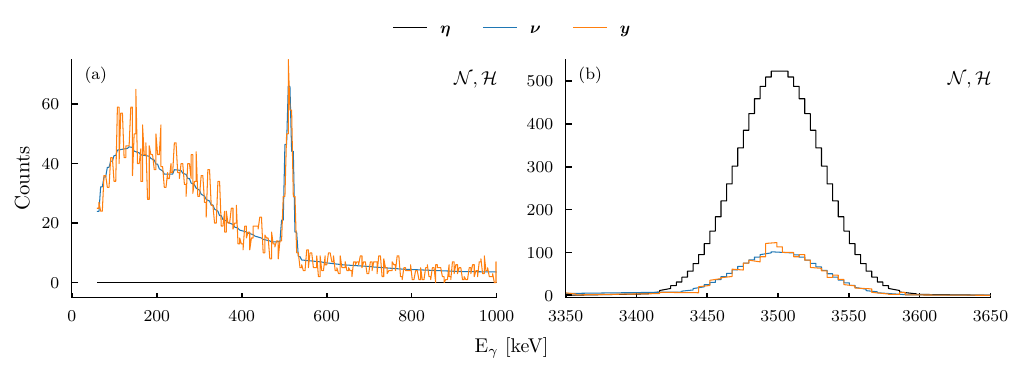}
    
\caption{A single peak at \SI{3.5}{MeV}, the folded spectrum $\vnu$ and a Poisson sample
    $\vn$. The spectrum has been split in two to make the finer details visible, with a lower $\Eg$ region (a) and
    a higher $\Eg$ region (b).
    The region in between (not shown) is comprised of Compton, single-escape and double-escape events.}
    \label{fig:peak0}
\end{figure*}

\begin{figure*}
    \centering
    \includegraphics{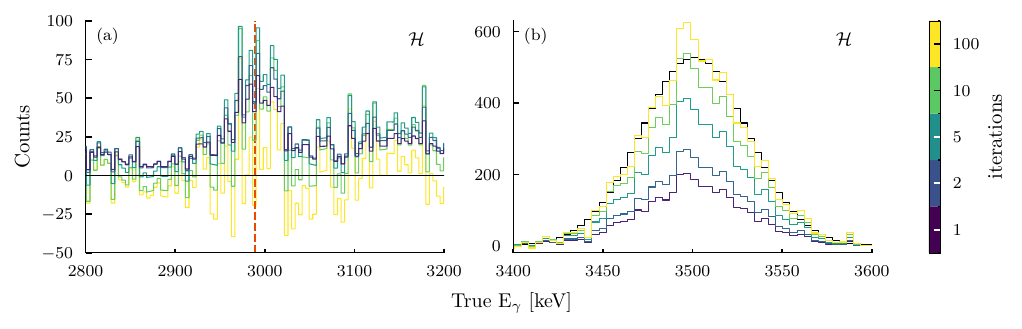}
    \caption{
The unfolded $\heta$ solution of FICS with increasing number of iterations (colored lines), compared to the true distribution $\veta$ (black line), shown for the lower (a) and upper (b) regions. The energy of the single escape peak ($\SI{3.5}{MeV} - \SI{511}{keV}$) is indicated by a dashed vertical red line. In the lower region (a), where the true $\veta = 0$, FICS overfits to noise immediately, while bins in the upper region (b) where $\veta > 0$ require more iterations to converge toward the peak, albeit with added noise. The spurious peak in the lower region is due to the single escape peak being misinterpreted as signal. For illustration, the step size was set to $0.3$ times the optimal step size [see Eq.~\eqref{eq:optimalstepsize}], allowing the small updates of FICS to be visible. For comparison, using the reduced step size means that iteration 100 here corresponds roughly to iteration 15 with the standard step size of 1.}
    \label{fig:iterations}
\end{figure*}

The inherent weaknesses of this iterative approach, particularly its susceptibility to noise and absence of physical constraints, are illustrated through the example of unfolding a single peak. This scenario is especially sensitive to noise and degeneracies in the response matrix. \Cref{fig:peak0} shows the setup: the true underlying distribution $\veta$ peaking at \SI{3.5}{MeV}, the folded distribution $\vnu$, and the Poisson-sampled data $\vn$ used as input. To isolate the effect of the algorithm, no background or contaminants are included.

\cref{fig:iterations} shows the evolution of the FICS solution $\heta$ over successive iterations. This demonstrates two critical weaknesses (these are expected, see~\cref{sec:richardson}):
\begin{description}
    \item[Noise amplification and spurious features] The iterative process does not
    distinguish between signal and noise in $\vn$. As a result, noise is
    integrated into the solution from the very first iteration. Spurious
    features, such as oscillations around zero and false peaks (particularly
    prominent in the lower energy region, left panel), converge rapidly—often
    faster than the actual signal peak (right panel). Early stopping in such a
    case yields an unsatisfactory compromise: the true peak is underestimated,
    while significant spurious noise structures remain. A notable example is the
    spurious peak near \SI{2.99}{MeV} (\SI{3.5}{MeV} - \SI{511}{keV}),
    corresponding to the single escape peak (marked by the red dashed line).
    We found that insufficient regularization frequently causes unfolding methods to
    misinterpret noise associated with strong response features (like escape
    peaks) as genuine peaks in $\veta$.
    
    \item[Unphysical negative values] FICS does not inherently restrict the solution space to non-negative values. While negative values can arise in purely mathematical solutions, they are unphysical. Their presence indicates the method's failure to confine the solution to a physically meaningful space. Simply rectifying (removing or redistributing negative values through some procedure) post-unfolding does not address this underlying deficiency. 
    Negative counts in the input data $\vn$ typically signal issues like improper background subtraction as discussed in~\cref{sec:bg}.
    Negative counts in the unfolded solution can also stem from an inaccurate response matrix.
\end{description}

\begin{figure*}
    \centering
    \includegraphics{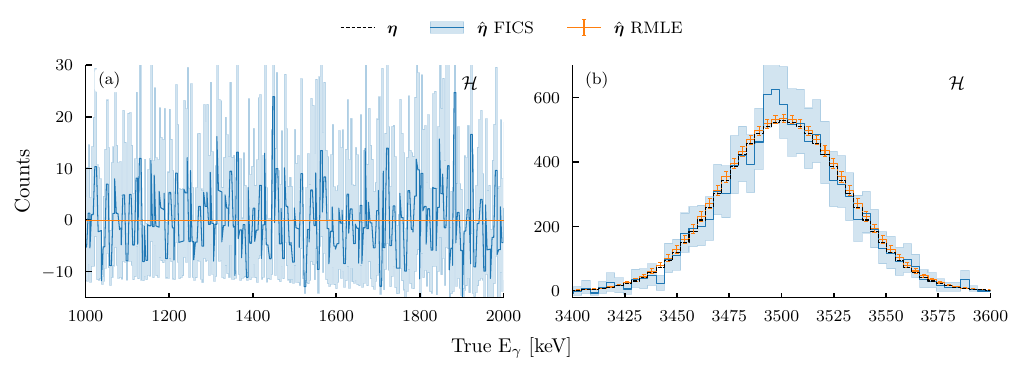}
    \caption{Comparison of the unfolded estimate $\heta$ to the true distribution $\veta$ (dashed black line), along with associated $2\sigma$ uncertainty bands in the lower (a) and upper (b) regions. The RMLE solution is completely smooth and closely tracks the true $\veta$, with their curves nearly overlapping; the corresponding uncertainty band is so narrow in the lower region (a) that it is not visible. In contrast, the FICS solution exhibits substantial noise and wider, more irregular uncertainty bands, reflecting overfitting to noise in the ensemble members. The CIs are marginal.}
    \label{fig:peak1}
\end{figure*}

\begin{figure*}
    \centering
    \includegraphics{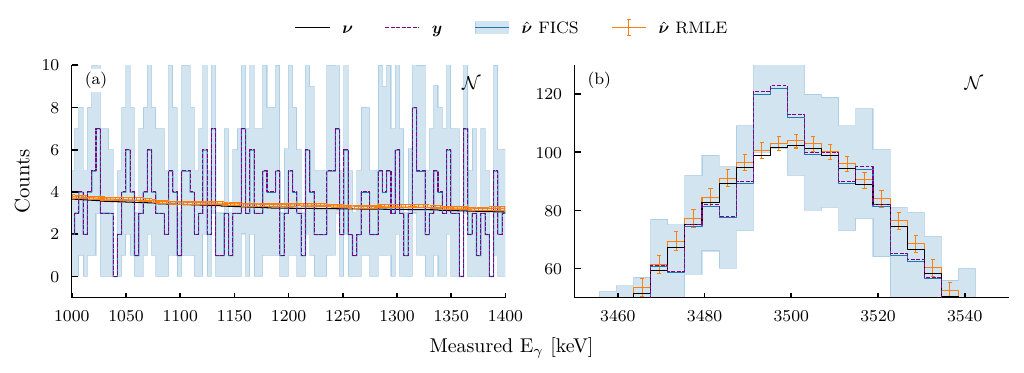}
\caption{Comparison of the refolded solutions $\hnu$ to the true $\vnu$ (black line) and to the observed data $\vn$ (dashed purple line), shown for the lower (a) and upper (b) regions. The refolded FICS solution perfectly fits the observed data $\vn$, while the refolded RMLE solution almost perfectly fits the true $\vnu$. The wider and fluctuating uncertainty bands of FICS reflect the incorporation of noise. The CIs are marginal.}
    \label{fig:peak2}
\end{figure*}

Comparing the final unfolded solutions $\heta$ from FICS and RMLE, shown in \cref{fig:peak1}, the FICS result is characterized by significant bin-to-bin
fluctuations, whereas the RMLE solution is smooth and more closely approximates
the true $\veta$. This reflects their differing approaches: FICS fits the noisy
data deterministically, while RMLE treats the data stochastically.
As described in~\cref{subsec:paramregular,sec:fluctuations}, not properly accounting for the different
sources of variation will lead to fluctuations in the solution. This difference is further emphasized by the refolded
solutions $\hnu$ in \cref{fig:peak2}. FICS's $\hnu$ almost perfectly
reproduces the noise profile of the input $\vn$, confirming overfitting. In
contrast, RMLE's $\hnu$ converges towards the smooth, true folded distribution
$\vnu$.

The uncertainty bands for FICS $\heta$ (calculated via Monte Carlo using the
method of Midtb{\o} \textit{et al.}~\cite{MIDTBO2021107795}) are wide and
fluctuating, reflecting the high variance introduced by fitting to the noise in
$\vn$. In contrast, RMLE estimates the regularized expectation value of $\heta$,
making it less sensitive to input noise and resulting in substantially narrower
uncertainty bands (calculated as described in~\cref{sec:uncquant}).%
\footnote{%
Note that the confidence intervals of $\hnu$ in the lower region, \cref{fig:peak2}(a), fail to cover the true $\vnu$ due to the ensemble spread discussed in~\cref{sec:coverage_tendency}.} %
This also
illustrates a point made in~\cref{subsec:mcresampling}: using $\hnu$ as a basis
for resampling is preferable once $\hnu$ has converged sufficiently close to
$\vnu$. If, instead, $\vn$ is used as the resampling basis—as in the approach of
Midtb{\o}—the noise in $\vn$ introduces both greater bias and higher variance in
the resulting ensemble.

This juxtaposition of $\heta$ and $\hnu$ underscores why relying solely on the
residuals $|\vn - \hnu|$ to assess unfolding quality is deceptive. FICS
demonstrates that achieving a low residual (a good fit to $\vn$) does not
guarantee an accurate estimation of the true underlying distribution $\veta$.
Indeed, convergence to the noise in $\vn$ prevents convergence to the true
$\veta$. Moreover, even perfect convergence to the true $\vnu$ would not yield a
unique $\heta$ due to the inherent degeneracy (ill-posedness) of the unfolding
problem (\cref{subsec:paramregular}). While the spuriousness of the FICS $\heta$
is relatively obvious in this simple single-peak test case, identifying such
artifacts in complex experimental spectra by visual inspection alone is
practically impossible.

%\erlend{G and compton(?) destroys high frequency components. However,
%for our case this is either irrelevant (due to eta), or not 
%relevant for low gamma since the sigma here is sufficiently low to
%preserve them.}

%END SECTION
%%%%%%%%%%%%%%%%%%%%%%%%%%%%%%%%%%%%%%%%%%%%%%%%%%%%%%%%%%%%%%%%%%

%%%%%%%%%%%%%%%%%%%%%%%%%%%%%%%%%%%%%%%%%%%%%%%%%%%%%%%%%%%%%%%%%%
%START SECTION
\FloatBarrier
\section{Summary and outlook}
\label{sec:sumout}

In this work, we introduced a novel method for unfolding gamma-ray spectra,
specifically tailored for the Oslo Method. This approach addresses key
limitations of traditional unfolding techniques, such as their tendency to
overfit noise and inability to provide reliable uncertainty estimates. By
combining regularized maximum log-likelihood optimization with Monte Carlo
simulations, the method offers more robust and transparent confidence intervals,
particularly for spectra characterized by low complexity.

We developed a theoretical framework to better understand the challenges of
unfolding, which helped guide the design of the method to reduce overfitting.
For simpler low complexity spectra, the regularization scheme proved highly effective,
producing narrow confidence intervals that accurately cover the true solutions.
This success arises from the ability to incorporate prior knowledge into the
regularization cost function.

For high-complexity spectra, the regularization schemes were less effective.
Non-statistical noise and the diversity of spectral structures lead to
difficulties in constraining the solutions. As a result, handling the
variability across different types of spectra remains a challenge in unfolding
gamma-ray spectra for Oslo Method applications.

Moving forward, the unfolding methodology can be improved by addressing
the challenges encountered in this work:

\begin{enumerate}

    \item \textit{New regularization schemes based on physical principles:}
    A significant obstacle for improved unfolding is the lack of
    general regularization schemes. 
    Developing regularization schemes grounded in physical principles, rather
    than relying solely on statistical methods, will better capture the
    underlying nature of the spectra.
    
    \item \textit{Develop adaptive regularization:}
    Barring physics-based regularization, combine standard regularization schemes
    to handle high-complexity spectra. The main problem here is finding schemes
    that combine, and that have parameters that are feasible to tune.
   
    \item \textit{Parameter tuning:} Implementing techniques to predict optimal regularization
    parameters based on the structure of the input spectra and nuclear systematics.

    \item \textit{Response function discrepancy correction:} Investigating how
    to systematically correct for discrepancies between modeled and true detector response
    functions. 

\end{enumerate}

We believe answering these problems would improve the method's applicability to 
more complex and varied gamma-ray spectra.

\acknowledgments

Simulations with RAINIER were performed on resources provided by Sigma2, the National Infrastructure for High Performance Computing and Data Storage in Norway (using ``Saga'' on Project No. NN9464K). 
A.~C.~L. and E.~L. gratefully acknowledge funding of this research by the Research Council of Norway, Project Grant No. 316116. 
A.~C.~L.,  E.~L. and A.~H.~M. acknowledge financial support from the Research Council of Norway through the Norwegian Nuclear Research Centre (project No. 341985). A.~K. was supported by the Research Council of Norway through the FRIPRO grant 323985 PLUMBIN’.
All the authors also  acknowledge continued support through dScience (Centre for Computational and Data Science) at the University of Oslo, Norway.
We are grateful to Dr. Maria Markova for stimulating discussions and inspiring comments.

%END SECTION
%%%%%%%%%%%%%%%%%%%%%%%%%%%%%%%%%%%%%%%%%%%%%%%%%%%%%%%%%%%%%%%%%%
\newpage
%%%%%%%%%%%%%%%%%%%%%%%%%%%%%%%%%%%%%%%%%%%%%%%%%%%%%%%%%%%%%%%%%%
%START SECTION
\appendix
\section{Mathematical background and proofs}
\label{app:math}
This section collects some linear–algebra material used in the paper. We recall only what we need—namely the four fundamental subspaces of a matrix \(\mathbf{R}\in\mathbb{R}^{m\times n}\): \(\operatorname{Ran}(\mathbf{R})\), \(\operatorname{Ker}(\mathbf{R})\), \(\operatorname{Ran}(\mathbf{R}^{\top})\), and \(\operatorname{Ker}(\mathbf{R}^{\top})\)—and fix notation for the rank–nullity identity and the standard orthogonality relations between these spaces. We also provide the Moore-Penrose pseudoinverse and the associated Penrose equations. Further, we present short proofs of the statements invoked in the main text, so the arguments there can be read without interruption. The presentation is deliberately minimal: it is not a self-contained primer, and for results or background not proved here we refer to standard references in linear-algebra, convex analysis and statistical analysis.

\subsection{Four Fundamental Subspaces}

\begin{enumerate}
    \item The range of $\mathbf{R}$, denoted $\textnormal{Ran}(\mathbf{R})$, is the set of all possible values that $\mathbf{R}\boldsymbol{\mu}$ can take for any vector $\boldsymbol{\mu}\in\mathbb{R}^{n}$. Formally, for a linear map $\mathbf{R}:\mathbb{R}^{n}\rightarrow \mathbb{R}^{m}$, it is defined as:
    \begin{align}
        \textnormal{Ran}(\mathbf{R})=\{\mathbf{R}\boldsymbol{\mu}\,|\,\boldsymbol{\mu}\in\mathbb{R}^n\}\,,
    \end{align}
    which in our case represents the set of all smeared means. This is also sometimes called the column space of $\mathbf{R}$.
    \item The null space of $\mathbf{R}$, denoted $\textnormal{Ker}(\mathbf{R})$, consists of all vectors $\boldsymbol{\lambda}\in\mathbb{R}^n$ that $\mathbf{R}$ maps to zero:
    \begin{align}
        \textnormal{Ker}(\mathbf{R})=\{\boldsymbol{\lambda}\in\mathbb{R}^n \,|\,\mathbf{R}\boldsymbol{\lambda}=0\}\,.
    \end{align}
    This space contains all directions in which $\mathbf{R}$ has no effect. When $\mathbf{R}$ is ill-conditioned, $\textnormal{Ker}(\mathbf{R})$ is non-trivial, meaning there are vectors other than zero that lie in this space.
    \item Let $\mathbf{R}^{\textnormal{T}}$ denote the transpose of $\mathbf{R}$. Then, $\textnormal{Ran}(\mathbf{R}^{\textnormal{T}})$ is the set of all linear combinations of the rows of $\mathbf{R}$:
    \begin{align}
        \textnormal{Ran}(\mathbf{R}^{\textnormal{T}})=\{\mathbf{R}^{\textnormal{T}}\mathbf{v}\,|\,\mathbf{v}\in\mathbb{R}^{m}\}\,,
    \end{align}
    and is called the row space of $\mathbf{R}$.
    \item The left null space of $\mathbf{R}$, denoted $\textnormal{Ker}(\mathbf{R}^{\textnormal{T}})$, consists of all vectors $\mathbf{v} \in \mathbb{R}^{m}$ such that $\mathbf{R}^{\textnormal{T}}\mathbf{v} = 0$. Formally, it is defined as:
    \begin{align}
        \textnormal{Ker}(\mathbf{R}^{\textnormal{T}}) = \{\mathbf{v} \in \mathbb{R}^{m} \, | \, \mathbf{R}^{\textnormal{T}}\mathbf{v} = 0\}\,.
    \end{align}
    This space contains all vectors that lie in the kernel of the transpose of $\mathbf{R}$, representing directions that are annihilated by $\mathbf{R}^{\textnormal{T}}$.
\end{enumerate}
Then, the \emph{Four Fundamental Subspaces Theorem} establishes the following orthogonality relations:
\begin{align}
    \textnormal{Ker}(\mathbf{R}) &= \textnormal{Ran}(\mathbf{R}^T)^\perp \,,
    \\
    \textnormal{Ker}(\mathbf{R}^{\textnormal{T}}) &= \textnormal{Ran}(\mathbf{R})^\perp \,.
\end{align}
These relations describe the fact that:
\begin{itemize}
    \item The null space of \( \mathbf{R} \) is the orthogonal complement of the range (row space) of \( \mathbf{R}^\textnormal{T} \). That is, every vector in \( \textnormal{Ker}(\mathbf{R}) \) is orthogonal to all vectors in \( \textnormal{Ran}(\mathbf{R}^\textnormal{T}) \).
    
    \item The null space of \( \mathbf{R}^T \) is the orthogonal complement of the range (column space) of \( \mathbf{R} \). This means that every vector in \( \textnormal{Ker}(\mathbf{R}^{\textnormal{T}}) \) is orthogonal to all vectors in \( \textnormal{Ran}(\mathbf{R}) \).
\end{itemize}

\subsection{Moore--Penrose Pseudoinverse}\label{app:mpinverse}
For a matrix $\mathbf{R}\in\mathbb{R}^{m\times n}$, the \emph{Moore--Penrose pseudoinverse} $\mathbf{R}^\dagger\in\mathbb{R}^{n\times m}$ is the unique matrix satisfying the \emph{Penrose equations}:
\begin{align}
    \mathbf{R}\mathbf{R}^\dagger\mathbf{R}&=\mathbf{R}, &
    \mathbf{R}^\dagger\mathbf{R}\mathbf{R}^\dagger&=\mathbf{R}^\dagger,\\
    (\mathbf{R}\mathbf{R}^\dagger)^\textnormal{T}&=\mathbf{R}\mathbf{R}^\dagger, &
    (\mathbf{R}^\dagger\mathbf{R})^\textnormal{T}&=\mathbf{R}^\dagger\mathbf{R}.
\end{align}
These imply that $\mathbf{R}\mathbf{R}^\dagger$ and $\mathbf{R}^\dagger\mathbf{R}$ are the orthogonal projectors onto $\operatorname{Range}(\mathbf{R})$ and $\operatorname{Range}(\mathbf{R}^\textnormal{T})$, respectively.  
If $\mathbf{R}$ has rank $r$ and singular value decomposition (SVD) $\mathbf{R} = \mathbf{U}\boldsymbol\Sigma\mathbf{V}^\textnormal{T}$, with 
\[
\boldsymbol\Sigma = \begin{bmatrix} 
\operatorname{diag}(\sigma_1,\dots,\sigma_r) & 0\\
0 & 0
\end{bmatrix},\quad \sigma_i>0,
\]
then
\[
\mathbf{R}^\dagger = \mathbf{V}\boldsymbol\Sigma^\dagger\mathbf{U}^\textnormal{T},\qquad 
\boldsymbol\Sigma^\dagger = \begin{bmatrix}
\operatorname{diag}(\sigma_1^{-1},\dots,\sigma_r^{-1}) & 0\\
0 & 0
\end{bmatrix}.
\]
This definition extends inversion to rank-deficient matrices: if $\mathbf{R}$ is invertible, $\mathbf{R}^\dagger=\mathbf{R}^{-1}$.

\subsection{Proof of theorem \ref{the:least squares}}
\label{app:proof_moorepenrose}
\emph{Proof:}
Let $\mathbf{P}_{\mathcal{R}} := \mathbf{R}\mathbf{R}^\dagger$ and
$\mathbf{P}_{\mathcal{K}} := \mathbf{I}-\mathbf{R}^\dagger\mathbf{R}$, the
orthogonal projectors onto $\operatorname{Range}(\mathbf{R})$ and
$\operatorname{Ker}(\mathbf{R})$, respectively, by the Penrose equations.  

The least-squares problem 
\[
\min_{\vmu} \|\mathbf{R}\vmu-\mathbf{y}\|_2
\]
is solved by projecting $\mathbf{y}$ onto $\operatorname{Range}(\mathbf{R})$, giving the minimal residual $\mathbf{p}=\mathbf{P}_{\mathcal{R}}\mathbf{y}$.  
One minimizer is 
\[
\vmu_0 = \mathbf{R}^\dagger\mathbf{y},
\]
since $\mathbf{R}\vmu_0 = \mathbf{R}\mathbf{R}^\dagger\mathbf{y} = \mathbf{p}$.  
If $\vmu$ is any other minimizer, then
\[
\mathbf{R}(\vmu-\vmu_0) = \mathbf{R}\vmu - \mathbf{R}\vmu_0 = 0,
\]
so $\vmu-\vmu_0\in\operatorname{Ker}(\mathbf{R})$.  
Thus every solution has the form
\[
\vmu = \vmu_0 + \boldsymbol\lambda,\quad \boldsymbol\lambda\in\operatorname{Ker}(\mathbf{R}).
\]
Finally, any $\boldsymbol\lambda\in\operatorname{Ker}(\mathbf{R})$ can be
written as $\boldsymbol\lambda=\mathbf{P}_{\mathcal{K}}\boldsymbol{v} =
(\mathbf{I}-\mathbf{R}^\dagger\mathbf{R})\boldsymbol{v}$ for some $\boldsymbol{v}\in\mathbb{R}^n$, proving
the claim:
\[
\vmu_{\text{LS}} = \mathbf{R}^\dagger\mathbf{y}+(\mathbf{I}-\mathbf{R}^\dagger\mathbf{R})\boldsymbol{v},\quad \boldsymbol{v}\in\mathbb{R}^n.
\]

\hfill$\square$

\subsection{Proof of corollary \ref{eq:corlstsq}}
\label{app:proof_cor_1}
\emph{Proof}
From Theorem \ref{the:least squares}, the least-squares solutions are
\[
\hat{\vmu}_{\mathrm{LS}}
    = \mathbf{R}^\dagger\mathbf{y} + \boldsymbol\lambda,\quad \boldsymbol\lambda \in \operatorname{Ker}(\mathbf{R}).
\]
The solution is unique if and only if the only possible $\boldsymbol\lambda$ is $\mathbf{0}$, i.e., when $\operatorname{Ker}(\mathbf{R})=\{\mathbf{0}\}$.  
Equivalently, $\mathbf{R}$ has full column rank (and for square $\mathbf{R}$, is invertible), in which case $\mathbf{R}^\dagger=\mathbf{R}^{-1}$ and 
\[
\hmu_{\mathrm{LS}} = \mathbf{R}^{-1}\mathbf{y}.
\]

\hfill$\square$

\subsection{Proof of proposition~\ref{prop:nullvectors}}\label{proofprop:negativecomponents}

%\begin{proposition}
%Let $\mathbf{R} \in \mathbb{R}^{m \times n}$ be a non-negative matrix, $R_{ij} %\geq0$ for all $i,j$, with no zero rows or columns, and suppose that $\mathbf{R}$ has a non-trivial null space. Then any non-zero vector $\boldsymbol{\lambda} \in \textnormal{Ker}(\mathbf{R})$ must have both positive and negative components; that is, $\boldsymbol{\lambda}$ oscillates between positive and negative values.
%\end{proposition}

\emph{Proof:} We will prove this by contradiction, considering two cases where $\boldsymbol{\lambda}$ is either entirely non-positive or entirely non-negative. The key to the proof is to use the orthogonality relation between $\textnormal{Ker}(\mathbf{R})$ and $\textnormal{Ran}(\mathbf{R}^{\textnormal{T}})$.

The orthogonality relation implies that for all $\boldsymbol{\lambda}\in\textnormal{Ker}(\mathbf{R})$ and for all $\mathbf{v}\in\textnormal{Ran}(\mathbf{R}^{\textnormal{T}})$, the inner product takes the form
\begin{align}
    \langle \boldsymbol{\lambda},\mathbf{v}\rangle =\boldsymbol{\lambda}^{\textnormal{T}}\mathbf{v}=0\,.
\end{align}
Suppose, for contradiction, that $\boldsymbol{\lambda}$ is either entirely non-positive or non-negative.

Without loss of generality, we first assume $\boldsymbol{\lambda}\geq 0$ component wise and $\boldsymbol{\lambda}\neq 0$. Since $\mathbf{R}^{\textnormal{T}}$ has non-negative entries (because $\mathbf{R}$ does), any vector $\mathbf{v}\in\textnormal{Ran}(\mathbf{R}^{\textnormal{T}})$ obtained from a non-negative $\mathbf{w}\in\mathbb{R}^{m}$ will also have non-negative components. Choose $\mathbf{w}$ with non-negative components and at least one positive component, and define $\mathbf{v}=\mathbf{R}^{\textnormal{T}}\mathbf{w}$. Then $\mathbf{v}\geq 0$ component wise and $\mathbf{v}\neq 0$, because $\mathbf{w}$ is not the zero vector and $\mathbf{R}^{\textnormal{T}}$ has no zero columns (since $\mathbf{R}$ has no zero rows).

Now, compute the inner product 
\begin{align}
    \langle \boldsymbol{\lambda},\mathbf{v}\rangle=\boldsymbol{\lambda}^{\textnormal{T}}\mathbf{v}=\boldsymbol{\lambda}^{\textnormal{T}}(\mathbf{R}^{\textnormal{T}}\mathbf{w})=(\mathbf{R}\boldsymbol{\lambda})^{\textnormal{T}}\mathbf{w}=0\,.
\end{align}
The inner product is a sum of non-negative terms
\begin{align}
    \langle \boldsymbol{\lambda},\mathbf{v}\rangle=\sum_{k=1}^{n}\lambda_k v_k \,.
\end{align}
Since $\lambda_k \geq 0$ and $v_k\geq 0$, each term $\lambda_k v_k\geq 0$. A sum of non-negative terms resulting in zero is only possible if $\lambda_k v_k =0$ for all $k=1,2\dots n$. This implies that for each $k$, either $\lambda_k =0$ or $v_k =0$. However, since $\boldsymbol{\lambda}\neq 0$, there exists at least one index $\lambda_k >0$. For such $k$, it must be that $v_k =0$.

Next, examine the case $v_k =0$. By definition
\begin{align}
    v_k = (\mathbf{R}^{\textnormal{T}}\mathbf{w})_k = \sum_{i=1}^{m} R_{ik}w_i\,.
\end{align}
Since by assumption $R_{ik}\geq 0$ and $w_i \geq 0$, the only way $v_k =0$ is if $R_{ik} w_i =0$ for all $i$. But $w_i\geq 0$ with at least one $w_i >0$, so $R_{ik} =0$ for all $i$ where $w_i >0$.

This would imply that the $k$-th column of $\mathbf{R}$ is zero, but this leads to a contradiction as $\mathbf{R}$ is assumed to have non-trivial columns. Hence, the assumption that $\boldsymbol{\lambda}\geq 0$ must be false. The case $\boldsymbol{\lambda}\leq 0$ is completely symmetrical, and also leads to a contradiction. Thus, $\boldsymbol{\lambda}$ must have both positive and negative components to yield $\langle \boldsymbol{\lambda},\mathbf{v}\rangle =0$.

\hfill$\square$

\subsection{Proof of~\cref{theorem:Tikhonov reg}}\label{prooftheorem:Tikhonov reg}

\emph{Proof:}
Define
\begin{align}
f_\alpha(\boldsymbol{\mu}) &= \|\mathbf{y}-\mathbf{R}\boldsymbol{\mu}\|^2+\alpha\|\boldsymbol{\mu}\|^2 .
\end{align}

\emph{1. Existence and uniqueness:}
We have
\begin{align}
\nabla f_\alpha(\boldsymbol{\mu}) &= -2\mathbf{R}^{\textnormal{T}}(\mathbf{y}-\mathbf{R}\boldsymbol{\mu})+2\alpha\boldsymbol{\mu},\\
\nabla^2 f_\alpha(\boldsymbol{\mu}) &= 2(\mathbf{R}^{\textnormal{T}}\mathbf{R}+\alpha\mathbf{I}).
\end{align}
For any nonzero $\mathbf{z}$,
\begin{align}
\mathbf{z}^{\textnormal{T}}(\mathbf{R}^{\textnormal{T}}\mathbf{R}+\alpha\mathbf{I})\mathbf{z}
&= \|\mathbf{R}\mathbf{z}\|^2+\alpha\|\mathbf{z}\|^2>0,
\end{align}
so $\mathbf{R}^{\textnormal{T}}\mathbf{R}+\alpha\mathbf{I}$ is symmetric positive definite and invertible. The unique minimizer therefore satisfies
\begin{align}
(\mathbf{R}^{\textnormal{T}}\mathbf{R}+\alpha\mathbf{I})\hat{\boldsymbol{\mu}}_\alpha &= \mathbf{R}^{\textnormal{T}}\mathbf{y},\\
\hat{\boldsymbol{\mu}}_{\alpha} &= (\mathbf{R}^{\textnormal{T}}\mathbf{R}+\alpha\mathbf{I})^{-1}\mathbf{R}^{\textnormal{T}}\mathbf{y}.
\end{align}

\emph{2. Limiting case $\alpha\to0^+$:}
Let $\mathbf{R}=\mathbf{U}\boldsymbol{\Sigma}\mathbf{V}^{\textnormal{T}}$ be a singular value decomposition with rank $r$ and singular values $\sigma_1,\dots,\sigma_r>0$, and let $(\mathbf{u}_i)$, $(\mathbf{v}_i)$ denote the corresponding singular vectors. Then
\begin{align}
\hat{\boldsymbol{\mu}}_\alpha
&= \mathbf{V}(\boldsymbol{\Sigma}^2+\alpha\mathbf{I})^{-1}\boldsymbol{\Sigma}\,\mathbf{U}^{\textnormal{T}}\mathbf{y}
= \sum_{i=1}^{r}\frac{\sigma_i}{\sigma_i^2+\alpha}\,(\mathbf{u}_i^{\textnormal{T}}\mathbf{y})\,\mathbf{v}_i.
\end{align}
Taking limits termwise, $\frac{\sigma_i}{\sigma_i^2+\alpha}\to 1/\sigma_i$ as $\alpha\to0^+$, hence
\begin{align}
\hat{\boldsymbol{\mu}}_\alpha \xrightarrow[\alpha\to0^+]{} \sum_{i=1}^{r}\frac{1}{\sigma_i}\,(\mathbf{u}_i^{\textnormal{T}}\mathbf{y})\,\mathbf{v}_i
= \mathbf{R}^{\dagger}\mathbf{y}=: \hat{\boldsymbol{\mu}}_{\textnormal{MNLS}}.
\end{align}
Moreover,
\begin{align}
\hat{\boldsymbol{\mu}}_\alpha-\hat{\boldsymbol{\mu}}_{\textnormal{MNLS}}
&= \sum_{i=1}^{r}\!\Bigl(\frac{\sigma_i}{\sigma_i^2+\alpha}-\frac{1}{\sigma_i}\Bigr)
(\mathbf{u}_i^{\textnormal{T}}\mathbf{y})\,\mathbf{v}_i\\
&= -\sum_{i=1}^{r}\frac{\alpha}{\sigma_i(\sigma_i^2+\alpha)}(\mathbf{u}_i^{\textnormal{T}}\mathbf{y})\,\mathbf{v}_i,
\end{align}
so, using $\frac{1}{\sigma_i^2+\alpha}\le \frac{1}{\sigma_i^2}$,
\begin{align}
\bigl\|\hat{\boldsymbol{\mu}}_\alpha-\hat{\boldsymbol{\mu}}_{\textnormal{MNLS}}\bigr\|
&\le \alpha\Bigl(\sum_{i=1}^{r}\frac{(\mathbf{u}_i^{\textnormal{T}}\mathbf{y})^2}{\sigma_i^{6}}\Bigr)^{\!1/2}
\xrightarrow[\alpha\to0^+]{}0.
\end{align}

\emph{(3) Noise dampening under the additive model:}
Insert $\mathbf{y}=\mathbf{R}\boldsymbol{\mu}+\boldsymbol{\epsilon}$ into the solution map to obtain
\begin{align}
\hat{\boldsymbol{\mu}}_\alpha(\mathbf{y})
&= \hat{\boldsymbol{\mu}}_\alpha(\mathbf{R}\boldsymbol{\mu})
+(\mathbf{R}^{\textnormal{T}}\mathbf{R}+\alpha\mathbf{I})^{-1}\mathbf{R}^{\textnormal{T}}\boldsymbol{\epsilon}.
\end{align}
With the SVD,
\begin{align}
\hat{\boldsymbol{\mu}}_\alpha(\mathbf{R}\boldsymbol{\mu})
&= \mathbf{V}(\boldsymbol{\Sigma}^2+\alpha\mathbf{I})^{-1}\boldsymbol{\Sigma}^2\mathbf{V}^{\textnormal{T}}\boldsymbol{\mu}\nonumber
\\
&= \sum_{i=1}^{r}\frac{\sigma_i^2}{\sigma_i^2+\alpha}\,(\mathbf{v}_i^{\textnormal{T}}\boldsymbol{\mu})\,\mathbf{v}_i,
\end{align}
such that
\begin{align}
\hat{\boldsymbol{\mu}}_\alpha(\mathbf{y})-\hat{\boldsymbol{\mu}}_\alpha(\mathbf{R}\boldsymbol{\mu})
&= \sum_{i=1}^{r}\frac{\sigma_i}{\sigma_i^2+\alpha}\,(\mathbf{u}_i^{\textnormal{T}}\boldsymbol{\epsilon})\,\mathbf{v}_i.
\end{align}
In particular, using the spectral norm and the elementary inequality
\begin{align}
\frac{t}{t^2+\alpha} \le \frac{1}{2\sqrt{\alpha}}
\qquad(\forall\,t\ge0,\ \alpha>0),
\end{align}
we obtain the bound
\begin{align}
\bigl\|\hat{\boldsymbol{\mu}}_\alpha(\mathbf{y})-\hat{\boldsymbol{\mu}}_\alpha(\mathbf{R}\boldsymbol{\mu})\bigr\|
&\le \frac{1}{2\sqrt{\alpha}}\,\|\boldsymbol{\epsilon}\|.
\end{align}
In contrast, the unregularized least-squares map employs gains $1/\sigma_i$, which may be arbitrarily large when $\sigma_i$ are small, hence regularization provides a smooth spectral cut-off and damps noise amplification.

\hfill$\square$

\hfill$\square$

\subsection{Effects on non-negativity constraint on the feasible set}\label{app:feasible set}

\begin{proposition}
Let $\mathbf{R}\in\mathbb{R}^{n\times n}$ satisfy $R_{ij}\ge 0$ for all $i,j$ and have no zero columns, and suppose $\operatorname{rank}(\mathbf{R})=r<n$ with $d=n-r>0$. Let $\{\mathbf{k}_1,\dots,\mathbf{k}_d\}$ be an orthonormal basis of $\ker(\mathbf{R})$ and set $K=[\mathbf{k}_1\ \cdots\ \mathbf{k}_d]\in\mathbb{R}^{n\times d}$. Fix a particular vector $\boldsymbol{\mu}_0\in\mathbb{R}^n$ with $\mathbf{R}\boldsymbol{\mu}_0=\boldsymbol{\nu}$, and parametrize
\[
\boldsymbol{\mu}=\boldsymbol{\mu}_0+\boldsymbol{\lambda},\qquad \boldsymbol{\lambda}=K\boldsymbol{\beta}=\sum_{i=1}^d \beta_i \mathbf{k}_i,\quad \boldsymbol{\beta}\in\mathbb{R}^d.
\]
Impose the componentwise nonnegativity constraint $\boldsymbol{\mu}\ge 0$, i.e.
\[
(\mu_0)_j+\sum_{i=1}^d \beta_i\,k_{i,j}\ \ge\ 0,\qquad j=1,\dots,n.
\]
Let the feasible set be
\[
\mathcal{B}=\{\ \boldsymbol{\beta}\in\mathbb{R}^d:\ \boldsymbol{\mu}_0+K\boldsymbol{\beta}\ge 0\ \}.
\]
If $\mathcal{B}\neq\varnothing$, then:
\begin{enumerate}
\item $\mathcal{B}$ is a convex polytope (a bounded convex polyhedron) in $\mathbb{R}^d$.
\item The set of feasible perturbations $\{\,\boldsymbol{\lambda}=K\boldsymbol{\beta}:\ \boldsymbol{\beta}\in\mathcal{B}\,\}$ is bounded. In particular, since the $\mathbf{k}_i$ are orthonormal, there exists $M<\infty$ such that
\[
\|\boldsymbol{\lambda}\|=\|\boldsymbol{\beta}\|\le M\qquad\text{for all }\boldsymbol{\beta}\in\mathcal{B}.
\]
\end{enumerate}
Consequently, the nonnegativity constraint reduces the affine family $\boldsymbol{\mu}_0+\ker(\mathbf{R})$ to a bounded convex polytope of (physically meaningful) nonnegative solutions.
Moreover, the boundedness in \emph{(1)}–\emph{(2)} follows from
\[
\ker(\mathbf{R})\cap \mathbb{R}^n_{+}=\{0\},
\]
which holds under $R_{ij}\ge 0$ and the no–zero–columns assumption.
\end{proposition}

\emph{Proof:}
Convexity: $\mathcal{B}$ is an intersection of halfspaces, hence convex and polyhedral. 

Recession cone: the recession directions of $\mathcal{B}$ are $\{\mathbf{d}\in\mathbb{R}^d: K\mathbf{d}\ge 0\}$. If there were a nonzero $\mathbf{d}$ with $K\mathbf{d}\ge 0$, then $\mathbf{v}:=K\mathbf{d}\in \ker(\mathbf{R})\cap \mathbb{R}^n_{+}$ would be nonzero. Under $R_{ij}\ge 0$ and no zero columns, any $\mathbf{v}\ge 0$ with $\mathbf{R}\mathbf{v}=0$ must satisfy $\mathbf{v}=0$ (each row has a nonnegative dot with $\mathbf{v}$ that can vanish only if $\mathbf{v}$ is supported on columns that are zero in that row; intersecting over all rows forces $\mathbf{v}$ to be supported on a column that is zero in every row, i.e.\ a zero column, which is excluded). Hence the recession cone is $\{0\}$ and $\mathcal{B}$ is bounded.

Bound on $\boldsymbol{\lambda}$: since $\boldsymbol{\lambda}=K\boldsymbol{\beta}$ with orthonormal columns of $K$, we have $\|\boldsymbol{\lambda}\|=\|\boldsymbol{\beta}\|$. Boundedness of $\mathcal{B}$ gives $\sup_{\boldsymbol{\beta}\in\mathcal{B}}\|\boldsymbol{\beta}\|<\infty$, yielding the stated $M$.

\hfill $\square$

\subsection{Richardson's method}
\label{sec:richardson}

Previous work on unfolding the gamma-ray spectra in this work has primarily focused on using an
iterative approach with a stopping criterion and a subtraction scheme to obtain
estimates for the underlying spectrum, see~\cref{app:fics}. As FICS
 in effect is a variant Richardson's iterative
method~\cite{richardson1911finite}, it warrants a deeper analysis of its
robustness.

While the Richardson method is transparent, convergent, and fast for
well-defined systems, the presence of noise and inclusion of background is
problematic. As usual in iteration methods, a stopping criterion is used to
regularize, and one is not forced to alter the problem via normal equations. However, as we will show, a stopping criterion is usually not
enough for the Richardson method. For rank deficient and ill-conditioned
systems, the situation only gets worse. 

The Richardson iteration method for solving $\mathbf{y}=\mathbf{R}\boldsymbol{\mu}$ is defined by:
\begin{align}\label{eq:Richardsoniteration}
    \boldsymbol{\mu}^{(k+1)}=\boldsymbol{\mu}^{(k)}+\omega(\mathbf{y}-\mathbf{R}\boldsymbol{\mu}^{(k)})\,,
\end{align}
where:
\begin{itemize}
    \item $\boldsymbol{\mu}^{(k)}$ is the approximation of the solution at iteration $k$
    \item $\omega$ is the relaxation parameter (step-size)
\end{itemize}
We note that this formulation assumes \textit{Gaussian} errors. However, our data model is Poisson-distributed, in which case the corresponding update step takes the form of a ratio rather than a difference
For the Richardson method to have any possibility of converging, the choice of $\omega$ is crucial:
\begin{theorem}
For a square matrix $\mathbf{R}\in\mathbb{R}^{n\times n}$ of full rank with singular values $\sigma_i >0$, the Richardson iteration converges to a unique solution for any initial guess $\boldsymbol{\mu}^{(0)}$ if and only if the relaxation parameter $\omega$ satisfies
\begin{align}
        0\,<\,\omega\,<\,\frac{2}{\sigma_{\textnormal{max}}}\,,
\end{align}
where $\sigma_{\textnormal{max}}$ is the largest singular value of $\mathbf{R}$.
\end{theorem}
To achieve the fastest convergence of the Richardson iteration, it is essential to select the optimal relaxation parameter $\omega$ within the convergence interval. The convergence rate is directly influenced by the spectral radius $\rho$ of the iteration matrix $\mathbf{I}-\omega\mathbf{R}$,\footnote{This form is easily found by recasting \cref{eq:Richardsoniteration}.} and minimizing this leads to faster convergence. For a full-rank matrix, the optimal relaxation parameter is given by:
\begin{align}
\label{eq:optimalstepsize}
    \omega_{\text{opt}}=\frac{2}{\sigma_{\text{min}} + \sigma_{\text{max}}}\,,
\end{align}
with corresponding spectral radius
\begin{align}
    \rho_{\text{opt}}=1-\omega_{\text{opt}}\sigma_{\text{min}}=1-\frac{2}{1+\textnormal{Cond}(\mathbf{R})}\,.
\end{align}

As demonstrated above, even in well-defined systems where the matrix is of full column rank, noise amplification is inevitable for ill-conditioned matrices. The hope to prevent this is to use early stopping by monitoring the residual $||\mathbf{y}-\mathbf{R}\boldsymbol{\mu}^{(k)}||$ and stop iterations when it stops decreasing or starts increasing, indicating that further iterations may be fitting to noise. Unfortunately, in the Richardson method, early stopping is simply not enough

In the presence of noise, the Richardson iteration method amplifies noise if the system is ill-conditioned. Specifically:
    \begin{enumerate}
        \item If $\mathbf{R}$ is of full column rank, all singular values are $\sigma_i >0$, but depending on the condition number, noise may be significantly amplified despite the residual norm decreasing.
        \item If $\mathbf{R}$ is rank-deficient, it has zero singular values, leading to non-unique solutions. In this case, the Richardson iteration cannot converge to a unique solution, and noise is amplified in the directions corresponding to zero singular values. The method will produce non-unique solutions dominated by noise, and early stopping fails to prevent this issue.
    \end{enumerate}

While early stopping based on the residual norm is often used to regularize iterative methods, for Richardson iteration this is generally insufficient. In full-rank but ill-conditioned systems, the condition number is large, $\rho_{\text{opt}}\to1$, which results in slow convergence and amplification of noise along directions associated with small singular values, even while the residual norm decreases. In rank-deficient systems, zero singular values cause non-uniqueness. The method cannot converge to a unique solution, and noise is projected into the null space of $\mathbf{R}$, producing solutions dominated by noise and often containing negative components.

These effects arise directly from the structure of the method. Since Richardson iteration has no built-in mechanism to constrain non-negativity or to regulate the influence of small or vanishing singular values, additional regularization would require modifications beyond early stopping. As a result, applying the method in practice requires considerable care.

\section{Spectral complexity profile measure}
\label{app:smoothness}

To quantify local variations in the target spectrum \(\veta\) relative to the
energy-dependent detector resolution \(\sigma_\gamma(\Eg)\), we introduce the scale-invariant smoothness measure
\[
s(\Eg) \;=\;
\sqrt{
\left[\frac{1}{\sigma_\gamma(\Eg)}\,\frac{d\veta}{dE_\gamma}\right]^{2} +
\left[\frac{1}{\sigma_\gamma^2(\Eg)}\,\frac{d^{2}\veta}{dE_\gamma^{2}}\right]^{2}
}\, .
\]
Large values of \(s(\Eg)\) mark sharp structure or rapid changes that exceed the
local Gaussian resolution.  
To probe piecewise smoothness we smooth \(s(\Eg)\) with a Gaussian kernel of
bandwidth \(w\,\sigma_\gamma(\Eg)\),
\[
s_w(\Eg) \;=\;
\frac{\sum_{E_k} \Gg^{w}(\Eg,E_k)\, s(E_k)}
     {\sum_{E_k} \Gg^{w}(\Eg,E_k)} ,
\]
where \(\Gg^{w}(\Eg,E_k)\) is the smearing kernel of
Eq.\,\eqref{eq:gaussiancenters} with \(\sigma_\gamma \mapsto w\sigma_\gamma\).
A larger \(w\) reduces noise at the cost of resolving fine structure.
Evaluating \(s_w(\Eg)\) on a mesh \(\{E_k\}_{k=0}^{M}\) gives the set
\(S=\{s_w(E_k)\}_{k=0}^{M}\).  Overall complexity is summarized by
\[
s_{\max/\text{med}} \;=\;
\frac{\max_k S_k}{\operatorname{median}_k S_k} ,
\qquad
s_{\mathrm{CV}} \;=\;
\frac{\sqrt{\operatorname{Var}[S]}}{\operatorname{mean} S}\, .
\]
The ratio \(s_{\max/\text{med}}\) highlights the single sharpest feature relative to
the typical one.  The coefficient of variation \(s_{\mathrm{CV}}\) measures how
evenly smoothness is distributed.  Combining these metrics suggests three
heuristic classes, shown in Fig.\,\ref{fig:complexity}. :
\begin{itemize}
  \item \textbf{Smooth}\,:  low \(s_{\max/\text{med}}\), low \(s_{\mathrm{CV}}\).  Uniformly smooth.
  \item \textbf{Pseudo-smooth}\,:  high \(s_{\max/\text{med}}\), low \(s_{\mathrm{CV}}\).  Mostly smooth with isolated sharp features.
  \item \textbf{Non-smooth}\,:  high \(s_{\max/\text{med}}\), high \(s_{\mathrm{CV}}\).  Complex structure with variable smoothness.
\end{itemize}
The fourth
logical quadrant, low \(s_{\max/\text{med}}\) and high \(s_{\mathrm{CV}}\), is
theoretically possible but has not been observed.  Such spectra would also be
labeled \emph{non-smooth}.

No fixed numerical cutoffs are supplied.  The scheme is intended as a
qualitative aid when tuning global or locally adaptive regularization methods.

\section{Convergence conditions}
\label{sec:convergence}
The qualitative result of the unfolding was found to be independent
of the specific gradient descent optimizer used. However, the rate of convergence can be highly dependent
on the particular spectrum; an optimizer with well-tuned hyperparameters may converge rapidly on
one spectrum but be orders of magnitude slower on another, even within the same dataset.\footnote{The reason for this behavior is related to the spectral radius. Optimizers with modifications that, in practice, result in an effective step size larger than the spectral radius tend to experience unstable convergence, or even divergence. On the other hand, if the effective step size is too small compared to the spectral radius, convergence becomes slow. Optimizers that only have minor modifications compared to standard gradient descent typically do not encounter this issue.}
In practice, the NAdam~\cite{nadam} optimizer, a variant of Adam~\cite{adam} that incorporates Nesterov Momentum, was
found to perform efficiently across all tested spectra when using a step size $d\tau$, set within the convergence interval:

\begin{equation}\label{eq:optLips}
    0<d\tau <1/L
   % d\tau = \frac{2}{\sigma_{\text{min}} + \sigma_{\text{max}}},
\end{equation}
where $L$ is the Lipschitz constant of the gradient of the loss function.
However, calculating $L$ directly is challenging as
it amounts to calculating the Hessian, which in our case is infeasible.
If we instead make the approximation to a quadratic loss, the Lipschitz constant is
given by the maximum singular value of the relevant response,
$L=2\sigma_{\text{max}}^{2}$. A complication occurs when we introduce a
regularization term as this changes the loss landscape. Adding a regularization
can lead to a larger Lipschitz constant, as the regularization term can
completely dominate the behavior of the loss function. Effectively, this leads to a
shrinking of the convergence interval as the upper bound is shifted and
\cref{eq:optLips} with $L=2\sigma_{\text{max}}^{2}$ can no longer guarantee
convergence.

When adding a regularization term to the loss function
the Lipschitz constant takes the form
\begin{align}
    L=L_{\ell}+\alpha L_{\Omega}\,.
\end{align}
While these are possible to calculate in theory, they are not available in practice. 
For the special case of Tikhonov regularization with a quadratic loss function,
the Lipschitz constant takes the form $L=2\sigma_{\text{max}}^2 +\alpha$, serving as a new upper bound to achieve convergence. For the regularization schemes we use, no simple
analytical expression is available. In practice, a sufficiently good step length can
be found by manually tuning the parameter until the loss curve is sufficiently steep without signs of divergence.

\section{Marginal and simultaneous confidence intervals}
\label{app:margsimci}

The Monte Carlo resampling method (\cref{sec:uncquant}) generates an ensemble of unfolded solutions $\boldsymbol{\eta}$, which characterizes the variability inherent in the RMLE method. Because this distribution typically exhibits non-negligible higher-order moments, summarizing the ensemble inevitably discards important information. Consequently, for subsequent steps in the Oslo Method, it is strongly recommended to propagate the full ensemble rather than any summary statistic.

However, summaries—particularly confidence intervals (CIs)—can offer intuitive measures of uncertainty. One must then distinguish between \emph{marginal} and \emph{simultaneous} confidence intervals.

A \textbf{marginal confidence interval} is constructed independently for each bin from the ensemble. Common approaches include the percentile method or bias-corrected and accelerated (BCa) intervals. While simple to compute, marginal intervals consider each bin in isolation and thus ignore correlations between bins. As a result, marginal intervals are inadequate for global inference tasks, such as testing whether a peak significantly differs from zero or comparing entire unfolded solutions $\heta^*$ to $\heta'^*$.

In contrast, a \textbf{simultaneous confidence interval} accounts for bin-to-bin correlations and provides global coverage. Specifically, a simultaneous CI describes an interval or region that contains the entire true function $\boldsymbol{\eta}$ with a given confidence level. Formally, a simultaneous interval ensures that, asymptotically, at least a fraction\footnote{Here, $\alpha$ denotes the significance level, not to be confused with a regularization parameter.} $1 - \alpha$ of the repeated samples will contain the complete function.

The simultaneous CI is inherently high-dimensional, corresponding geometrically to an $N$-dimensional confidence ellipsoid enclosing a proportion $1 - \alpha$ of the probability mass. Compared to marginal 
intervals, simultaneous CI are always wider. Practical methods to derive simultaneous CIs typically reduce this region into interpretable intervals, using methods such as:
\begin{description}
\item[Supremum norm intervals] Computed by evaluating the maximum absolute deviation across bins and determining a uniform margin that simultaneously covers all bins. This approach yields conservative intervals.
\item[Studentized supremum norm intervals] Similar to the supremum norm method, but deviations are standardized by local variance estimates. This results in tighter intervals that better reflect spatially varying uncertainty.
    \item[Bonferroni intervals] Constructed by adjusting the confidence level of each marginal interval to $1 - \frac{\alpha}{N}$, ensuring that the global coverage probability is at least $1 - \alpha$ via the Bonferroni correction.
\end{description}
In practice, we find that the studentized supremum norm and the Bonferroni-based simultaneous confidence intervals yielded nearly identical results. Unless otherwise stated, all confidence intervals presented in this work are marginal. Although these are easier to compute and interpret, it is important to exercise caution when drawing global inferences from marginal intervals, as they do not account for bin-to-bin correlations. Researchers should be mindful of these limitations when interpreting the results.

\section{Coverage}
\label{sec:appcoverage}

For each physical bin \(i\), we denote by \(\eta_i\) the (unknown) true intensity.
If a confidence interval \(\operatorname{CI}_i\) is constructed for that bin, the
\emph{coverage probability}
\begin{equation}
  p_i \;=\; \operatorname{Pr}\!\left(\,\eta_i \,\in\, \operatorname{CI}_i \right)
  \label{eq:def_covprob}
\end{equation}
quantifies how often, in repeated experiments, the interval contains the true value. The \textit{confidence level} is the prescribed coverage probability that we want our constructed confidence interval to have.

The \textit{empirical} or \textit{actual} coverage probability of the CI can 
be estimated using Monte Carlo simulations. Given a fixed underlying
spectrum \(\boldsymbol{\eta}\), we simulate \(M\) independent reconstructions, each
producing a confidence interval \(\operatorname{CI}_i^{(m)}\) for bin \(i\). 
For each simulation, we define a Bernoulli random variable
\begin{equation}
  I_i^{(m)} \;=\;
  \mathbbm{1}\!\bigl\{\eta_i \,\in\, \operatorname{CI}_i^{(m)}\bigr\},
  \label{eq:indicator}
\end{equation}
which equals one if the interval contains the true value and zero otherwise.
The estimate for the actual coverage probability is then given by
\begin{equation}
  \hat p_i
  \;=\;
  \frac{1}{M}\sum_{m=1}^{M} I_i^{(m)} ,
  \qquad
  \mathbb{E}\!\bigl[\hat p_i\bigr] = p_i .
  \label{eq:coverage_rate}
\end{equation}
Since \(I_i^{(m)}\) is Bernoulli distributed, the standard error of \(\hat p_i\) is
\begin{equation}
  \operatorname{SE}\!\left(\hat p_i\right)
  \;=\;
  \sqrt{\frac{\hat p_i\bigl(1-\hat p_i\bigr)}{M}} .
  \label{eq:coverage_se}
\end{equation}

\section{Experimental data used for the Oslo Method}
\label{app:exp}
The aim of the Oslo Method is to obtain nuclear level densities and gamma-ray transmission coefficients. 
The method consists of five main steps:
\begin{enumerate}
    \item Use experimental data to construct the $\matt{P}$ ($E_{\mathrm{in}}, E_\gamma$) matrix of
    $\gamma$-ray spectra for each initial excitation-energy bin $E_{\mathrm{in}}$.
    
    \item Unfold the  ($E_{\mathrm{in}}, E_\gamma$) matrix using response
    functions of the $\gamma$-detector array to obtain a spectrum of full-energy
    peaks, the \textit{all-generation} matrix. This unfolding has usually been done with the method
    described in Ref.~\cite{GUTTORMSEN1996371}.
    \item Extract the distribution of the \textit{first-generation} gamma rays for each initial excitation-energy bin through an iterative subtraction technique, thus obtaining the first-generation
    matrix $P(E_{\mathrm{in}}, E_\gamma)$~\cite{GUTTORMSEN1987518}. 
    \item
    Decompose the first-generation matrix into two vectors, namely, the nuclear
    level density \mbox{$\rho(E_{\mathrm{in}}-E_\gamma)$} and the gamma-ray transmission
    coefficient $\mathcal{T}(E_\gamma)$, under the ansatz that the
    first-generation matrix can be written as $P(E_{\mathrm{in}}, E_\gamma) \propto
    \rho(E_{\mathrm{in}}-E_\gamma) \times \mathcal{T}(E_\gamma)$~\cite{SCHILLER2000498}.
    %\begin{equation}
    %    \matt{P} = \vecc{\rho} \otimes \vecc{\mathcal{T}}
    %\end{equation}
    %\begin{equation}
    %    p(e_i, e_f) \propto \rho(e_f) \cdot \mathcal{T}(e_i)
    %\end{equation}
    
    \item Normalize $\rho(E_{\mathrm{in}}-E_\gamma)$ and $\mathcal{T}(E_\gamma)$ to
    auxiliary data~\cite{SCHILLER2000498,Larsen2011}.
\end{enumerate}
%\anders{Just to clarify notation: should the symbol $P(E_{\mathrm{in}}, E_\gamma)$ represent \textit{the distribution of the first-gen matrix} or \textit{the first-gen matrix} itself? In the above list it is used for both, so it's a bit unclear if \textit{the distribution of the matrix} and \textit{the matrix} are the same thing or not.} 
%\cecilie{It is the distribution of first-generation $\gamma$ rays, so this was a typo. Hope it's more clear now!}
This work focuses on step 2, i.e., the unfolding of complex gamma-ray spectra. But here we first briefly describe the way the experimental data are obtained (step 1).

A large variety of nuclear reactions can be used for the Oslo Method.
To reach high excitation energies in the nuclei of interest, the following reactions have been
used:
\begin{itemize}
    \item Charged-particle reactions such as  ($^{3}$He,$\alpha$), ($d,p$) and ($p,p'$) in normal kinematics; see, e.g., Refs.~\cite{Nyhus2010,GUTTORMSEN2021136206,Larsen2013}.
    \item The (d,p) reaction in inverse-kinematics, e.g., Ref.~\cite{Ingeberg2020}.
    \item $\beta^-$ decay, for \textit{the beta-Oslo Method},   e.g.,\ Refs.~\cite{Spyrou2014,Liddick2016,Larsen2018}.
\end{itemize}
We remark that for the beta-Oslo Method, the initial excitation energy is
inferred from total absorption spectrometry, which poses additional challenges
for the unfolding. For the purpose of discrete spectroscopy using a total
absorption spectrometer like the SuN detector~\cite{SIMON201316}, machine
learning has very recently been applied to such data
sets~\cite{DEMBSKI2024169026}. Moreover, a wide range of gamma-ray detectors
have been utilized for applications of the Oslo Method, such as NaI:Tl
scintillators (CACTUS~\cite{CACTUS1990}, SuN~\cite{SIMON201316}), LaBr$_3$:Ce
scintillators (HECTOR$^+$~\cite{GIAZ2013910}, OSCAR~\cite{oscarfabio}), and
high-purity Ge detectors with anti-Compton shields such as
STARLiTeR~\cite{Simon2016}.
As a consequence, there is large diversity of gamma-ray spectra from different experiments and
nuclei.

\begin{figure}
    %\centering
    \includegraphics[width=1.\columnwidth]{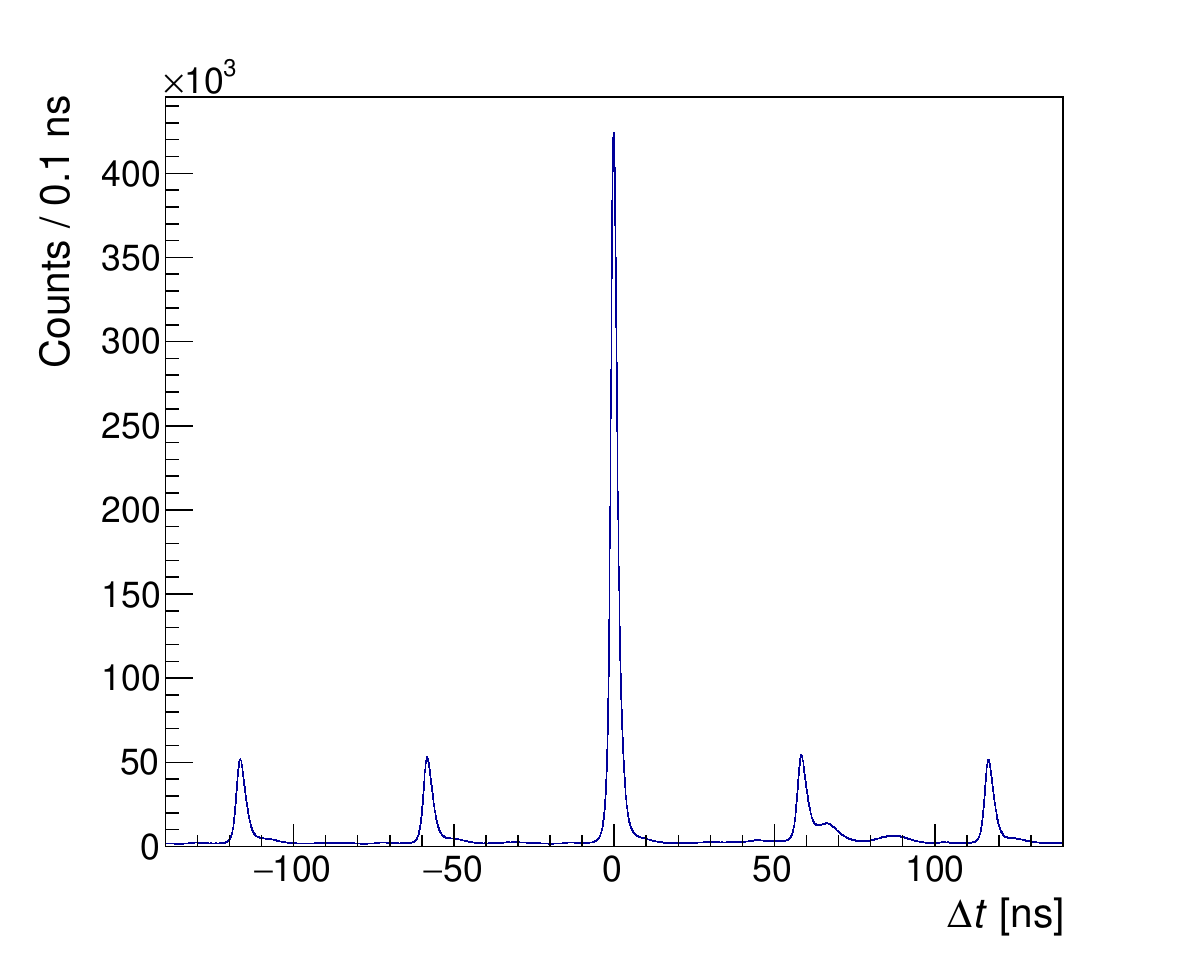}
    \caption{The time spectrum for all 30 OSCAR detectors from the $^{28}$Si($p,p'\gamma$) reaction. The
    large peak at \SI{0}{ns} is gated for the total counts, and the lower peaks for the background.}
    \label{fig:si28_time}
\end{figure}
In this work we focus on experimental data for which the initial excitation energy is determined from information independent from the gamma rays, i.e., from the charged-particle ejectiles. 
Here we use a calibration run on natural Si as an example, using a proton beam of 16 MeV. 
As natural Si mainly consists of $^{28}$Si (92\%), the data set is dominated by the $^{28}$Si($p,p'\gamma$)$^{28}$Si reaction.
This data set was taken with OSCAR~\cite{oscarfabio} for gamma-ray detection and the Silicon Ring (SiRi)~\cite{Siri} for proton detection, with SiRi in backward angles 
($126$--$140^\circ$ with respect to the beam direction), and using digital electronics for the data acquisition where all events are time stamped (XiA Pixie-16 digitizers, 14-bit and 500 MHz sample rate for OSCAR and 16-bit and 250 MHz sample rate for SiRi). 
SiRi consists of eight $\Delta E$--$E$ telescope modules, with a thin front detector of $\approx130$ $\mu$ that is segmented into eight strips, which cover $\approx 2^\circ$ each. 
The front detector is accompanied by a thick $E$ detector of thickness $\approx 1550$ $\mu$m, where the protons are stopped. 
OSCAR is built of 30 large-volume LaBr$_3$:Ce detectors distributed on a spherical frame.

\begin{figure*}
    %\centering
    %\includegraphics[width=2.\columnwidth]{figures/Si28_matrices.pdf}
    %\includegraphics[width=2.\columnwidth]{figures/Si28_projections.pdf}
    \includegraphics{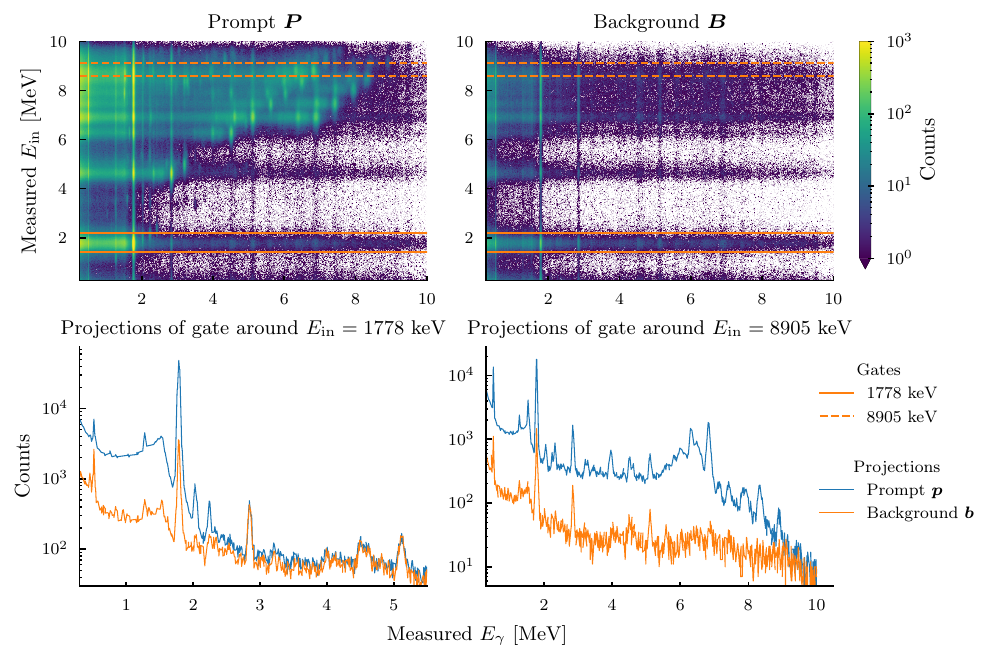}
    \caption{Data from a calibration run using the $^{28}$Si($p,p'\gamma$) reaction, where gamma rays were measured with OSCAR. (Top left) Gates on prompt events; (top right) Gates on background events; 
    (bottom left) gamma spectra for gates on prompt and background events for $E_{\mathrm{in}} \approx 1778$ keV; (bottom right) gamma spectra for gates on prompt and background events for $E_{\mathrm{in}} \approx 8900$ keV.}
    \label{fig:si28}
\end{figure*}

In Fig.~\ref{fig:si28_time}, the time spectrum of the $^{28}$Si data set is shown. 
The reference time (the ``start'') is the proton detection in SiRi, while the ``stop'' is the signal in one of the OSCAR detectors. 
The figure shows the summed time spectrum, i.e., the time spectrum for all the 30 OSCAR detectors. 
The prompt peak has a time difference between the start and stop signals of zero nanoseconds and has a resolution of $\approx 2.4$ ns FWHM. 
Due to the pulsed beam from the cyclotron, other peaks show up in the time spectrum, for which the gamma was detected before the proton, such as the peak at $\approx -59$ ns, or after the proton, such as the peak at $\approx 59$ ns. 
These peaks are used to estimate the background contribution (random background plus beam-induced background) in the prompt peak, so that the spectra are incremented 
%\anders{Should this be `event counts are incremented'?} 
when the time difference between the $\Delta E$ and OSCAR detector is within $\Delta t = [-5.0,5.0]$ ns and decremented if  $\Delta t = [-62,-52]$ ns. %\anders{Is ``randoms'' common terminology?} 
%\erlend{en random er kompositt. den flate bakgrunn er uniform, isotrop, flat, cosmic strårling. beam induced randoms, toppen.}
%\cecilie{I added ``(random background plus beam-induced background)'' in the text.}
The resulting initial excitation energy versus gamma-ray energy matrix of $^{28}$Si with the gate on the prompt time peak is shown in Fig.~\ref{fig:si28}a.
Here, the detected proton energy has been converted to initial excitation energy in $^{28}$Si using the reaction kinematics.
Correspondingly, the matrix generated with the gate on the random time peak is shown in Fig.~\ref{fig:si28}b. 

To better visualize the background contribution, we have made projections of both the prompt and background gamma-ray spectra for two different initial-excitation-energy regions as marked by the black lines in Fig. ~\ref{fig:si28}a and b, and shown in Fig. ~\ref{fig:si28}c and d, respectively. 
The gamma-ray spectrum for the initial-excitation-energy gate corresponding to the first 2$^{+}$ level in $^{28}$Si (Fig. ~\ref{fig:si28}c) clearly shows the 1779-keV line, which is the full-energy peak corresponding to the decay to the ground state. 
We observe that the background spectrum contains many of the same features as the prompt one, however in the gamma-energy region of interest, i.e., the region up to about the maximum initial excitation energy in the gate, the background has significantly less counts than the prompt gamma-ray spectrum, about a factor of 10 or so. 
This is what is expected from the time spectrum in Fig.~\ref{fig:si28_time}.
For the spectra obtained by gating on $E_{\mathrm{in}} \approx 8905$ keV, the same behavior is seen. 

Before 2019 and the commissioning of OSCAR, the CACTUS array was the workhorse
in the nuclear physics experiments at the Oslo Cyclotron Laboratory (OCL). Also, the data acquisition system was
analogue, with a  ``master gate'' to control the event building. This master
gate was a logic signal generated either from one of the 64 $\Delta E$ detectors
giving a signal (an OR signal of all the 64 strips), or taking the overlap of
the $\Delta E$ and $E$ detectors requiring both a $\Delta E$ strip and an $E$
detector giving signals, or just using an OR of the eight $E$ detectors. This
same master gate was used as the ``start'' in the time-to-digital converters for
the NaI:Tl detectors of CACTUS.

\begin{figure*}
    \centering
    \includegraphics{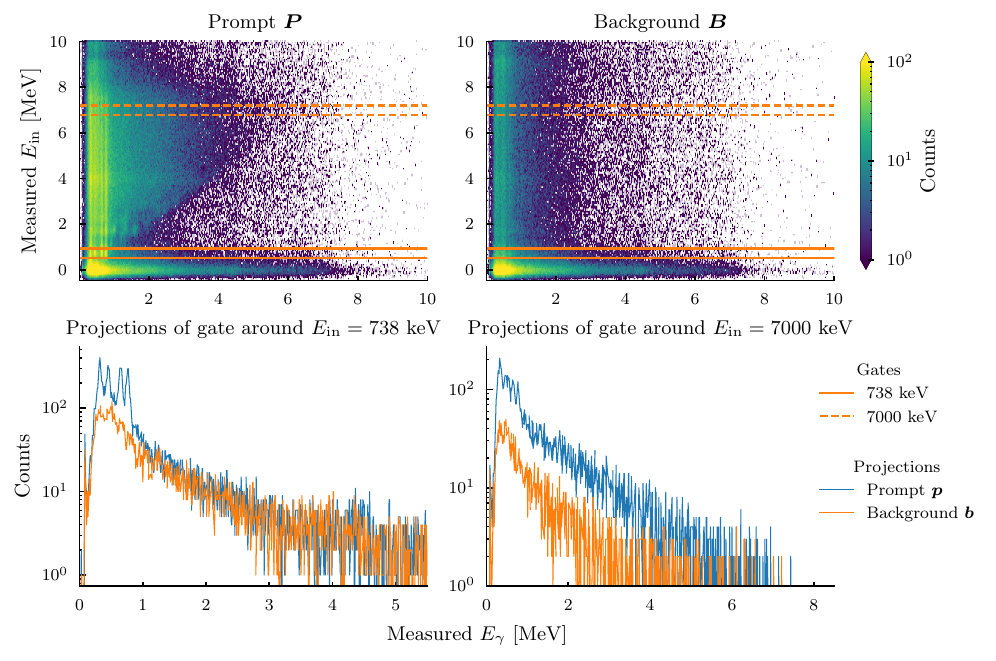}
    \caption{Data from the $^{186}$W($\alpha,\alpha'\gamma$) reaction published
    in Ref.~\cite{Larsen2023}, where gamma rays were measured with CACTUS.
    (Top left) Gates on prompt events; (top right) Gates on background events; (bottom left)
    gamma spectra for gates on prompt and background events for $E_{\mathrm{in}}
    \approx 738$ keV; (bottom right) gamma spectra for gates on prompt and background events
    for $E_{\mathrm{in}} \approx 7000$ keV.}
    \label{fig:w186}
\end{figure*}

An  example of a CACTUS experiment with  poor statistics is the $^{186}$W($\alpha,\alpha'\gamma$) data shown in Fig.~\ref{fig:w186} and published in Ref.~\cite{Larsen2023}. 
%\anders{This sentence sounds a bit strange to me. Maybe change to ``[\ldots] a CACTUS experiment with rather poor statistics [\ldots]''? Or change to ``poor-statistics experiment''?}
%\cecilie{Updated.}
In this experiment, the start was generated from the $E$ detectors only, and as they are quite thick detectors, the timing properties are not as good as for the thin $\Delta E$ detectors.
In addition, CACTUS had a significantly worse time resolution than OSCAR, as signals from the NaI:Tl crystals have a slower rise time than the LaBr$_3$:Ce crystals. 
Due to the worse time resolution, the time window set on the prompt events was quite broad, $\Delta t = [-21.6,21.6]$ ns, and so the gate on the randoms was broad as well, $[110.4,153.6]$ ns.

The prompt $(E_{\mathrm{in}},E_\gamma)$ events for the $^{186}$W($\alpha,\alpha'\gamma$) reaction are shown in Fig.~\ref{fig:w186}a, and the background matrix is shown in Fig.~\ref{fig:w186}b.
We observe that the background component is quite reasonable also in this case for the lower initial excitation energies, as seen from the projection around the second 2$^+$ level at $737.97$ keV in Fig.~\ref{fig:w186}c. 
However, as shown in Fig.~\ref{fig:w186}d for the gate at higher initial excitation energies around 7 MeV, the background component becomes comparable to the prompt spectrum for  $E_\gamma \approx 5.5$ MeV and above. 
For such cases, the unfolding might become even more challenging, especially for methods where the background is not input to the unfolding but where the unfolding is performed directly on the background-subtracted spectra.
Inevitably, the low signal-to-noise ratio will lead to large uncertainties in the resulting unfolded spectrum. 
The noise in the data is one of the major complicating factors for the unfolding problem, as discussed in this work.

\section{Regularization systematics}
\label{sec:regularization_systematics}

We benchmark the RMLE method on both low- and high-complexity spectra, varying
total number of counts, and applying a range of regularization types and
strengths. While synthetic discrete spectra are trivial to construct, realistic
quasi-continuous or high-complexity spectra require gamma-cascade simulations
using codes such as RAINIER\cite{RAINIER} or DICEBOX\cite{DICEBOX}. A
\ce{^{120}Sn}-like spectrum is used as an example of a high-complexity spectrum,
while a \ce{^{166}Ho}-like spectrum represents a quasi-continuous spectrum. 
Although the simulations are not optimized for realism, they adequately support the analysis presented here.
For clarity of presentation we unfold only one-dimensional spectra.

The Kullback–Leibler (KL) divergence $D_{\text{KL}}(\heta||\veta)$ is a natural
choice for evaluating the discrepancy between the fitted distribution $\heta$ and
the actual distribution $\veta$, but it becomes numerically unstable when either
distribution contain vanishingly small elements. As a robust alternative we
employ the first-order Wasserstein (earth-mover)
distance\cite{kantorovich1942translocation, kantorovich1960mathematical},
denoted $W_1(\heta, \veta)$,
which measures the minimum \enquote{work} required to 
transform one distribution into the other. When comparing distributions with different total counts, normalization
is applied to ensure an unbiased comparison.

A practical strategy for selecting the regularization strength $\alpha$ is the
\textit{L-curve} diagnostic. An L-curve plots the residual (fidelity) cost --- here 
$D_{\text{KL}}(\hnu||\vn)$ --- on the abscissa and the regularization cost 
on the ordinate for a sweep of $\alpha$. Ideally, the resulting curve forms an L-shape, where the point of maximum curvature --- the corner of the L --- balances data fidelity and solution regularization.

Because our data is simulated,  $\veta$ is known \textit{a priori}. This allows us
to validate the L-curve diagnostic by computing independent \enquote{cost curves}
using $W_1\left(\heta, \veta \right)$ for the residual.
In an ideal scenario the
minimum residual cost coincides with the L-curve corner.

\begin{figure*}
  \centering
\includegraphics{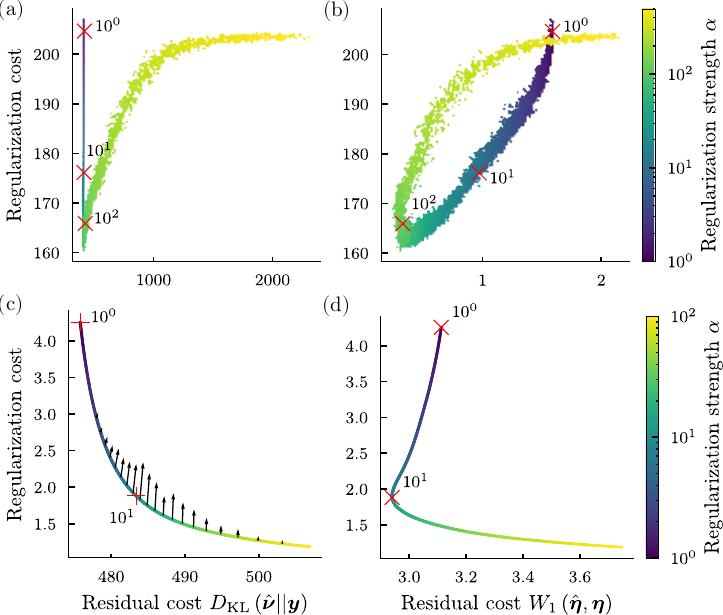}
  \caption{%
    (a) L-curve for a single-peak spectrum with sparsity penalty. A pronounced minimum occurs near $\alpha \approx 10^2$, and the scattered shape reflects a non-convex loss landscape. 
    (b) Corresponding Wasserstein cost curve for a peaked spectrum; the optimal value $\alpha \approx 10^2$ aligns with the L-curve elbow in panel~(a). 
    (c) L-curve for a \ce{^{166}Ho}-like spectrum with smoothness penalty, whose curvature peaks near $\alpha \approx 10$ without a sharp corner. 
    (d) Wasserstein cost curve for the \ce{^{166}Ho} case; its minimum near $\alpha \approx 10$ coincides with the curvature peak in panel~(c).
  }
  \label{fig:lcurve_sparsity}
\end{figure*}

Figure~\ref{fig:lcurve_sparsity}(a, b) thus confirm that, for a low‑complexity peak with sparsity regularization, the L‑curve correctly pinpoints the optimal sparsity regularization strength ($\alpha \approx 10^2$ in this example).

%\begin{figure}[h]
%    \centering
%    \includegraphics{figures/curvature.pdf}
%  \caption[\ce{^{166}Ho} cost curve, smoothness penalty]{%
%    Wasserstein cost curve for the \ce{^{166}Ho} case.  
%    The minimum at $\alpha \approx 10$ coincides with the curvature maximum in Fig.~\ref{fig:lcurve_sparsity}(c).}
%    \label{fig:curvature:Ho}
%\end{figure}

For the continuous \ce{^{166}Ho}‑like spectrum with smoothness regularization, the curvature criterion again selects the optimal $\alpha$ as the cost minimum, shown in~\cref{fig:lcurve_sparsity}(c, d). % and Fig~\ref{fig:curvature:Ho}.
The L-curve lacks
an obvious elbow, but the point of largest curvature still signals 
the best regularization strength.

\begin{figure*}
    \centering
    \includegraphics[]{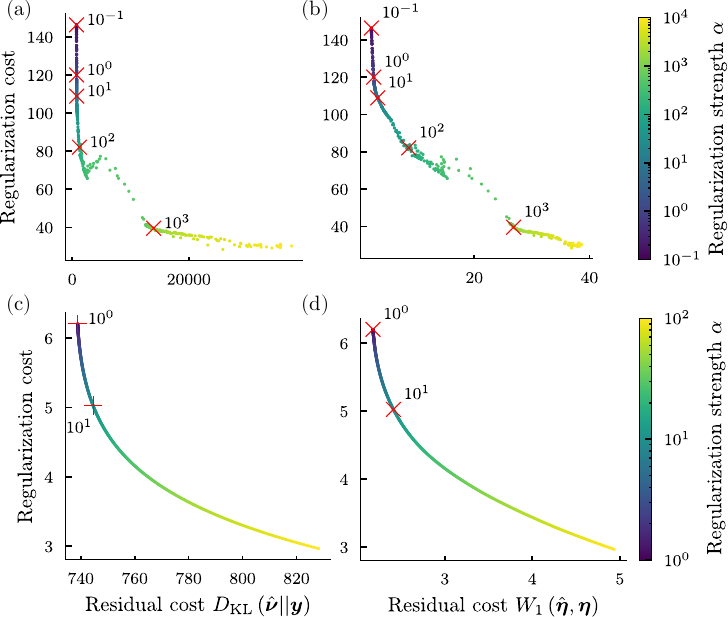}
\caption{%
(a) L-curve with sparsity penalty. A superficial bend appears near $\alpha \approx 3\times10^{\,2}$, but this is a false optimum, see panel~(b). 
(b) Cost curve for sparsity penalty. Residual increases monotonically with~$\alpha$, confirming the absence of a valid optimum. 
(c) L-curve with smoothness penalty. The curvature is nearly monotonic; no clear corner is present. 
(d) Cost curve for smoothness penalty. Residual worsens steadily; no suitable regularization strength is found. 
Panels (a)–(d) thus show L-curves (left) and corresponding cost curves (right) for a high-complexity \ce{^{120}Sn} spectrum under sparsity (top row) and smoothness (bottom row) regularization. While the L-curves suggest possible optima, the cost curves reveal that neither sparsity nor smoothness regularization yields a meaningful trade-off between residual and penalty.%
}\label{fig:lcurve_sn}
\end{figure*}

High‑complexity spectra---such as the \ce{^{120}Sn} benchmark---behave differently.  As shown in \cref{fig:lcurve_sn}(a,c), their L‑curves may contain inflection‑like features that could be mistaken for corners.  However, the cost curves in \Cref{fig:lcurve_sn}(b,d) demonstrate that these features do not correspond to true minima.  In other words,
the L‑curve criterion fails, and regularization offers no clear benefit.

\begin{figure}[h]
    \centering
    \includegraphics[width=\linewidth]{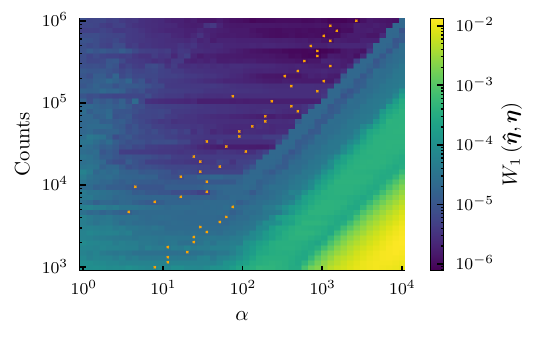}
    \caption{Residual cost $W_1\!\left(\heta,\veta\right)$ for a single‑peak
    spectrum as a joint function of the total number of detected counts
    (vertical axis) and the sparsity strength~$\alpha$ (horizontal axis).  At low
    statistics the landscape is steep: the optima (orange dots) are narrowly confined and
    highly sensitive to~$\alpha$.  As the number of counts increases the surface
    flattens, indicating that data statistics progressively outweigh the
    influence of the penalty term.}
    \label{fig:discrete_error_mesh}
\end{figure}

Finally, \Cref{fig:discrete_error_mesh} explores how counting statistics
modulates the sensitivity to regularization.  With fewer counts the residual cost
exhibits a narrow trough, making the solution acutely dependent
on~$\alpha$.  As statistics improve, the trough widens and shallows, implying
that the data themselves begin to dominate the optimization and tolerate
stronger penalties without a loss of fidelity. The situation is analogous for
\ce{^{166}Ho} with smoothness regularization.

%\paragraph*{Summary}  
For spectra of low spectral complexity, the L‑curve---interpreted via its point of
maximum curvature---is a reliable, data‑driven tool for selecting the
regularization strength, and its recommendation is corroborated by the
Wasserstein cost curves.  In contrast, high complexity spectra lack a single
optimal penalty parameter; any fixed-value regularization either over‑ or
under‑constrains the solution.  Moreover, increased counting statistics
systematically reduce the sensitivity of the reconstruction to regularization
choices, highlighting the practical value of high‑statistical‑quality data when
unfolding complex gamma-spectra.

\FloatBarrier
\section{FICS}
\label{app:fics}

\newcommand{\hv}[1]{\hat{\vecc{#1}}}
\newcommand{\hvu}{\hv{u}}
\newcommand{\hvf}{\hv{f}}
\newcommand{\hvuk}{\hvu^{(k)}}
\newcommand{\itat}[2]{{#1}^{\left(#2\right)}}
\newcommand{\itk}[1]{\itat{#1}{k}}
\newcommand{\hvfk}{\itk{\hvf}}

The \textit{Folding Iteration with Compton
Subtraction} method (FICS), first described in~\cite{GUTTORMSEN1996371},
consists of an iterative component that solves the inverse problem, followed by
a correction to the unfolded solution called \textit{Compton subtraction}.
The complete method is summarized in~\cref{alg:fics}.
The iterative component implements Richardson's method for Gaussian error, with constant step size $\omega=1$,
and early stopping. For a response matrix $\matt{R}$, the update to the current solution $\hvuk$
is given by
\begin{equation}
\hvu^{(k+1)} = \hvu^{(k)} + \vecc{y} - \matt{R}\hvu^{(k)},
\end{equation}
which is identical to~\cref{eq:Richardsoniteration}. Guttormsen et al.~\cite{GUTTORMSEN1996371} observed
that the solution diverged, necessitating early stopping. Furthermore, they noted
that the solution contained spurious fluctuations, requiring the iterations to be stopped
before these fluctuations dominated the spectrum. The stopping criteria (not detailed in~\cite{GUTTORMSEN1996371})
combine two terms: the residuals between the folded solution $\hv{f}^{(k)}=\matt{R}\hvuk$ and the data $\vecc{y}$,
and a cost term based on the fluctuations in $\hv{u}^{(k)}$. The residual cost is expressed as a $\chi^2$:
\begin{equation}
c_{\text{res}}^{(k)} = \sum_i \frac{\left(y_i - f_i^{(k)} \right)^2}{|y_i|}.
\end{equation}
The fluctuations cost uses the $\ell_1$ norm of the difference between the solution $\hvu^{(k)}$ and
its smoothed version under a smoothing operator $S$:
\begin{align}
\itk{c_{\text{fluct}}} = \sum_i \left| \hat u_i^{(k)} - S\left(\hat{u}^{(k)}\right)_i \right|.
\end{align}
The total cost combines these terms with a weighting factor:
\begin{equation}
\itk{c_{\text{tot}}} = \alpha \itk c_{\text{res}} + (1-\alpha)\itk c_\text{fluct}
\end{equation}
where $0 \leq \alpha \leq 1$. The iteration terminates when \mbox{$\itat{c_\text{tot}}{k+1} > c_\text{tot}^{k}$}.

Guttormsen et. al.~\cite{GUTTORMSEN1996371} recognized that this scheme remained insufficient, as the solution continued to exhibit spurious fluctuations.
Specifically, they observed that the fluctuations in $\hvuk$ did not correspond to those in
the data $\vecc{y}$. The Compton subtraction method was developed to address this discrepancy.
The method relies on physical assumptions about the Compton component,
particularly its smoothness: we expect this component to be free of fluctuations. To derive the method,
recall from~\cref{eq:Dcomponents} that the discrete matrix $\matt{D}$ comprises the following
discrete components:

\begin{align}
    \mathbf{D}=\mathbf{P}_{\text{f}}+\mathbf{P}_{\text{s}}+\mathbf{P}_{\text{d}}+\mathbf{P}_{\text{a}}+\mathbf{P}_{\text{c}}\,.
\end{align}

Let $\matt{P}_{\text{p}} =\mathbf{P}_{\text{s}}+\mathbf{P}_{\text{d}}+\mathbf{P}_{\text{a}}$ denote the discrete peaks
excluding the full energy peak (corresponding to $w(i)$ in~\cite{GUTTORMSEN1996371}). The data $\vecc{y}$ can then be expressed in terms of $\vecc{x}$ as
\begin{equation}
\label{eq:f0}
\vecc{y} = \Gl\matt{D}\vecc{x} = \Gl\left(\mfe + \matt{P}\text{p} + \mcom\right)\vecc x =
\vecc f_{\text{f}} + \vecc f_{\text{p}} + \vecc f_{\text{c}}\, ,
\end{equation}
for some smearing operator $\Gl$.
From $\hvuk$, we derive an estimate $\hv{f}_{\text{c}}$ of the Compton component
by rearranging~\cref{eq:f0} (corresponding to Eq. (17) in~\cite{GUTTORMSEN1996371}):
\begin{equation}
\label{eq:f1}
\hv{f}_{\text{c}} = \vecc{y} - \underbrace{\Gl\mfe\hvuk}_{\text{Smeared full energy peaks}} 
- \underbrace{\Gl\mpeak\hvuk}_{\text{Smeared discrete peaks}}.
\end{equation}
We then modify $\hv{f}{_\text{c}}$ based on its expected physical properties by applying a smoothing operator $S$, attempting to better approximate the true $\vecc f_{\text{c}}$. The final solution is obtained by subtracting both the smoothed Compton
estimate and the discrete peak estimate from the data:
\begin{equation}
\label{eq:f2}
\Gl\mfe \hv{u}_{\text{sol}} := \vecc{y} - \underbrace{S\left(\hv{f}_\text{c}\right)}_{\text{Smoothed Compton}} -  \underbrace{\Gl\mpeak\hvuk}_{\text{Smeared discrete peaks}}.
\end{equation}

This formulation differs from Eq. (18) in~\cite{GUTTORMSEN1996371} due to two fundamental issues.
The first stems from an error propagated from~\cite{sukosd_spectrum_1995}. S\"uk\"osd et al.
correctly identified that the response matrix is ill-conditioned, primarily due to
Gaussian smoothing, necessitating regularization. They attempted to
unfold to an $\eta$-space with higher resolution than the experimental spectrum.
However, as demonstrated in~\cref{sec:unfoldingspaces}, this is impossible without strong and robust regularization.

While S\"uk\"s\"od et al.\ observed oscillations when unfolding at full
resolution \(\sigma_\gamma\), they found that unfolding with \(\sigma =
0.5\sigma_\gamma\) produced optimal results, based on analysis of a synthetic
spectrum with discrete peaks. However, unfolding discrete peaks implicitly
incorporates prior information about the solution’s discrete nature. This
additional prior information serves as a regularizer, enabling unfolding of
discrete peaks to resolutions beyond the experimental resolution. In effect, it
is the \emph{human analyst}, not the method, that is performing the regularized
optimization.

For continuous spectra, comprising numerous overlapping Gaussians, this approach fails.
The solution space becomes too degenerate to identify the true solution uniquely,
and unfolding with $\sigma < \sigma_\gamma$ is impossible without substantial additional regularization.
Any choice of $\sigma < \sigma_\gamma$ leaves residual degeneracy.
Therefore, the only viable choice for the response is $\matt{R}=\Gg\matt{D}$, contrary to standard FICS.

The second issue, related to the first, concerns an ambiguity in the interpretation of $\hvu_\text{sol}$:
whether it represents sharp peaks (\enquote{delta peaks}) or a spectrum with Gaussian correlation. For the former interpretation,
~\cref{eq:f2} is formally correct but practically unsolvable due to the non-invertibility of $\Gl$. For the latter,
the appropriate equation becomes
\begin{equation}
\label{eq:f3}
   \mfe \hv{u}_{\text{sol}} := \vecc{y} - S\left(\hv{f}_\text{c}\right) -  \Gl\mpeak\hvuk\,.
\end{equation}
In this case, $\hvuk$ and $\hvu_\text{sol}$ exist in distinct spaces: the former comprising sharp peaks,
the latter exhibiting Gaussian correlation.

FICS can accommodate any smoothing operator $S$, but Gaussian smoothing is the
most natural choice. Without any significant loss of generality, we define the smearing operation as a Gaussian smoothing
$S\left(\hv{f}_\text{c}\right) = \Gk\hv{f}_\text{c}$, where $\kappa$ determines the resolution. Combining~\cref{eq:f1} and~\cref{eq:f3}
yields
\begin{align}
    \mfe \hvu_\text{sol} &= \vecc{y} - \Gk\hv{f}_\text{c} - \vecc{y} + \Gl\mfe\hvuk + \hv{f}_\text{c}.
\end{align}
Since $\mfe$ is diagonal, we can rearrange terms to obtain
\begin{align}
\label{eq:f30}
    \hvu_\text{sol} &= \mfe^{-1}\Gl\mfe \hvuk + \mfe^{-1}(\matt{I} - \Gk)\hv{f}_\text{c}.
\end{align}
Had we instead employed definition~\cref{eq:f2}, and assuming $\Gl$ were invertible, we would obtain
\begin{align}
\label{eq:f4}
    \hvu_\text{sol} &\stackrel{!}{=} \mfe^{-1}\Gl^{-1}\Gl\mfe \hvuk + \mfe^{-1}\Gl^{-1}(\matt{I} - \Gk)\hv{f}_\text{c}\notag\\
    &\stackrel{!}{=} \hvuk + \mfe^{-1}\Gl^{-1}(\matt{I} - \Gk)\hv{f}_\text{c}.
\end{align}

In both cases, $\matt{I} - \Gk$ extracts the residual between the original and smoothed spectrum.
Thus, Compton subtraction can be interpreted as augmenting the original solution $\hvuk$ with modulated
residual fluctuations, where the modulation factor is the full energy peak probability. For~\cref{eq:f30},
$\hvuk$ undergoes smearing from a sharp peak to experimental resolution via $\Gl$,
consistent with the definition of $\hvu_\text{sol}$.

Assuming sufficient convergence of the iteration procedure, we have $\vecc{y} = \matt{R}\hvuk + \vecc{\delta}$
for some residual $\vecc{\delta}$. Furthermore, under an additive noise model as in~\cref{sec:discretization},
we have $\vecc{y} = \matt{R}\vmu+\vecc{\epsilon}$, yielding
\begin{equation}
\label{eq:deltaterm}
    \vecc{\delta} = \matt{R}\left(\vmu - \hvuk\right) + \vecc{\epsilon},
\end{equation}
meaning the residual is the sum of the noise $\vecc{\epsilon}$ and of the difference between the true solution $\vmu$ and the
unfolded solution $\hvuk$.
Under the convergence condition,~\cref{eq:f1} becomes
\begin{align}
    \hv{f}_{\text{c}} &= \left(\Gl\matt{D} - \Gl\mfe - \Gl\mpeak\right)\hvuk + \vecc{\delta}\notag\\
    &= \underbrace{\Gl\mcom\hvuk}_{\text{Smeared Compton}} + \vecc\delta,
\end{align}
by the definition of $\matt{D}$. Combining with~\cref{eq:f30} and~\cref{eq:deltaterm} we obtain\footnote{
A third possibility involves unfolding with $\matt{R} = \matt{D}$, which would apparently yield
\begin{align}
    \hvu_\text{sol} &\stackrel{!}{=}\left[\matt{I} + \mfe^{-1}(\matt{I} - \Gk)\mcom\right]\hvuk\notag
    +\mfe^{-1}\left(\matt{I} - \Gk\right)\vecc{\delta}\,.
\end{align}
However, unfolding with $\matt{D}$ alone introduces a commutation error that invalidates both the
folding and this expression; see~\cref{sec:unfoldingspaces}.
}
\begin{align}
\label{eq:f6}
    \hvu_\text{sol} &=\left[\overbrace{\mfe^{-1}\Gl\mfe}^{\text{Full energy component}} + 
    \overbrace{\mfe^{-1}(\matt{I} - \Gk)\Gl\mcom}^{\text{Compton residual}}\right]\hvuk\notag\\
    &\quad+ \underbrace{\mfe^{-1}\left(\matt{I} - \Gk\right)\left[\matt{R}\left(\vmu - \hvuk\right)+\vecc{\epsilon}\right]}_{\text{Difference residual and noise}} .
\end{align}

The Compton subtraction method thus represents an affine transformation of $\hvuk$. It combines a smoothed version of the full energy peak with the residual of the smoothed Compton component, but critically, it also reintroduces both the noise component and the deviation from the true underlying signal.

A fundamental limitation of FICS arises from its convergence behavior. Instead
of converging towards the ideal expectation value $\vmu$, the iteration
converges to a solution $\vecc{x}$ that satisfies the observation equation
$\vecc{y} = \matt{R}\matt{x}$. Because the observed data $\vecc{y}$ inevitably
contain noise, seeking a solution that perfectly fits $\vecc{y}$ leads to
overfitting, where the result conforms to the noise rather than the signal. This
susceptibility to noise overfitting is a well-known characteristic of
unregularized iterative methods like Richardson iteration
(\cref{sec:discretization}), and it persists in FICS despite the mitigation
strategies of early stopping and Compton subtraction.
Indeed, the Compton subtraction step exacerbates this issue by reintroducing
noise via the second and third terms of~\cref{eq:f6}.

This noise-preserving behavior appears consistent with a stated objective of the original method \cite{GUTTORMSEN1996371}:
\begin{quote}\enquote{Our new Compton subtraction method (u) gives a much smoother spectrum with the same fluctuations as the observed spectrum (r).}\end{quote}
Given that earlier methods like the stripping method\cite{Trautmann1982,Radford1987,Love1989,Waddington1989} often yield large fluctuations, this goal of FICS is laudable.
However, preserving the observed spectrum's fluctuations is both ambiguous and fundamentally problematic from an unfolding perspective. As detailed in~\cref{sec:fluctuations}, this terminology conflates two distinct concepts: \emph{stochastic variation} (undesirable noise) and \emph{spectral complexity} (desirable signal features). The primary aim of unfolding is precisely to reconstruct the spectral complexity while rigorously minimizing the influence of stochastic variation. Consequently, the unfolded spectrum \emph{should not} replicate the high-frequency, bin-to-bin variations present in the observed spectrum. Doing so merely ensures that the final result inherits stochastic noise, a direct consequence of the noise reintroduction mechanism demonstrated in~\eqref{eq:f6}.

This critique may appear to contradict numerous studies (e.g., \cite{GUTTORMSEN1996371,PhysRevC.83.034315,PhysRevC.73.064301,PhysRevC.76.044303,PhysRevC.79.024316})
that demonstrate FICS's effectiveness. However, these studies share a common methodological limitation:
they evaluate $\hvu_\text{sol}$ not by statistical comparison to the true solution $\vmu$ or $\veta$,
but by comparing the folded spectrum $\matt{R}\hvu_\text{sol}$ to the raw spectrum $\vecc{y}$.
As detailed in~\cref{sec:discretization,subsec:paramregular}, this approach is inadequate
because infinitely many solutions satisfy $\vecc{y}=\matt{R}\vecc{x}$, all yielding identical
results under folding.

Nevertheless, within the broader context of the Oslo Method, the unfolding appears robust:
forward modeling using input nuclear level density and gamma strength function successfully
recovers both quantities~\cite{PhysRevC.83.034315}. However, this primarily validates
the Oslo Method's reliability rather than FICS specifically. The Oslo Method's robustness
likely stems from the strong physical constraints inherent in the first generation method,
combined with the dimension reduction achieved through decomposition into
unnormalized level density and transmission coefficient. To that end, the FICS solution 
appears sufficient as a point estimate.

\newpage
%\begin{align}
%    \mfe \hvuk_\text{sol} &= \vecc{y} - \matt{G}\vecc{y} + \matt{G}\mfe\hvuk + \matt{G}\mpeak\hvuk - \mpeak\hvuk\\
%    &= (\matt{I} - \matt{G})\vecc{y} + (\matt{G}\mfe - (\matt I-\matt G)\mpeak )\hvuk
%\end{align}

\begin{algorithm}[H]
\caption{Folding Iteration with Compton Subtraction (FICS). The algorithm follows~\cite{GUTTORMSEN1996371} more
closely rather than the main text, as that is what's generally used.}
\label{alg:fics}
\newcommand{\algbf}[1]{\texttt{#1}}
\begin{algorithmic}[1]
\Require
    \Statex \quad Observed spectrum $\vn$
    \Statex \quad Discrete detector response $\matt{D}$
    \Statex \quad Discrete smoothing operator $\operatorname{S}_{\text{D}}$
    \Statex \quad Fluctuation smoothing operator $\operatorname{S}_{\text{fluct}}$
    \Statex \quad Compton smoothing operator $\operatorname{S}_{\text{compton}}$
    \Statex \quad Discrete detector response components
    \Statex \qquad $\mfe, \mse, \mde, \map, \mcb$
    \Statex \quad Fluctuation weight $0 \leq \alpha \leq 1$
    \Statex \quad Maximum iterations limit $k_\text{max}$
\Ensure
    \Statex \quad Unfolded gamma-ray spectrum $\hat{\vecc{u}}_{\text{FICS}}$

\State \algbf{Initialize}
    \Statex \quad Smooth the discrete response $\matt{D}_\text{smooth} \leftarrow \operatorname{S}_{\text{D}}\left(\matt{D}\right)$
    \Statex \quad Set initial guess $\hat{\vecc{u}} \leftarrow \vn$
    \Statex \quad Set iteration counter $k \leftarrow 0$
    \Statex \quad Set previous cost $c_\text{tot} \leftarrow \infty$

\While{not converged}
    \State \algbf{Fold the spectrum}
    \Statex \qquad Compute the folded trial spectrum: 
    \Statex \qquad \(\itk{\hat{\vecc{f}}} \leftarrow \matt{D}_{\text{smooth}}\itk{\hat{\vecc{u}}}\)
    
    \State \algbf{Update the unfolded spectrum}
    \Statex \qquad Update the trial function using the difference between observed and folded spectra:
    \Statex \qquad \(\itat{\hat{\vecc{u}}}{k+1} \leftarrow \itk{\hat{\vecc{u}}} + (\vn - \itk{\hat{\vecc{f}}})\)

    \State \algbf{Compute fluctuation cost}
    \Statex \qquad $\hat{\vecc{u}}_{\text{smooth}} \leftarrow 
    \operatorname{S}_{\text{fluct}}\left(\itat{\hat{\vecc{u}}}{k+1}\right)$
    \Statex \qquad $c_{\text{fluct}} \leftarrow \sum_i |\itat{\hat{u}}{k+1}_i - \left(\hat{u}_{\text{smooth}}\right)_i|$
    \State \algbf{Compute residual cost }\: $\left(\chi^2\right)$
    \State \qquad $\itat{c_{\text{res}}}{k+1} \leftarrow \sum_i \frac{\left(y_i - \itat{\hat{f}}{k+1}_i\right)^2}{y_i}$
    \State \algbf{Compute total cost as a weighted sum}
    \State \qquad $\itat{c_{\text{tot}}}{k+1} \leftarrow \alpha \itat{c_{\text{res}}}{k+1} + (1-\alpha) \itat{c_{\text{fluct}}}{k+1}$
    
    \State \algbf{Check for convergence}
    \Statex \qquad \textbf{If} $ \itat{c_{\text{tot}}}{k+1} > \itat{c_{\text{tot}}}{k+1} $: \textbf{Break}
    \Statex \qquad \textbf{If} $ k+1 > k_{\text{max}}$: \textbf{Break}
    \State \algbf{Update states}
    \Statex \qquad $k \leftarrow k + 1$
\EndWhile

\State \algbf{Compton Subtraction}
\State \quad Estimate the Compton background by removing the other
    components from $\vn$
\Statex \qquad $\vecc{f}_{\text{discrete}} \leftarrow \left(\mfe + \mse + \mde + \map\right)\itk{\hat{\vecc{u}}}$
\Statex \qquad $\vecc{f}_{\text{compton}} \leftarrow \vn -\vecc{f}_{\text{discrete}} $
\State \quad Smooth the estimated Compton
\Statex \qquad $\tilde{\vecc{f}}_{\text{compton}} \leftarrow \operatorname{S}_{\text{compton}}\left(\vecc{f}_{\text{compton}}\right)$
\State \quad Estimate the full-energy component by removing the other
    discrete components and our estimate for Compton from $\vn$
\Statex \qquad $\vecc{u}_{\text{fe}} \leftarrow \vn - \tilde{\vecc{f}}_{\text{compton}} -  \left(\mse + \mde + \map\right)\itk{\hat{\vecc{u}}}$
\State \quad Correct for probability for full energy
\Statex \qquad $\vecc{u}_{\text{fe}} \leftarrow \vecc{u}_{\text{fe}} / \vfe$
\State \quad Correct for detector efficiency
\Statex \qquad $\vecc{u}_{\text{fe}} \leftarrow \vecc{u}_{\text{fe}} / \vecc{\varepsilon}$
\Statex \qquad $\hat{\vecc{u}}_{\text{FICS}} \leftarrow \hat{\vecc{u}}_{\text{fe}}$
\State \textbf{Return}
    \Statex \quad Unfolded gamma-ray spectrum $\hat{\vecc{u}}_{\text{FICS}}$
    
\end{algorithmic}
\end{algorithm}

\FloatBarrier

\section{Notes on implementation}
The RMLE method is implemented using the \texttt{Jax}\cite{jax} Python package, which
provides just-in-time compilation and GPU support, making the optimization
process feasible. We experimented with other methods, such as expectation
maximization, and software packages designed for inverse problems, but they did
not scale well to the millions of variables our problem requires. In contrast, using
\texttt{Jax} allows the optimization to complete in just a few seconds
on a desktop GPU. \texttt{Jax} also targets the CPU. Relatively small unfoldings can be
accomplished on the CPU, but a GPU is strongly recommended for large matrices,
especially for uncertainty propagation.

The complete implementation of the RMLE algorithm is available as open-source
software in the \texttt{OMpy} package\cite{lima_rmle_2025,MIDTBO2021107795}, permanently archived at Zenodo
(doi:\href{https://doi.org/10.5281/zenodo.17594267}{10.5281/zenodo.17594267}).
Although the \texttt{OMpy} package is relatively extensive, the MC RMLE algorithm
itself, as described in this paper, can be implemented in roughly one
hundred lines of \texttt{Jax}-based Python code.

\medskip
%Here is a citation~\cite{Bire82}
%END SECTION
%%%%%%%%%%%%%%%%%%%%%%%%%%%%%%%%%%%%%%%%%%%%%%%%%%%%%%%%%%%%%%%%%%

%END SECTION
%%%%%%%%%%%%%%%%%%%%%%%%%%%%%%%%%%%%%%%%%%%%%%%%%%%%%%%%%%%%%%%%%%

%%%%%%%%%%%%%%%%%%%%%%%%%%%%%%%%%%%%%%%%%%%%%%%%%%%%%%%%%%%%%%%%%%
%BIBLIOGRPHY
\FloatBarrier
\bibliography{biblio}

\end{document}